%% file: xfelo_25Mar2019.tex
\documentclass[aps,twocolumn,floatfix,superscriptaddress,showpacs]{revtex4-1}
\usepackage{graphicx}
\usepackage{amssymb,amsmath}

\usepackage{nicefrac}

\begin{document}
\widetext

\title{Scientific Opportunities with an X-ray Free-Electron Laser Oscillator}

\author{Bernhard Adams} 
\affiliation{Incom Incorporated, 294 Southbridge Road, Charlton, MA 01507, USA}

\author{Gabriel Aeppli}
\affiliation{Paul Scherrer Institut, 5232 Villigen PSI, Switzerland}
\affiliation{Department of Physics, ETH Z\"urich, Z\"urich CH-8093, Switzerland}
\affiliation{Institut de Physique, EPFL, Lausanne CH-1015, Switzerland}

\author{Thomas Allison}
\affiliation{Stony Brook University, Stony Brook, NY 11794, USA}

\author{Alfred Q. R. Baron}
\affiliation{Materials Dynamics Laboratory, RIKEN SPring-8 Center, RIKEN, 1-1-1 Kouto, Sayo, Hyogo 679-5148 Japan}

\author{Phillip Bucksbaum} 
\affiliation{Stanford PULSE Institute, SLAC National Accelerator Laboratory, Menlo Park, California 94025, USA }
\affiliation{Department of Physics, Stanford University, Stanford, California 94305, USA }
\affiliation{Department of Applied Physics, Stanford University, Stanford, California 94305, USA}

\author{Aleksandr I. Chumakov} 
\affiliation{ESRF - The European Synchrotron, 71 avenue des Martyrs, CS 40220, 38043 Grenoble Cedex 9, France}

\author{Christopher Corder}
\affiliation{Stony Brook University, Stony Brook, NY 11794, USA}

\author{Stephen P. Cramer} 
\affiliation{Department of Chemistry, UC Davis, One Shields Ave, Davis, CA 95616, USA}

\author{Serena DeBeer} 
\affiliation{Max Planck Institute for Chemical Energy Conversion, Stiftstr. 34-36, 45470 M\"ulheim an der Ruhr, Germany}

\author{Yuntao Ding}
\affiliation{SLAC National Accelerator Laboratory, 2575 Sand Hill Road, Menlo Park, CA 94025, USA}

\author{J\"org Evers} 
\affiliation{Max-Planck-Institut f\"ur Kernphysik, Saupfercheckweg 1, 69117 Heidelberg, Germany}

\author{Josef Frisch} 
\affiliation{SLAC National Accelerator Laboratory, 2575 Sand Hill Road, Menlo Park, CA 94025, USA}

\author{Matthias Fuchs} 
\affiliation{Department of Physics and Astronomy, University of Nebraska - Lincoln,  Lincoln, NE 68588, USA}

\author{Gerhard Gr\"ubel}
\affiliation{Deutsches Elektronen-Synchrotron DESY, Notkestr. 85, 22607 Hamburg, Germany}
\affiliation{The Hamburg Centre for Ultrafast Imaging, Luruper Chaussee 149, 22761 Hamburg, Germany}

\author{Jerome B. Hastings}
\affiliation{SLAC National Accelerator Laboratory, 2575 Sand Hill Road, Menlo Park, CA 94025, USA}

\author{Christoph M. Heyl}
\affiliation{Deutsches Elektronen-Synchrotron DESY, Notkestr. 85, 22607 Hamburg, Germany}
\affiliation{JILA, National Institute of Standards and Technology and University of Colorado, Department of Physics, 440 UCB, Boulder, Colorado 80309, USA }
\affiliation{Department of Physics, Lund University, PO Box 118, 22100 Lund, Sweden}

\author{Leo Holberg}
\affiliation{Department of Physics, Stanford University, Stanford, California 94305, USA }

\author{Zhirong Huang}
\affiliation{SLAC National Accelerator Laboratory, 2575 Sand Hill Road, Menlo Park, CA 94025, USA}

\author{Tetsuya Ishikawa}
\affiliation{RIKEN SPring-8 Center, Kuoto 1-1-1, Sayo, Hyogo 679-5148, Japan}

\author{Andreas Kaldun}
\affiliation{Stanford PULSE Institute, SLAC National Accelerator Laboratory, Menlo Park, California 94025, USA }

\author{Kwang-Je Kim}
\affiliation{Advanced Photon Source, Argonne National Laboratory, Argonne, Illinois 60439, USA}

\author{Tomasz Kolodziej}
\affiliation{National Synchrotron Radiation Centre SOLARIS, Jagiellonian University, 30-392 Krak\'ow, Poland}

\author{Jacek Krzywinski}
\affiliation{SLAC National Accelerator Laboratory, 2575 Sand Hill Road, Menlo Park, CA 94025, USA}

\author{Zheng Li}
\affiliation{Stanford PULSE Institute, SLAC National Accelerator Laboratory, Menlo Park, California 94025, USA }
\affiliation{Center for Free-Electron Laser Science, DESY, Notkestr. 85, 22607 Hamburg, Germany}

\author{Wen-Te Liao}
\affiliation{Department of Physics, National Central University, Taoyuan City 32001, Taiwan}

\author{Ryan Lindberg}
\affiliation{Advanced Photon Source, Argonne National Laboratory, Argonne, Illinois 60439, USA}

\author{Anders Madsen}
\affiliation{European XFEL GmbH, Holzkoppel 4, 22869 Schenefeld, Germany}

\author{Timothy Maxwell}
\affiliation{SLAC National Accelerator Laboratory, 2575 Sand Hill Road, Menlo Park, CA 94025, USA}

\author{Giulio Monaco}
\affiliation{Department of Physics, University of Trento, Via Sommarive 14, 38123 Povo, Italy}

\author{Keith Nelson}
\affiliation{Department of Chemistry, Massachusetts Institute of Technology, 77 Mass Avenue, Cambridge, MA 02139, USA}

\author{Adriana P\'alffy}
\affiliation{Max-Planck-Institut f\"ur Kernphysik, Saupfercheckweg 1, 69117 Heidelberg, Germany}

\author{Gil Porat}
\affiliation{JILA, National Institute of Standards and Technology and University of Colorado, Department of Physics, 440 UCB, Boulder, Colorado 80309, USA }

\author{Weilun Qin}
\affiliation{IHIP, Peking University, Beijing 100871, China}

\author{Tor Raubenheimer}
\affiliation{SLAC National Accelerator Laboratory, 2575 Sand Hill Road, Menlo Park, CA 94025, USA}

\author{David A. Reis}
\affiliation{Stanford PULSE Institute, SLAC National Accelerator Laboratory, Menlo Park, California 94025, USA }
\affiliation{Department of Applied Physics, Stanford University, Stanford, California 94305, USA}

\author{Ralf R\"ohlsberger}
\affiliation{Deutsches Elektronen-Synchrotron DESY, Notkestr. 85, 22607 Hamburg, Germany}
\affiliation{The Hamburg Centre for Ultrafast Imaging, Luruper Chaussee 149, 22761 Hamburg, Germany}

\author{Robin Santra}
\affiliation{Center for Free-Electron Laser Science, DESY, Notkestr. 85, 22607 Hamburg, Germany}
\affiliation{Department of Physics, University of Hamburg, Jungiusstr. 9, 20355 Hamburg, Germany}
\affiliation{The Hamburg Centre for Ultrafast Imaging, Luruper Chaussee 149, 22761 Hamburg, Germany}

\author{Robert Schoenlein}
\affiliation{SLAC National Accelerator Laboratory, 2575 Sand Hill Road, Menlo Park, CA 94025, USA}

\author{Volker Sch\"unemann}
\affiliation{Department of Physics, Technische Universit\"at Kaiserslautern, 67663 Kaiserslautern, Germany}

\author{Oleg Shpyrko}
\affiliation{Department of Physics, University of California, San Diego 9500 Gilman Drive, La Jolla, CA 92093, USA}

\author{Yuri Shvyd'ko}
\affiliation{Advanced Photon Source, Argonne National Laboratory, Argonne, Illinois 60439, USA}

\author{Sharon Shwartz}
\affiliation{Department of Physics, Bar-Ilan University, Ramat-Gan, 52900, Israel}

\author{Andrej Singer}
\affiliation{Department of Physics, University of California, San Diego 9500 Gilman Drive, La Jolla, CA 92093, USA}

\author{Sunil K. Sinha}
\affiliation{Department of Physics, University of California, San Diego 9500 Gilman Drive, La Jolla, CA 92093, USA}

\author{Mark Sutton}
\affiliation{The Centre for the Physics of Materials, (CPM) and the Physics Department, McGill University, 3600 rue University, Montréal (Qu\'ebec) H3A 2T8, Canada}

\author{Kenji Tamasaku}
\affiliation{RIKEN SPring-8 Center, Kuoto 1-1-1, Sayo, Hyogo 679-5148, Japan}

\author{Hans-Christian Wille}
\affiliation{Deutsches Elektronen-Synchrotron DESY, Notkestr. 85, 22607 Hamburg, Germany}

\author{Makina Yabashi}
\affiliation{RIKEN SPring-8 Center, Kuoto 1-1-1, Sayo, Hyogo 679-5148, Japan}

\author{Jun Ye}
\affiliation{JILA, National Institute of Standards and Technology and University of Colorado, Department of Physics, 440 UCB, Boulder, Colorado 80309, USA }

\author{Diling Zhu}
\affiliation{SLAC National Accelerator Laboratory, 2575 Sand Hill Road, Menlo Park, CA 94025, USA}

\date{\today}

\vspace*{8mm}

\begin{abstract}
An X-ray free-electron laser oscillator (XFELO) is a new type of hard X-ray source that would produce fully coherent pulses with meV bandwidth and stable intensity.  The XFELO complements existing sources based on self-amplified spontaneous emission (SASE) from high-gain X-ray free-electron lasers (XFEL) that produce ultra-short pulses with broad-band chaotic spectra. This report is based on discussions of scientific opportunities enabled by an XFELO during a workshop held at SLAC on June 29 - July 1, 2016 \cite{Kolodziej2016}. 
\end{abstract}

\maketitle


\section{Introduction}

\begin{figure}[b]
\begin{center}
     \includegraphics[width=\columnwidth]{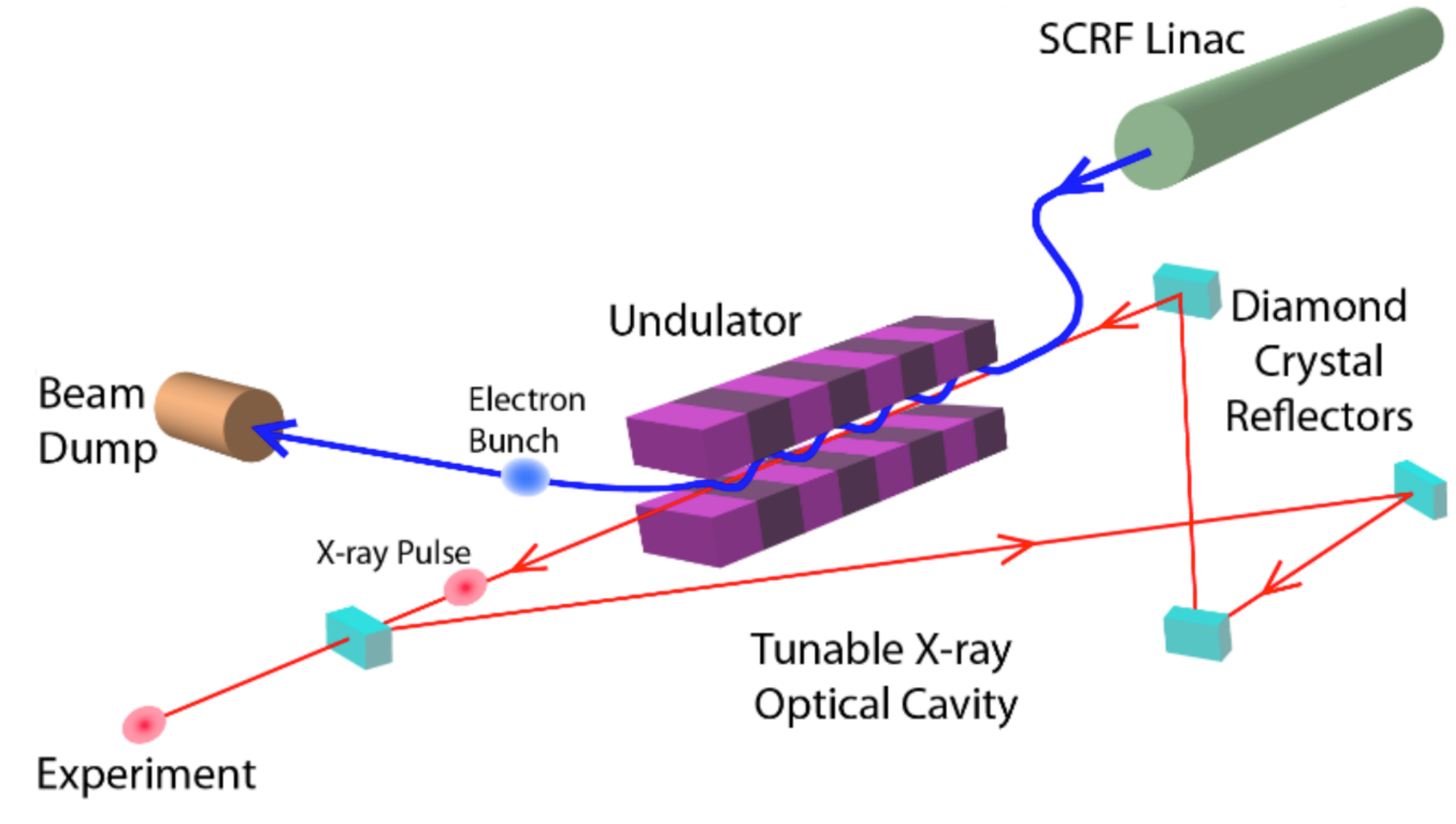}
\end{center}
\caption{A schematic illustration of an XFELO. Four crystals form a closed X-ray cavity via Bragg reflection.}
\label{Fig1}
\end{figure}

An XFELO is a low-gain device, in which an X-ray pulse that circulates in a cavity formed by diamond crystal Bragg mirrors is amplified every time it overlaps with an electron bunch in the undulator, as illustrated in Fig.\,\ref{Fig1} \cite{Kim2008,Kim2009,Lindberg2011}. Due to its high reflectivity and excellent thermo-mechanical properties, diamond is the preferred material for the Bragg crystals employed to form the X-ray cavity \cite{Shvydko2017}. An XFELO will work at any photon energy for which the Bragg reflectivity of diamond is sufficiently high and the bandwidth is sufficiently broad so that the initial exponential gain of the intra-cavity pulse energy can be sustained for a reasonable set of electron beam and undulator parameters. This range is expected to extend from 5 to 25 keV \cite{Lindberg2011}.  The photon energy can be continuously tuned for a given setting within a range of about 5$\%$ by changing the Bragg angle and adjusting the crystal positions so that the cavity roundtrip time remains fixed.  

The defining property of an XFELO is the narrow spectral bandwidth, which can be as small as a few meV, whereby the (almost transform-limited) temporal duration of the XFELO pulses is about one ps. In contrast, the self-amplified spontaneous emission (SASE) from high-gain X-ray FELs \cite{Kondratenko1980,Bonifacio1984,Huang2007} is characterized by the ultra-short temporal duration, which could be as small as a fraction of a femtosecond, while the spectral bandwidth of SASE pulses is a few tens of eV.  Thus, an XFELO has a high spectral photon density, while an X-ray SASE FEL has a high temporal photon density. Additionally, an XFELO produces fully coherent and highly stable pulses, while SASE pulses are spectrally chaotic with significant pulse-to-pulse fluctuations in intensity and spectrum.  In view of the progress in beam dynamics and crystal optics needed for an XFELO \cite{Kim2012}, together with the recent demonstration that diamond crystals survive the high intra-cavity X-ray power density \cite{Kolodziej2018}, an XFELO appears to be technically feasible now.

Several high-gain SASE X-ray FELs have been operating since 2009 when the LCLS was successfully commissioned \cite{Emma2010}. The second XFEL worldwide, SACLA in Japan, commenced its operation in 2011 \cite{Ishikawa2012}. The European XFEL \cite{Altarelli2007a,Madsen2017,Weise2017,Tschentscher2017} is the first hard X-ray FEL based on a high-energy super-conducting electron linac with MHz repetition rate, and it is expected that several machines of this type will become available within the next decade \cite{Schoenlein2017,Zhu2017,Sekutowicz2013}.  Although these facilities are currently intended for SASE applications, an XFELO branch could significantly widen their scientific scope.  

\subsection{Basic performance of an XFELO}
\begin{figure}
\begin{center}
   \includegraphics[width=\columnwidth]{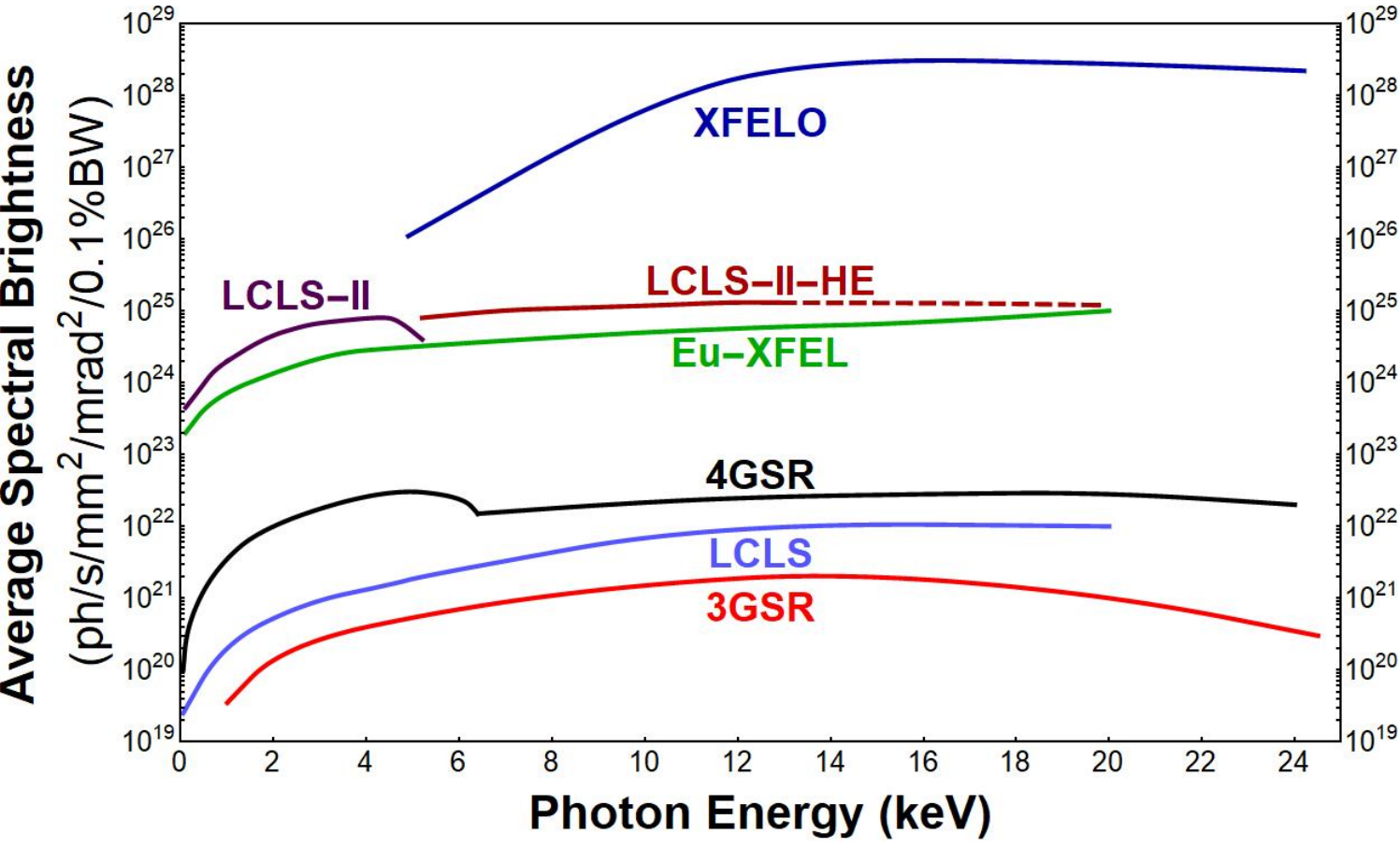}
\end{center}
\caption{Brightness of XFELO \cite{Qin2017a} and other advanced X-ray sources.  The curves for LCLS and 4GSR (4th generation (ultralow emittance) storage ring, 2 - 6 GeV electron energy, 0.2 - 1.1 km circumference) are from Ref. \cite{Schoenlein2017}.   
}
\label{Fig2}
\end{figure}

The spectral brightness of an XFELO compared to other advanced X-ray sources is shown in Fig.\,\ref{Fig2}. The XFELO is assumed to be driven by an 8 GeV, 1 MHz superconducting linac, as planned for the LCLS-II-HE \cite{Schoenlein2017} and other projects \cite{Zhu2017}, with an optimized injector \cite{Qin2017a}.  The linac system for the European XFEL can drive a pulsed XFELO, or can in principle be converted for CW operation \cite{Sekutowicz2013}.
 
 As an example, at 14.4 keV (the transition energy of the $^{57}$Fe M\"ossbauer resonance) the number of photons per pulse is expected to be 1$\times 10^{10}$ within a pulse length of 680 fs (FWHM) and a spectral bandwidth of 3 meV (FWHM). This amounts to an average spectral flux of about 3$\times 10^{15}$ photons/sec/meV or 1.5$\times 10^{10}$ photons/sec/$\Gamma_0$, which corresponds to 1.5$\times 10^{4}$ photons/pulse/$\Gamma_0$ where $\Gamma_0$ = 4.7 neV is the natural linewidth of the $^{57}$Fe M\"ossbauer resonance. These numbers are four orders of magnitude larger than those observed at the best third-generation synchrotron radiation sources to date.

The bandwidth of an XFELO closely corresponds to the Fourier transform limit and therefore the pulse is expected to be temporally coherent, without the complicated spike structure characteristic of SASE spectra. In order to compare intensities with those of other sources in commonly used units, the XFELO peak and average spectral fluxes are 7 $\times10^{25}$ and 5 $\times 10^{19}$ [ph/(s\,0.1$\%$ BW)], respectively. The pulses are transversely coherent, so the spectral brightness may be obtained by dividing the spectral flux by the coherent phase space area, $\lambda$/2, for each dimension.  The peak and average spectral brightnesses are then 3.8$\times$10$^{34}$ and 2.6$\times 10^{28}$ [ph/(s\,mm$^2$\,mrad$^2\,0.1\%$BW], respectively.

The XFELO brightness is higher than SASE mainly due to the narrower spectral bandwidth of XFELO:  SASE at LCLS-II-HE will produce $\sim 5 \times 10^{10}$ photons per pulse at 13 keV, five times more than XFELO, but with a relative bandwidth of 0.06$\%$, broader by a factor of 2.5$\times 10^3$ compared to XFELO \cite{Schoenlein2017}. The average spectral brightness of an XFELO is three orders of magnitude higher than SASE, assuming the same repetition rate for both. The very large bandwidth ratio means that even the peak brightness of an XFELO is an order of magnitude higher than that of SASE, despite the shorter pulse length of SASE. The dashed line in Fig.\,\ref{Fig2} indicates that the SASE could be extended to 20 keV if the electron beam emittance of LCLS-II-HE can be improved.  The European XFEL with its high electron energy (17.5 GeV) can cover this range. Beyond 25 keV, we can compare the XFELO with the undulator radiation from the 4th generation (diffraction limited) storage rings (4GSR) employing the multi-bend achromat (MBA) lattice \cite{Cai2012}. We find that XFELO brightness is more than five orders of magnitude higher than that of a hard x-ray 4GSR. 

For comparison, the SASE brightness can be improved significantly with the self-seeding scheme \cite{Geloni2011,Amann2012}. A two-stage self-seeding XFEL with a tapered undulator and a 17.5 GeV pulsed superconducting linac at European XFEL can produce an average brightness within a factor of ten smaller than that of the XFELO at 9 keV \cite{Chubar2016}. However, its performance drops steeply at higher photon energy; at 14.4 keV the brightness is less than that of the XFELO by a factor of 500. In addition, the pulse-to-pulse intensity fluctuations cannot be removed in a self-seeded case. 

The XFELO spectral brightness is larger by about seven orders of magnitude compared to that of existing third-generation storage rings (3GSR), the brightness of which is about 10$^{20}$ - 10$^{21}$\,[ph/(s\,mm$^2$\,mrad$^2\,0.1\%$BW)] in the hard X-ray region. The spectral flux is about four orders of magnitude higher for an XFELO than that of the existing 3GSRs. 

We emphasize two significant features of the XFELO:  (1) The pulse intensity is much more stable than in case of a high gain SASE source with shot-to-shot fluctuations less than 1$\%$. This is because the statistical fluctuations inherent in the SASE process are absent in an XFELO and the pulse fluctuation originating from the electron bunch fluctuation is reduced since an XFELO output pulse is formed by the combined action of many, $>\sim$100, electron bunches; (2) The XFELO source is intrinsically narrow ($\sim$ meV) bandwidth, and is therefore well matched to high-resolution experiments. This drastically reduces the heat-load on the optics as compared to filtering the bandwidth from a broader-band SASE source.

An XFELO may also be possible in a bypass of an ultralow-emittance storage ring that is diffraction limited at energies of hard X-rays \cite{Lindberg2013}.  Large-circumference storage rings such as PEP-X and PETRA IV with MBA lattices can produce the required low emittance. However, a storage-ring based XFELO must be operated in a pulsed mode, thus reducing the effective repetition rate by two orders of  magnitude compared to a linac based one. First studies in this direction look promising and will certainly be featured on one of the next workshops dedicated to XFELO science and technology.

\subsection{Extended capabilities}
\label{subsec:extended}

The basic XFELO scheme can in principle be extended, for example, to reach higher power, higher photon energies, and, if sufficiently stabilized, even for generating frequency combs at the energies of hard x-rays, as described in the following.

\subsubsection{Amplification and harmonic generation}
\label{sec:mopa}
 
\begin{figure}
\begin{center}
    \includegraphics[width=\columnwidth]{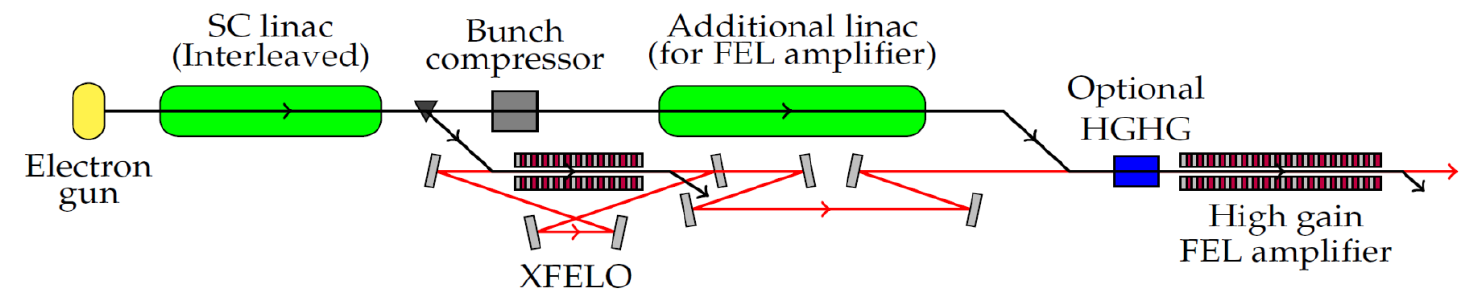}
\end{center}
\caption{A concept for an integrated XFEL oscillator/amplifier facility. The e-gun produces two types of interleaved electron bunches, one for XFELO and one for XFEL amplifier. These bunches are interleaved and accelerated in a super-conducting linac and then separated into the oscillator branch and the amplifier branch. The latter bunches are compressed and may be accelerated further in an additional linac section. A part of the XFELO output pulses serves experiments requiring high-coherence and $\sim$meV bandwidth. The other part of the XFELO output pulses are delayed and overlap with the compressed amplifier bunches to produce $\sim$eV bandwidth X-rays. An optional harmonic generation stage before the amplifier can extend the spectral coverage to higher photon energies.} 
\label{Fig3}
\end{figure}

With the basic XFELO scheme shown in Fig.\,\ref{Fig1}, photon energies up to about 25 keV can be covered.  However, the photon energy coverage can be extended in several ways.  One way is to run the XFELO at higher harmonics \cite{Dai2012}. Another way is to use the XFELO output as an input seed to a high-gain harmonic generation (HGHG) system \cite{Yu1991} consisting of a modulator, a magnetic chicane, and an amplifier at a harmonic photon energy, as illustrated in Fig.\,\ref{Fig3}.  The HGHG section will be driven by another electron beam with higher energy and higher current. Such a system was studied for extending the photon energy to 40-60 keV \cite{Qin2017b}. The interest in such high energy photon beams has been raised by the MaRIE project \cite{Sheffield2017}.  The HGHG configuration can also be operated at the fundamental yielding a master oscillator-power amplifier (MOPA) configuration.  By driving the amplifier section with an ultrashort electron beam, the MOPA can produce ultrashort pulses comparable to SASE from, for example, LCLS, except that the pulse will be fully coherent and without fluctuation in intensity. The tapering section in the amplifier will work well due to the high degree of coherence.  
 
\subsubsection{X-ray comb generation}

\begin{figure}
\begin{center}
    \includegraphics[width=\columnwidth]{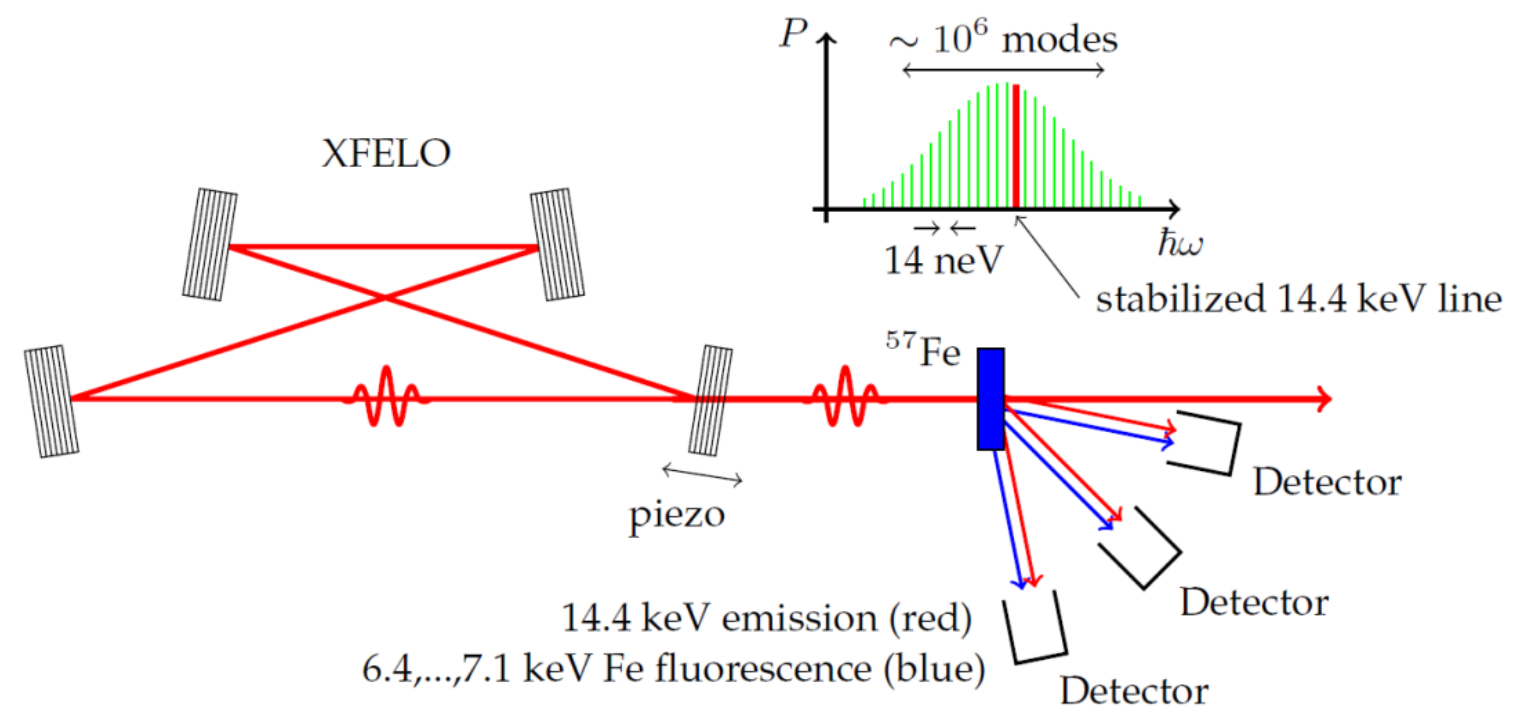}
\end{center}
\caption{Schematic of the cavity-stabilization scheme. A nuclear-resonant sample (here
$^{57}$Fe) is placed into the XFELO output, and the nuclear-resonant and K-shell electronic
fluorescence are monitored as function of cavity tuning with a piezoelectric actuator.
A feedback loop keeps one of longitudinal modes of the XFELO on resonance with the
sample. Adopted from \cite{Adams2015}}
\label{Fig4}
\end{figure}

It should be possible to phase lock the successive XFELO output via stabilizing the X-ray cavity to an auxiliary laser beam. An alternative approach would be to reference the X-ray cavity to a narrow nuclear resonance line such as the 14.4 keV transition in $^{57}$Fe \cite{Adams2015}, as shown schematically in Fig.\,\ref{Fig4}. The XFELO output will then form an X-ray spectral comb. The comb consists of about one million spectral lines of a few nano-eV bandwidth, each containing about 1000 photons per pulse. Further discussion and applications of X-ray comb generation are given in section \ref{sec:xraycomb}.

\subsection{Science with an XFELO}

The unique characteristics of an XFELO will provide several levels of scientific opportunities. It is useful to categorize the scientific applications of an XFELO as follows, although the boundaries between the categories are somewhat fluid:

\subsubsection{Extending the parameter space of established techniques}

Inelastic X-ray scattering (IXS) has been successful at third generation facilities in exploring the various states of matter. With an XFELO providing about 4 orders of magnitude higher spectral flux (7 orders in brightness) than the existing third generation sources, the reach of the technique could be greatly increased in directions that had previously only been dreamed of, e.g. thoroughly investigating materials in extreme conditions such as found at the center of the earth, and beyond, determining the phonon dispersion in complex materials of practical importance, and a wide range of strongly correlated systems, with unprecedented resolution in energy and momentum transfer.

Nuclear resonance scattering (NRS) experiments \cite{Gerdau1999,Gerdau2000,Roehlsberger2004} are typically limited by the counting statistics due to the low spectral flux at the resonance frequency. The XFELO will enhance the power of NRS tremendously by providing a 4 order-of-magnitude increase in spectral flux, enabling the study of complex systems with excellent energy resolution, in particular to address important problems in dilute iron containing systems in chemistry, biology and materials science, in particular when isotopic enrichment is not possible.

The combination of high spectral brightness and high stability will enable X-ray photon correlation spectroscopy (XPCS) \cite{Madsen2015} to be extended to larger angles (shorter length scales) and shorter time-scales.  This will allow access to dynamics at nanometer length scales in disordered materials as well as higher-order correlations in weakly ordered systems, e.g. characterizing short-range ordering and preferential orientations. 

\subsubsection{Broadening the user community}

An XFELO can stimulate interest in communities that have not been frequent users of the third generation synchrotron radiation facilities. For example, XFELO based M\"ossbauer spectroscopy will enable the biological community to monitor iron-centered reactions inside cells, providing important insights into many aspects of infection pathways, for example.

The interest in basic nonlinear optical phenomena at X-ray wavelengths, such as second harmonic generation and parametric down conversion, has significantly increased in recent years.  A XFELO can elevate these studies from demonstration experiments to tools of practical importance, e.g., to open the study of squeezed and entangled states at x-ray energies. Entangled photonic states may lead to radiation damage free studies of protein structure by ghost imaging.  Further, the XFELO should provide access to higher order correlation functions in space and time enabling the elucidation of details of electronic correlations in materials of practical importance.

\subsubsection{Opening of new fields}

An XFELO may enable the establishment of concepts in quantum optics and extreme metrology in the x-ray regime that are unthinkable presently. 
Stabilizing the pulse-to-pulse distance to a small fraction of the wavelength would give rise to an X-ray spectral comb consisting of a sequence of neV spectral peaks (see section \ref{sec:xraycomb}).  It could provide unprecedented X-ray metrology for fundamental physics. It could also revolutionize nuclear solid state physics, ultimately enabling 2-dimensional nuclear resonant spectroscopies, just as the optical comb did for atomic physics. The possibility of pulse-to-pulse coherence of x-ray pulses from a XFELO will thus open a new era for phase-coherent probing and steering of dynamical processes in atoms, molecules and solids. 

The remainder of the paper is organized into the following sections, each devoted to a particular method and its scientific applications that will most prominently benifit from an XFELO: section \ref{sec:IXS} presents the case for inelastic X-ray scattering, section \ref{sec:XPCS} for X-ray photon correlation spectroscopy, section \ref{sec:NRS} for nuclear resonance scattering, section \ref{sec:NLO} for non-linear optics and section \ref{sec:xraycomb} for the X-ray spectral comb. Section \ref{sec:conclusion} features a discussion and conclusion.   

\section{Inelastic x-ray scattering (IXS)}
\label{sec:IXS}

The potential of the XFELO for future high-resolution inelastic X-ray scattering (IXS) cannot be overstated: meV-resolution IXS experiments, which are often flux limited at present synchrotron radiation sources, will see an improvement in spectral flux at the XFELO by about four orders of magnitude. Naively, this reduces a 1-day measurement at present facilities to less than a minute, and leads one to consider extreme extensions of present work.   However, more so, such a dramatic increase means completely new approaches will become possible, and while we can, and do, speculate about such methods, in fact unanticipated applications can have huge impact. Thus, here we only scratch the surface of what may become possible.

\begin{figure}
\begin{center}
    \includegraphics[width=\columnwidth]{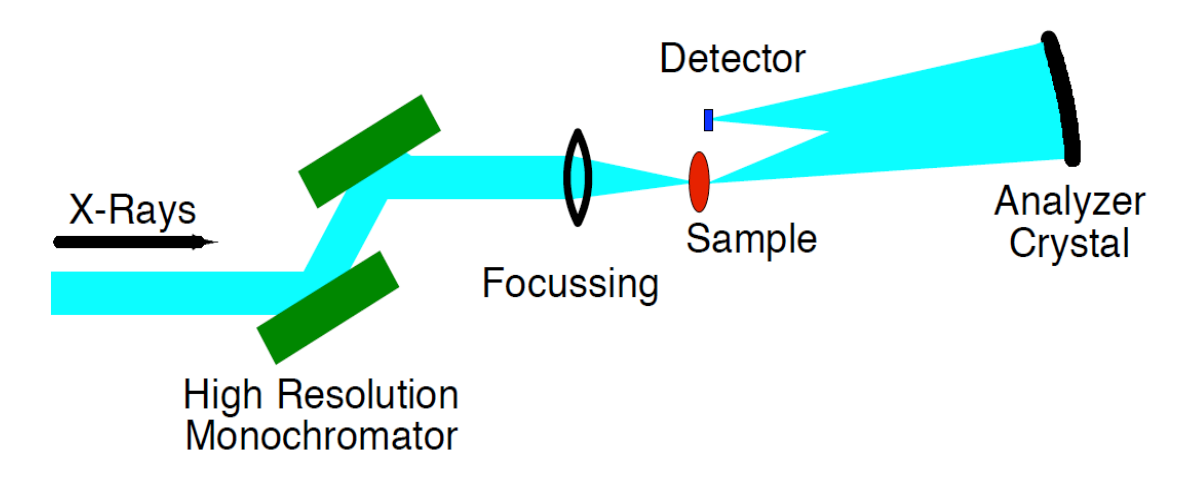}
\end{center}
\caption{Schematic of a typical spectrometer for the types of IXS experiments discussed in this section.  Energy scans are usually done by introducing a relative change between incident beam and the analyzer, allowing measurement of $S(Q,\omega)$ or related quantities.  Note that in some cases the analyzer crystal might be replaced by a nuclear analyzer, or even nuclear absorption within the sample, or the geometry may change from a spherical analyzer to some sort of post-sample collimation setup with flat analyzers.}
\label{Fig5}
\end{figure}

IXS, as a field, covers a large (and growing) range of science and techniques, see \cite{Schuelke2007,Baron2015}, providing often unique access to equilibrium materials properties and excited state structure. In the present section, we focus on spectrometer-based IXS applications where measurements take place in a "triple axis" (monochromator-sample-analyzer) or similar geometry (see Fig.\,\ref{Fig5}), with energy domain analysis.  This leads to measurement of the dynamic structure factor, $S(Q,\omega)$, or a closely related quantity.  We emphasize this style of investigation as it is well suited to the narrow bandwidth of the XFELO.   In particular, noting that the intrinsic bandwidth is expected to be $\sim$5 meV, we emphasize experiments using similar, or better, resolution. 

\subsection{Comparison of XFELO and SASE for IXS applications}

\begin{figure}
\begin{center}
    \includegraphics[width=\columnwidth]{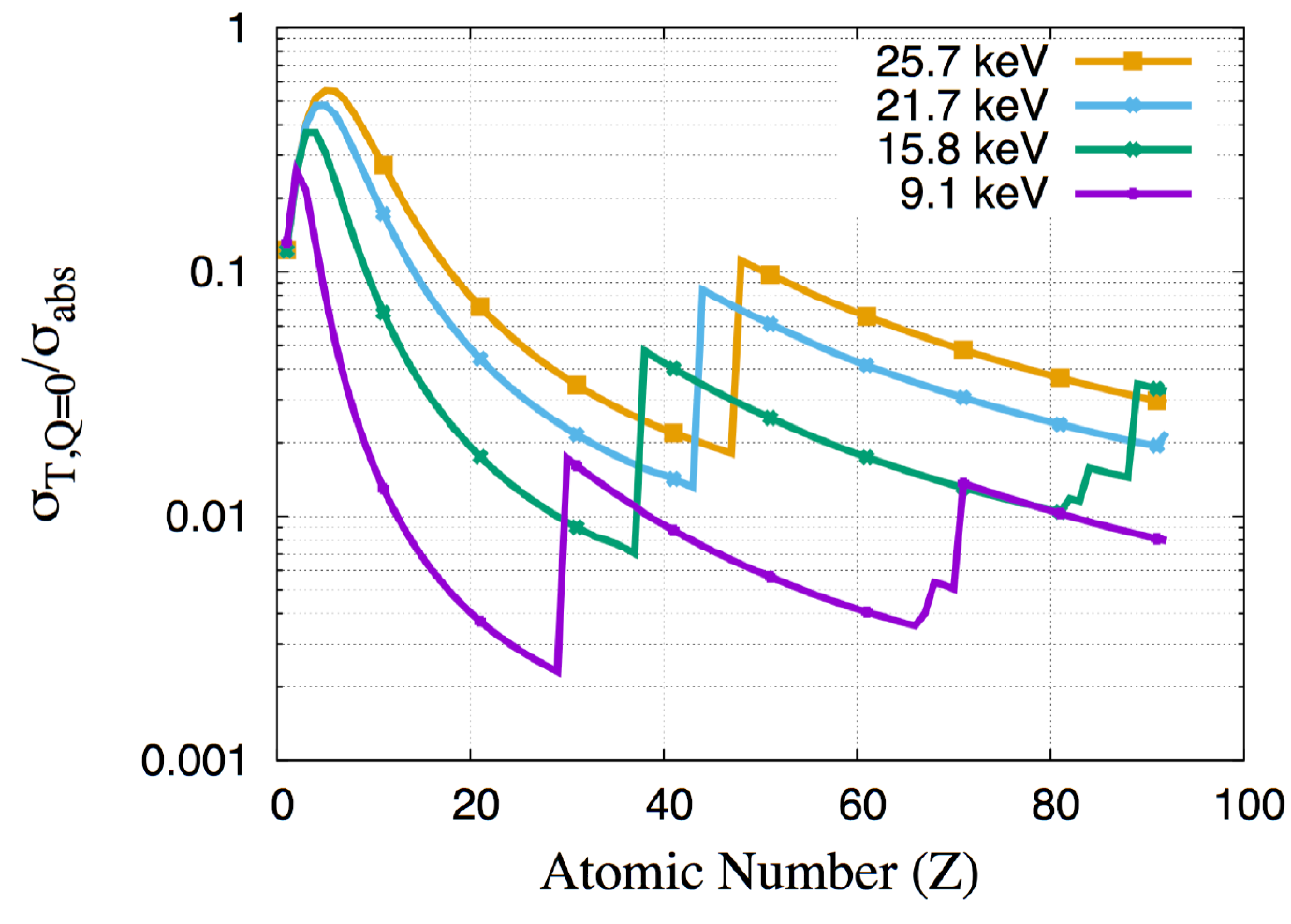}
\end{center}
\caption{Ratio of the Thomson scattering cross section to the photoelectric absorption cross section as a function of atomic number for several X-ray energies.  Higher numbers indicate improved scattering and (up to a factor of energy) reduced radiation damage per scattering event.  There is generally strong improvement at higher energy.  When coupled with the improved ability to penetrate into sample environments, access to more resonances and larger momentum space, there is a very strong case for attaining higher energies, as is possible at the XFELO.  (One notes similar issues have pushed synchrotron X-ray crystallography to higher, $\sim$\,30 keV, energies when reasonably possible.)}
\label{Fig6}
\end{figure}

The main parameter of interest for most IXS work is exactly the spectral density, so, to a first approximation, some of the applications we discuss here for an XFELO are possible, at least in principle, with SASE.   However, with its much higher (2 - 3 orders of magnitudes) spectral flux and higher spectral brightness, the XFELO is preferable for IXS, and, indeed, for many X-ray scattering experiments.  The extension from the $\sim$13 keV limit expected for SASE from LCLS-II-HE, to $>\sim\,$25 keV as expected for the XFELO, will, in nearly all experiments, improve signal rates, reduce radiation damage, and improve the ability to penetrate into sample environments (see Fig.\,\ref{Fig6}).   The extended energy range also will make available many more edges for resonant scattering experiments, and nuclei for nuclear-resonant scattering experiments, especially with access to the 14.4 keV resonance of $^{57}$Fe. The extended energy range also allows significantly more flexibility in optical schemes for non-resonant inelastic x-ray scattering \cite{Schuelke2007}, permitting one to directly carry over some of the most productive methods from SR sources.  Also, the XFELO will generally be superior to the XFEL as it provides better stability (without the fluctuations typical of the SASE) and a smaller bandwidth that is, at the start, more appropriate for high-resolution experiments.  This is worth emphasizing: if one needs a few-meV beam, as is discussed for experiments below,  then to get the same spectral intensity from a seeded SASE source (such as the European XFEL or LCLS-II-HE) one must start with 3 orders of magnitude more X-ray power.  In an environment where even factors of two in power can make critical differences in X-ray optics designs, the benefit of a factor of $\sim$1000 improvement cannot be overstated.

The discussion here will be separated into scientific targets and methodological ones. These viewpoints are complementary, as the targeted approach allows easier connection to a broader scientific audience while the methodological approach facilitates extension to other experiments.

\subsection{Scientific goals}

\subsubsection{A window into the center of the Earth: Composition, Temperature, Formation}

The most direct information about the deep interior of the earth comes through seismic measurements. These are interpreted in terms of a standard model that provides the density and sound velocity inside the earth \cite{Dziewonski1981}.  However, it is crucial to interpret that model in terms of chemical composition and temperature, which are presently unknown in detail: temperature uncertainties approach 20$\%$ and while it is generally agreed that there are lighter elements in the predominantly iron core, which elements remains a matter of discussion as the result has significant implications for planetary formation.   IXS is now being used, worldwide, to generate a catalogue of measured sound velocities as a function of temperature and density in extreme, high-pressure and high-temperature, conditions.   The most extreme conditions achieved to date are about half of those in the Earth's center \cite{Nakajima2015,Sakamaki2016}.   As pressure and temperature increase, these measurements become increasingly difficult: sample sizes decrease, and it becomes hard to stabilize the samples at desired temperatures ($\sim$ 6000K or more) over the few hour time scales needed for IXS at current instruments.  However, smaller samples and faster scans with improved data will become straightforward at the XFELO.  In addition, measurements will be possible with improved energy resolution  ($\sim$0.1 meV, or better), as is highly desirable to investigate the viscosity of the liquid outer core. This will advance understanding of the structure and formation of the earth, other planets, and, indeed, the evolution of the solar system.

\subsubsection{Probing inter-system coupling to comprehend complex materials}

Modern materials science leverages inter-system interaction to optimize functional materials - taking advantage of the interactions between lattice, electronic, and magnetic degrees of freedom to control material response or create new properties.  Examples include using structure or lattice properties to modify/control electronic, thermal or dielectric properties (e.g. superconductivity, thermoelectricity, ferroelectricity) or using electronic response to modify/control magnetic behavior (multiferroicity).  Thus, it is necessary to understand the interaction of atomic, orbital and spin, structure and dynamics, to tailor materials properties, such as higher temperature superconductivity, and to develop new recording media as well as more efficient materials for batteries and thermoelectric applications. An XFELO, with the potential to measure IXS with extremely high energy resolution in the range of 10 to 100 $\mu$eV, offers the possibility to directly investigate coupling to the lattice as is manifested in phonon line-widths, see e.g. \cite{Aynajian2008}.   Further, high-resolution resonant IXS (RIXS) will allow complementary investigations of electron-phonon coupling \cite{Ament2011,Deveraux2016}.  Also precise phonon intensity measurements (using non-resonant scattering) across multiple Brillouin zones can allow for the identification of changes in electronic structure induced by phonon motion, an important aspect of electron-phonon coupling.  Finally, we note that one can directly probe electronic excitations (d-d, crystal field, band-structure, band gaps) by non-resonant IXS, and both electronic and magnetic excitations using high-resolution resonant scattering.  All these fields will benefit directly from improved resolution and intensity possible at the XFELO.

\subsubsection{Dynamics of thin films and interfaces}

Many of the most intriguing properties of materials appear in thin films or at interfaces: examples include the possibility to make the interface between two insulators superconducting \cite{Reyren2007}, or the achievement of high ($\sim$100K) superconductivity in single layers \cite{Ge2014}.  While X-rays are well known as a surface and interface probe, especially in a grazing incidence geometry, such methods have been very difficult or impossible for inelastic studies because of lack of flux and small cross sections.   The XFELO will make studies of phonons in these systems feasible.

\subsubsection{Atomic dynamics of disordered materials}

IXS at the XFELO will allow significant strides in our understanding of disordered materials.  These include such technologically valuable materials as the extremely tough clear glasses used for touch screens, metallic glasses, and some of the new exceptionally light and strong disordered metallic alloys that are interesting structural materials. Basic science questions about disordered materials remain: the cross-over from continuum dynamics to atomistic dynamics in liquids remains an area of active work, as do liquid-liquid phase transitions, along with such issues as understanding the nature of the glass transition and the character of the boson peak.  In this context, non-resonant, meV-resolved IXS is a unique frequency-domain probe, overcoming the kinematic limits of neutron scattering. Starting with the first work on liquids over 20 years ago \cite{Sette1995}, the method has greatly contributed to our understanding of disordered materials.  However, present IXS work is limited in energy resolution ($\sim$1 meV) and momentum transfer ($\sim$1\,nm$^{-1}$).  The XFELO, owing to its higher spectral intensity, will allow simultaneous improvement of the energy resolution and access to low ($<$\,1\,nm$^{-1}$) momentum transfers, allowing a detailed investigation of the previously difficult to access crossover region, see e.g. \cite{Monaco2001}.

There is also a strong interest in probing the transverse dynamics of disordered materials \cite{Scopigno2003}.  It is increasingly appreciated that this has a large impact in liquids, being relevant specifically in the cross-over region mentioned above: the change from continuum limit (where transverse dynamics are largely suppressed due to small shear viscosity) to atomistic behavior where atom-atom scattering leads to strong, but short-lived, excitations, can be seen in calculations of transverse correlation functions.   However, at present, there is no known way to experimentally probe only transverse dynamics using conventional X-ray based scattering methods.  At the very least, one expects that the very high flux and high resolution at an XFELO will allow precision studies to help isolate the effect of transverse dynamics on the longitudinal dynamical structure factor.  

\subsubsection{Phonons in Hydrogen-based superconductors}

The discovery that hydrogen sulfide superconducts at up to $\sim$200 K at pressures of $\sim$200 GPa \cite{Drozdov2015} has created substantial excitement.  It had long been predicted that, due to the high energy of phonons in hydrogen-based materials, extremely strong electron-phonon coupling could allow high-temperature superconductivity.  However, while record-breaking transition temperatures have recently been observed in hydrogen sulfide, there remains substantial interest in exactly how this occurs.   The dramatically improved flux of the XFELO should allow for the measurement of the phonon density of states, including the high-energy hydrogen phonons that are expected to be critical for the high T$_C$. The phonon density of states is expected to be crucial information to compare against calculations to understand the nature of the superconductivity in what is now the world's highest temperature superconductor.  Given the extreme conditions, and resulting small sample quantities ($\sim$1 nanogram), and the very tiny scattering cross section for hydrogen, such investigations are probably only possible by employing IXS at a source such as the XFELO.  More broadly, the high intensity of the XFELO will allow one to probe the coherent dynamics of hydrogen containing materials with very high energy resolution.

\subsection{Methodology}

\subsubsection{Extremely High-Resolution Non-Resonant IXS}
Pushing the resolution of non-resonant IXS experiments to the range of 0.1 meV or even 0.01 meV will extend investigations of disordered materials and inter-system coupling, such as electron-phonon coupling, as discussed above.  We see this as a major potential area of gain at the XFELO since other sources simply lack sufficient flux to practically do such experiments, especially as they will require simultaneous improvement in energy resolution and momentum-transfer resolution. This will require development of new IXS spectrometers (some published possibilities include spectrographs \cite{Shvydko2013,Shvydko2015}, echo-type spectrometers \cite{Shvydko2016,Shvydko2017b} and nuclear analyzers \cite{Chumakov1996,Baron2013}, but see also the discussion in \cite{Baron2015}).

\subsubsection{High-Resolution Resonant IXS (HR-RIXS)}
Resonant IXS (RIXS) is one of the frontiers in IXS at present machines, and we see great possible benefits from pursuing this at an XFELO.  The method, especially as demonstrated in the soft X-ray region, but also for some of the hard X-ray region (specifically, recent work with Ir-containing samples \cite{Kim2012b}), shows sensitivity to electronic excitations, magnetic excitations, more complicated mixed spin-charge excitations, and also phonons.  However, improved resolution is needed, and such experiments will become possible with higher flux available at the XFELO.  This will also require significant optics development, probably of non-silicon optical components, see, e.g., \cite{Yavas2017}.  

\subsubsection{Non-Resonant IXS (NRIXS) for Electronic Excitations}
High-resolution studies of electronic excitations are presently severely flux limited.  Such experiments investigate the motion/transition of (a fraction of) one electron per unit cell making them much lower count rate experiments than phonon measurements (where many electrons move together). Thus, aside from one test experiment \cite{Baron2015}, most experiments have been carried out with $\sim$50-100 meV resolution.   These measurements are interesting for investigating localized d-d excitations and their extended excitation relatives, orbitons, and, in general, for momentum resolved investigation of any electronic excitation, gap, or band-structure \cite{Larson2007}.  The improved spectral density of the XFELO will allow these experiments to be readily carried out with improved resolution, moving count rates into a similar regime as some phonon experiments today.  Notably, improved resolution will also allow closer approach to the elastic/phonon lines. One particular application would be the identification/understanding of the successive steps in reaction pathways where measurements of the precise energy (at the few-meV level) of d-d excitations can help identifying the transient spin states. 

\subsubsection{Coherent Wave Fields (Standing Waves)}
The extremely high brilliance of the XFELO offers the potential of probing phonons and electronic excitations in the presence of a standing wave created by Bragg reflection in a crystalline sample, by specular reflection from a surface, or even using a beam prepared by an interferometer.  This allows access to off-diagonal components of the response  \cite{Schuelke1982,Gan2013} leading to more information, e.g., about plasmons  \cite{Schuelke1991} or phonon polarization vectors \cite{Kohl1985}.  From another perspective, it allows tailoring of the wave field to probe specific parts or layers in the sample, allowing, for example, investigation of specific sections of multilayers, or perhaps single layers on top of strongly reflecting substrates or crystals.  

\subsubsection{Transforming to Real Space and Time}
The spectroscopic work described here is naturally discussed in $(q,\omega)$ space where a well-defined momentum transfer indicates that a plane-wave state is being probed, as it is very much appropriate for most periodic crystalline systems.  However, in some cases, such as disordered materials, or localized anharmonic response in crystalline materials, it may be more appropriate to describe the response in terms of the van-Hove correlation function \cite{vanHove1954}. In some cases this can be obtained via transform of the dynamical structure factor \cite{Iwashita2017}.  This technique would benefit from being able to quickly measure over a large range of  $(q,\omega)$ space, which could be enabled by an XFELO. 

\subsubsection{Other Approaches}
There are some other areas of potential interest that are worth noting for future reference.  These include
\begin{itemize}
\item{Pump-Probe Experiments,}
\item{High resolution Microscopy,}
\item{High resolution variants of emission spectroscopy, X-ray Raman, Compton scattering.}
\end{itemize}
The application for pump-probe experiments is clear, as the number of photons per pulse will be quite large.   One can easily imagine various types of triggered chemical processes (reaction pathways) for which synchronized spectroscopic information could be extremely valuable, even with an X-ray pulse duration of a few picoseconds.  A focused beam, as should be possible at the few-nm level, would allow both site-specific spectroscopic probes, and open new avenues to locally probe extended-wave excitations (such as phonons).  However, one should bear in mind that preceding studies must be done to determine thresholds for acceptable levels of radiation damage in such experiments.  Finally, there are a variety of spectroscopic measurements (emission spectroscopy, X-ray Raman, Compton scattering) that are now carried out using X-rays in setups with $\sim$eV resolution, and the high flux of the XFELO should allow improvement in resolution to the few meV scale.  These will not see as large an increase with the XFELO as the higher-resolution work, but would improve.  In particular, there have been tests of a (relatively) low-energy, high-resolution, version of Compton scattering \cite{Huotari2000} that, with the XFELO, might allow one to achieve 3D electron momentum density maps  \cite{Tanaka2001} on small samples of complex materials with a resolution in momentum space rivaling that of angular-resolved photoelectron spectroscopy (ARPES).  This would permit bulk 3D probing of Fermi-surface topology as would be interesting for nearly any metallic or correlated material.

\section{X-ray Photon Correlation Spectroscopy}
\label{sec:XPCS}

Understanding the properties of a transforming material requires a non-equilibrium statistical mechanics description and necessitates a study of the time evolution in both structure and the resulting properties. Typically, this is performed by measuring the response of the material to an applied stimulus. Often this response function (or susceptibility) is determined by the thermodynamic fluctuations in the material, since the applied stimulus (or field) biases the fluctuations and drives the system towards a new (quasi) equilibrium or a steady state. This simple concept, although hard to carry through in practice, underlies much of the recent advancement in non-equilibrium statistical mechanics. The power of X-ray photon correlation spectroscopy (XPCS) is that it can be employed to directly measure fluctuations in the structure of materials, both in equilibrium and out of equilibrium.

An XFELO offers several critically important improvements for XPCS relative to other sources.  Many of these are implied in the XFELO's higher brightness, as to a first approximation signal rates in XPCS scale with brightness: the expected $\sim$3 orders improvement compared to a high-rate SASE source, and the 5-6 order improvement compared to 4th generation storage rings implies a similar huge improvement in signal. This translates immediately into a significantly enhanced time resolution, because for a given signal-to-noise ratio of the correlation signal, the highest accessible fluctuation rate scales quadratically with the coherent flux \cite{Falus2006,Shpyrko2014}. Moreover, the narrow bandwidth of the XFELO source will facilitate work at higher momentum transfers, where work is now limited by relatively short coherence length of present setups (see section \ref{sec:XPCScomp}). But also, when comparing specifically with SASE sources, the XFELO is expected to be extremely stable, with the $\sim$100$\%$ shot-to-shot fluctuations from SASE reduced to the 1$\%$ level. This will make all time-correlation measurements remarkably easier (see section \ref{sec:XPCStime}).  
Below we provide both an introduction to some of the issues, and some specific example where significant improvement may be expected.

\subsection{XPCS at small angles}

In soft condensed matter systems such as polymers, colloidal suspensions and gels, the measurement of long time constants is important for understanding ageing, creep, and many viscoelastic properties. These are, in some sense, what makes soft matter soft and are either key to their applications or in the case of ageing, what limits their longevity.  It is clear that an XFELO, with its high coherent flux and stable beam conditions, will greatly increase the range of times and types of materials that can be measured efficiently by XPCS. Simply extending current measurements to more fluid systems with faster dynamics at shorter length scales will have a big impact. References \cite{Sinha2014, Shpyrko2014, Sakurai2017} review some of the current work in this direction.

As mentioned, signal-to-noise arguments show that every factor of 10 in beam brightness leads to a potential factor of 100 gain in time resolution for correlation spectroscopy. The unparalleled brightness of a 1MHz XFELO means a tremendous improvement in time resolution. One caveat, however, will be radiation damage issues but novel experimental schemes like X-ray speckle visibility spectroscopy (XSVS) outlines that this is can be alleviated \cite{Verwohlt2018}. The damage decreases significantly with photon energy and the increased brightness of the XFELO enables XPCS with higher energy photons where the brightness is still comparable or higher than for third generation synchrotrons at 8 keV. 

\subsection{XPCS at large angles}

One of the largest impacts of X-ray diffraction has been to measure structures with atomic resolution on length scales up to macroscopic dimensions. Currently, large angle XPCS \cite{Ruta2012} has been severely limited by constraints related to small longitudinal coherence, which limits the diffraction volume, often by factors of hundreds. This is where the XFELO could have the largest impact for XPCS. Not only will the increased brightness mean that shorter correlation times can be accessed, but the increase in longitudinal coherence (from microns to millimeters) gives access to much larger volumes and consequently higher count rates. 

XPCS's ability to directly measure fluctuations in microstructures allows one to address many classical issues in non-equilibrium thermodynamics such as work hardening and tempering. Intermittency is another aspect associated with microstructure. For instance, quenching through martensitic phase transitions leads to a build-up of strain often released in a cascade of changes resembling an avalanche. Intermittent dynamics provides a particular challenge to time resolved measurements as it is hard to synchronize measurements to the event. With the megahertz repetition rates and stability of the XFELO, one can imagine developing area detectors to work like a digital scope and have a flexible triggering mechanism to capture this kind of transients.

\subsection{Comparison of XPCS using SASE from high-gain XFEL vs. XFELO}
\label{sec:XPCScomp}

Another source for XPCS that will become important in the future is SASE from a high-gain XFEL. It has been demonstrated, that coherent contrast (speckle) from a single SASE pulse with less than 100 fs width can be measured. This pulse width is short enough to freeze the atomic motions and the pulse intensity used does not damage the sample \cite{Hruszkewycz2012,Perakis2018}. This demonstrates that a split-pulse XPCS technique is feasible. Split-pulse XPCS works by measuring the contrast (or visibility) of the summed intensity of a pair of pulses as a function of delay between the pulses. If the sample does not change structure between the pair of pulses, the contrast of the summed image remains constant whereas if it does the contrast is reduced. Thus, measuring the contrast as a function of pulse delay is an alternative way to measure the time correlation function \cite{Gutt2009}. This provides a path to extend XPCS measurements to a time resolution of one or two times the pulse duration, i.e. giving sub-picosecond measurements. An appropriate X-ray beam splitter has now been demonstrated and first demonstration experiments performed \cite{Osaka2016,Roseker2018}. With a fast enough area detector \cite{Allahgholi2015a}, this split-pulse technique will also work at an XFELO. The greatly improved longitudinal coherence of an XFELO over SASE leads to increased throughput of the split-pulse delay line and yields enhanced signals, especially near Bragg peaks. This opens the possibility of measuring time correlation functions in correlated electron systems, as discussed above, on the relevant picosecond time scales.

SASE can also be used to perform XPCS in sequential mode \cite{Lehmkuehler2015}, where one speckle pattern is measured per pulse. The time resolution in this mode is limited by the spacing between pulses and not their duration. With SASE, jitter somewhat limits the measurement, but the expected stability of an XFELO will significantly improve the data quality.  All in all, an XFELO will make a major impact with XPCS measurements covering times scales from sub-picoseconds to hours and wave vector ranges from 10$^{-6}$ to 10 \AA$^{-1}$.

\subsection{Beyond time correlations}
\label{sec:XPCStime}

The discussion above is a straightforward extension of current XPCS techniques, i.e. simply taking advantage of the increased brightness to make faster and/or better measurements. The exemplary calculation given below points out some new possibilities. A disordered sample can be considered as made of $N$ scattering volume elements $\rho_i(\vec{r},t)$:
\begin{equation}
\rho_i(\vec{r},t) = \sum_{i = 1}^{N}\rho_i(\vec{r} - \vec{R}_i(t))
\end{equation}
and the scattered intensity is proportional to
\begin{equation}
I(\vec{q},t) = \left| \sum_{i = 1}^{N}\rho_i(\vec{q}) e^{-i \vec{q}\cdot \vec{R}_i} \right|^2
\end{equation}
Using a random walk type of argument, it can be shown that the correlation between intensities at two positions, separated in either or both space and time, is   
\begin{equation}
\frac{\langle I\,I' \rangle}{\langle I \rangle\,\langle I' \rangle} =
\left( 1 - \frac{1}{N}\right) 
( 1 + |g_1|^2) + 
\frac{\langle \rho_{i}^{2}\,\rho_{j}^{'2}\rangle_{ij}}
{N\,\langle \rho_{i}^{2} \rangle_{i}\,\langle \rho_{j}^{'2} \rangle_{j}}
\end{equation}
where $I = I(\vec{q},t)$ and $I' = I(\vec{q}',t')$ and $g_1 = \langle \rho \rho' \rangle/\sqrt{\langle I \rangle\langle I' \rangle}$ with 
$\langle \rho \rho' \rangle = N \langle \rho_{i} \rho'_{j} \rangle \langle \exp[i(\vec{q}\cdot\vec{R}_i - \vec{q}'\cdot\vec{R}'_j)]\rangle$.
The last equation assumes that there is no correlation between positions of volume elements and their type, which is a good approximation for random samples. 

This calculation gives several insights into XPCS. First, for large $N$, the correlations only depend on $|g_1|^2$ and cannot be distinguished from Gaussian fluctuations. Also if the phase factors fully average over $\pm\pi$, the correlations reduce to their incoherent values as measured in conventional X-ray scattering. When $|\vec{q}|$ is small, averaging over phases reduces to $e^{-q^2 \langle R(t) - R'(t') \rangle} = e^{-D q^2 |t - t'|}$ and allows measurements of the diffusion constant $D$. These considerations typically apply to conventional XPCS types of measurements.

The $N$ dependence of the XPCS signal provides an exciting opportunity to obtain further information. The enhanced coherent intensity of an XFELO will allow measurable scattering from smaller diffraction volumes or smaller values of $N$. Thus by measuring correlations from different wave vectors and for varying $N$, detailed information about the distribution of particle types (including shapes and sizes) can be obtained. This is particularly interesting for large angle XPCS where diffraction depends the orientation of the particles and the orientation distributions become accessible. Methods of varying $N$ include scanning the diffraction volume across the sample and/or varying the beam size.

\begin{figure}
\begin{center}
    \includegraphics[width=0.8\columnwidth]{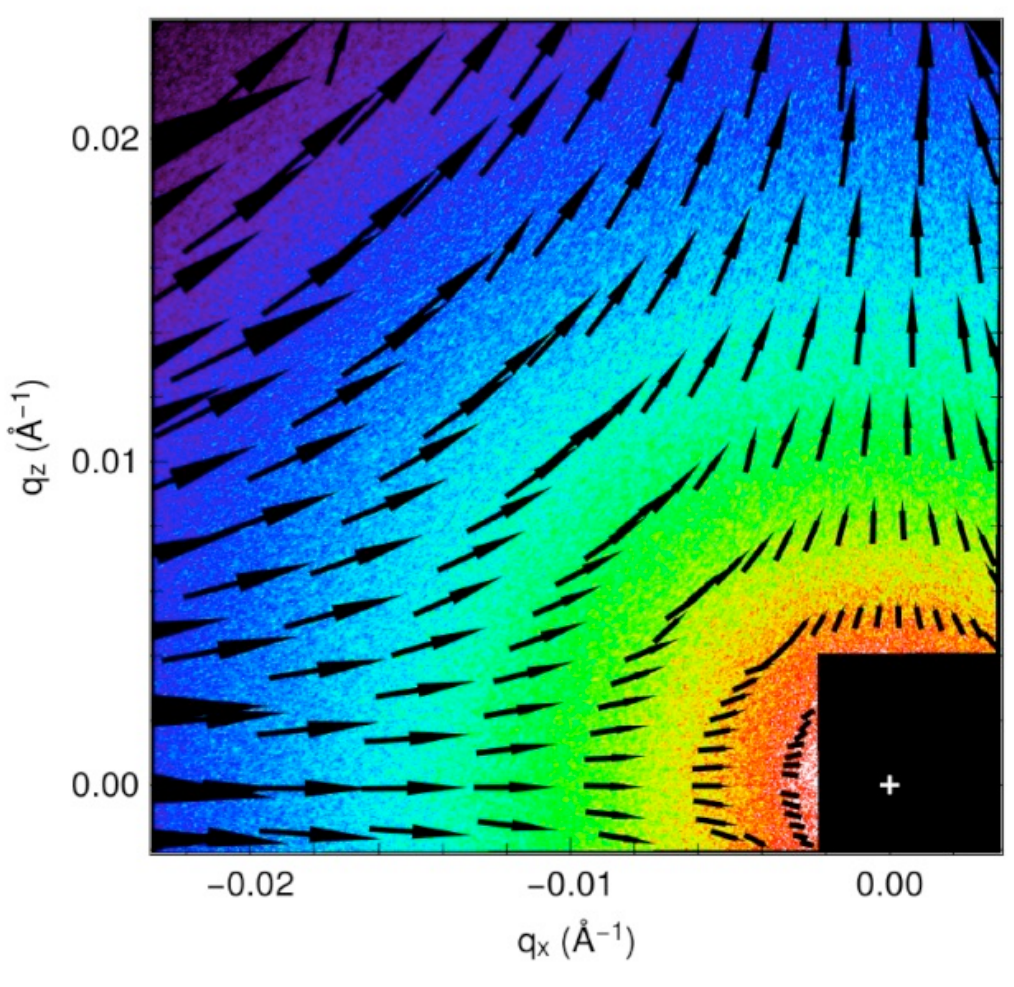}
\end{center}
\caption{The strain field of an ethylene-propylene polymer with 16$\%$ silica filler, at 60$\%$ elongation, 3000 s after the stretch. The deformation shown was extracted from the change in speckle positions for patterns 10.5\,s apart. The arrows show that shift, scaled by a factor of 200, superimposed on the coherent SAXS pattern. The X-ray beam was 20\,$\mu$m by 20\,$\mu$m on the the sample. The deformation is well described by $\delta q_x/q_x$ = -0.00049 and $\delta q_y/q_y$ = 0.00063.}
\label{Fig7}
\end{figure}

Another source of information from $q - q$ correlations, is access to sensitive measurements of strain. One can consider a speckle pattern, where the random phases interfere, as a sensitive probe of local order. Then, just as in polycrystalline materials, small displacements (or strain) can lead to systematic shifts of the speckle. Fig.\,\ref{Fig7} shows the result of doing this on an elastic rubber demonstrating how speckle tracking allows precision measurements of the strain tensors \cite{Lhermite2017}. The speckle size governs how accurately the shift can be measured. Under standard conditions it leads to strain sensitivity of a few parts per million, even in highly disordered materials. It is important to emphasize that this strain pattern results from analyzing only two speckle patterns. With the XFELO, strain measurements from two pulses will give time resolution in the microsecond range. Using the split-pulse technique allows for much faster measurements of this kind provided the individual speckle patterns can be resolved on an area detector. 

\section{Nuclear Resonance Scattering}
\label{sec:NRS}

\begin{figure}
\begin{center}
    \includegraphics[width=\columnwidth]{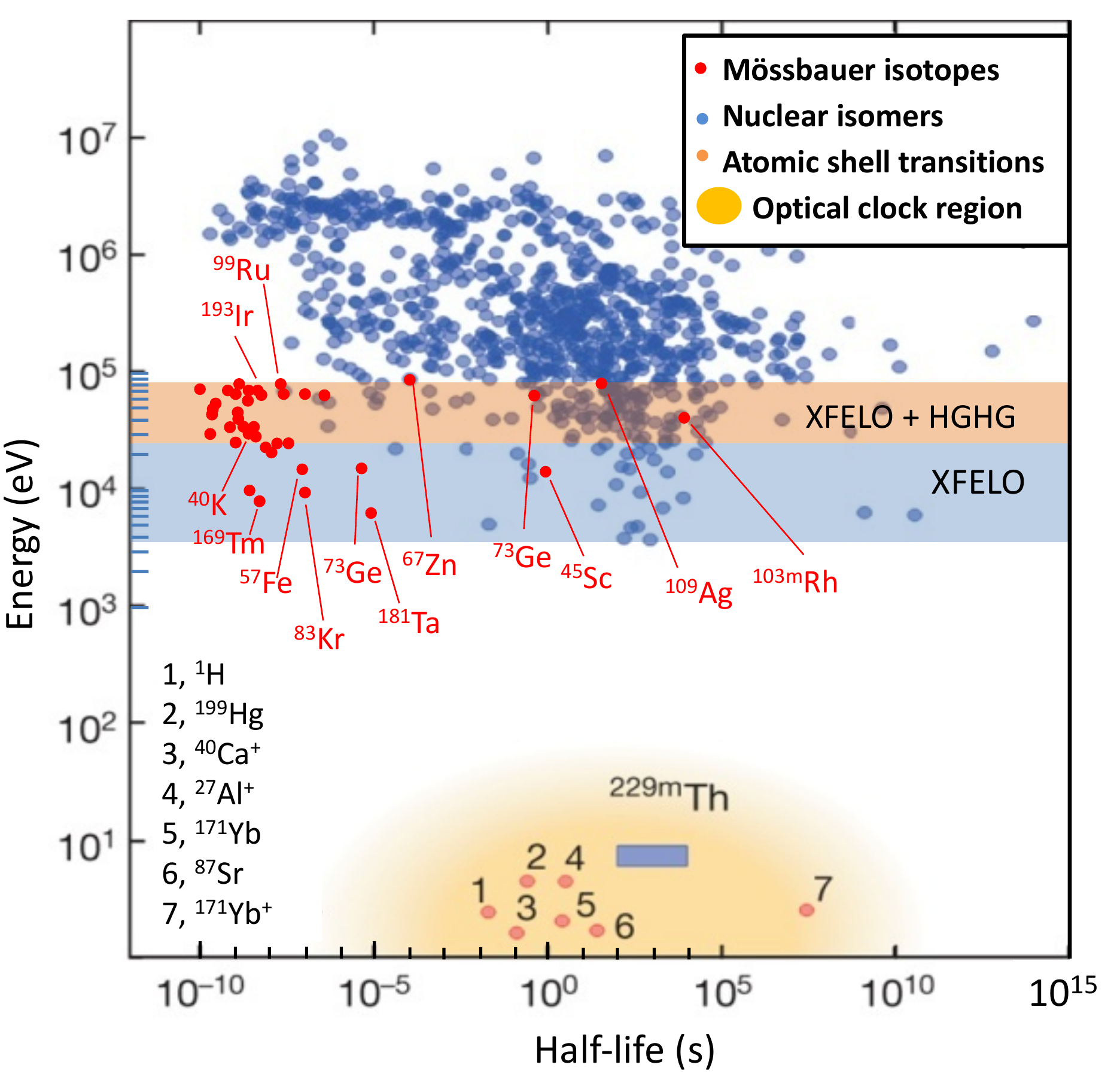}
\end{center}
\caption{Diagram of M\"ossbauer isotopes and nuclear isomers together with atomic and nuclear clock transitions in the parameter space of transition energy and half-life. Graphics adapted from Fig. 1 in \cite{vonderWense2016}. The shaded regions covers the energy range of the XFELO (light blue) and the XFELO + HGHG extension (light brown).}
\label{Fig8}
\end{figure}

The XFELO will open up completely new possibilities in the field of nuclear resonance scattering (NRS) for isotopes with resonance energies between 5 and 25 keV (see Fig.\,\ref{Fig8}).  Due to the narrow resonance linewidths of M\"ossbauer transitions, NRS will benefit from the extremely intense, narrow-bandwidth radiation from the XFELO in several ways: The anticipated hard X-ray spectral flux of 3$\times 10^{9}$ ph/sec/neV is more than 4 orders of magnitude larger than at existing 3rd generation synchrotron radiation sources. For isotopes like $^{57}$Fe one expects several thousand photons per pulse at a MHz repetition rate within the resonance linewidth.  This allows one to push M\"ossbauer science beyond the single photon regime, opening new perspectives for X-ray quantum optics and nonlinear science with nuclear resonances. The full transverse coherence of the radiation will allow for efficient focusing to extremely small spot sizes in the range of 10 nm, enabling one to combine NRS with high-resolution imaging techniques. Moreover, a frequency stabilized XFELO would enable a hard X-ray frequency comb with pulse-to-pulse coherence for unique applications in X-ray coherent control and extreme metrology. The transform-limited pulse length of about 1 ps also facilitates the study of non-equilibrium processes in pump-probe experiments, enabling one to probe transient dynamics with very high spatial, temporal and, in particular, isotopic resolution. 

\subsection{Dynamics in Condensed Matter}

Nuclear forward scattering (NFS) \cite{Hastings1991} probes magnetic and electronic degrees of freedom and their dynamics via the respective hyperfine interactions of the nuclei in the sample.  The short and intense XFELO pulses will allow for single-shot M\"ossbauer spectroscopy with time resolutions in the ps-regime, giving access to non-equilibrium states populated after impulsive stimuli such as heat, optical or magnetic and electric field pulses. After resonantly probing the sample by the ps-short XFELO pulses, the temporal evolution of the delayed NFS signal reveals unique information about the dynamic and hyperfine state at the probe moment and the relaxation dynamics thereafter.  For example, in optical-pump/NRS-probe experiments it will be possible to monitor the response of optically-excited, iron containing systems such as molecular switches or magneto-optical nanomaterials \cite{Pineider2013}.

Nuclear inelastic X-ray scattering (NIS) \cite{Seto1995,Sturhahn1995}, a.k.a. nuclear resonant vibrational spectroscopy (NRVS), is an established method to determine the equilibrium vibrational dynamics of M\"ossbauer isotopes in the sample via the corresponding partial density of phonon states.   The XFELO is especially interesting for NIS as it provides sufficient flux to allow higher resolution, or access to complicated or dilute systems.  The information provided is complementary to conventional IXS methods (see section \ref{sec:IXS}) and is especially useful as providing $q$-integrated information that might be supplemented by targeted studies using IXS.   In addition, the high flux will allow combined X-ray fluorescence analysis / NIS analysis, providing chemical-species-specific information about atomic motions - a conventional crystal analyzer selects the fluorescence from a specific oxidation state or spin state of the resonant isotope (e.g. only Fe$^{+2}$ from some mixture of Fe$^{+2}$ and Fe$^{+3}$) and a NIS analysis measuring only the radiation through the analyzer would be used to determine the vibrational density of states of that species.  This  would be useful, for example, for investigating the intermediate states of the nitrogen fixation cycle as they are encountered in the Fe-S clusters of nitrogenase enzymes \cite{Hoffman2014}. Moreover, the high flux would allow observation of very weak transitions, such as high frequency Fe-H stretching modes, that is beyond the range of current sources.

A very appealing application of NIS at an XFELO emerges from the fact that NIS essentially provides a snapshot of lattice dynamics at the very moment when the probing X-ray pulse hits the sample \cite{Shenoy2008}.  This will allow for studies of non-equilibrium lattice dynamics where a pump pulse selectively populates vibrational excitations, whereas the probe pulse detects the spectrum of these excitations at the moment of probing. An immediate application will be the study of the transient dynamics of heat-transfer and energy dissipation on nanometer length scales \cite{Siemens2010} that are largely unexplored at present.

An XFELO will significantly advance those research applications that demand very high-energy resolution.  Its extremely high spectral brightness will allow for an efficient implementation of high-resolution spectrometers based on the nuclear resonance scattering processes that provide an energy resolution in the neV - $\mu$eV regime like time-domain interferometry (TDI) \cite{Baron1997}, high-speed Doppler shifting \cite{Roehlsberger1997} and the nuclear lighthouse effect \cite{Roehlsberger2000}. A large body of unexplored dynamical properties of condensed matter in the regime from neV to meV is related to multi-atom correlations on mesoscopic length scales from nanometers to micrometers. This applies to fundamental and applied aspects in the physics of glasses and mesoscopically structured materials that exhibit disorder and density fluctuations on these length scales. In all of these cases the material properties are affecting propagating excitations like phonons, leading to new dynamical properties that have no counterpart in homogeneous bulk materials. The vibrational frequencies of mesoscopically-structured materials are substantially smaller than those of crystalline materials and the first Brillouin zone extends to much smaller $q$-values. The corresponding range in phase space is not covered by existing methods. Thus, the XFELO would close the existing energy and momentum gap of various inelastic scattering techniques in the neV-meV energy range. In combination with focusing techniques, these methods can achieve spatial resolutions approaching the 10 nm regime, allowing one to obtain detailed information about matter under extreme conditions of pressure, temperature and external fields.

At the high-end limit of energy resolution, the XFELO will make it practical to carry out studies of general samples (not containing M\"ossbauer isotopes) with neV resolution by employing nuclear scatterers as analyzers.   Proof-of-principle experiments \cite{Baron1997,Masuda2009} have been conducted at present day machines, but the applications have been severely limited due to insufficient flux. The improvement in flux at the XFELO would allow these methods to be applied to general samples: one can think of probing liquid diffusion in broad classes of materials and jump diffusion in solids.  While there is some overlap with neutron spin echo (NSE), the X-ray techniques extend to smaller energy transfers (longer times) than NSE and allow access to much smaller ($< \sim$ 0.1 mm) samples, also making experiments compatible with extreme environments.
Eventually, a high-resolution analysis of the symmetry of the intermediate scattering function determined via TDI will enable one to distinguish between classical and quantum correlations in a sample \cite{Castrignano2018}.  Much of the technology to do this is available now, with the main contribution needed being a source with a much higher spectral brightness. In this respect, an XFELO operated as a hard-x-ray comb laser (see section \ref{sec:xraycomb}) would be the ultimate source for such experiments.

\subsection{Biology and Chemistry}

\begin{figure}
\begin{center}
    \includegraphics[width=\columnwidth]{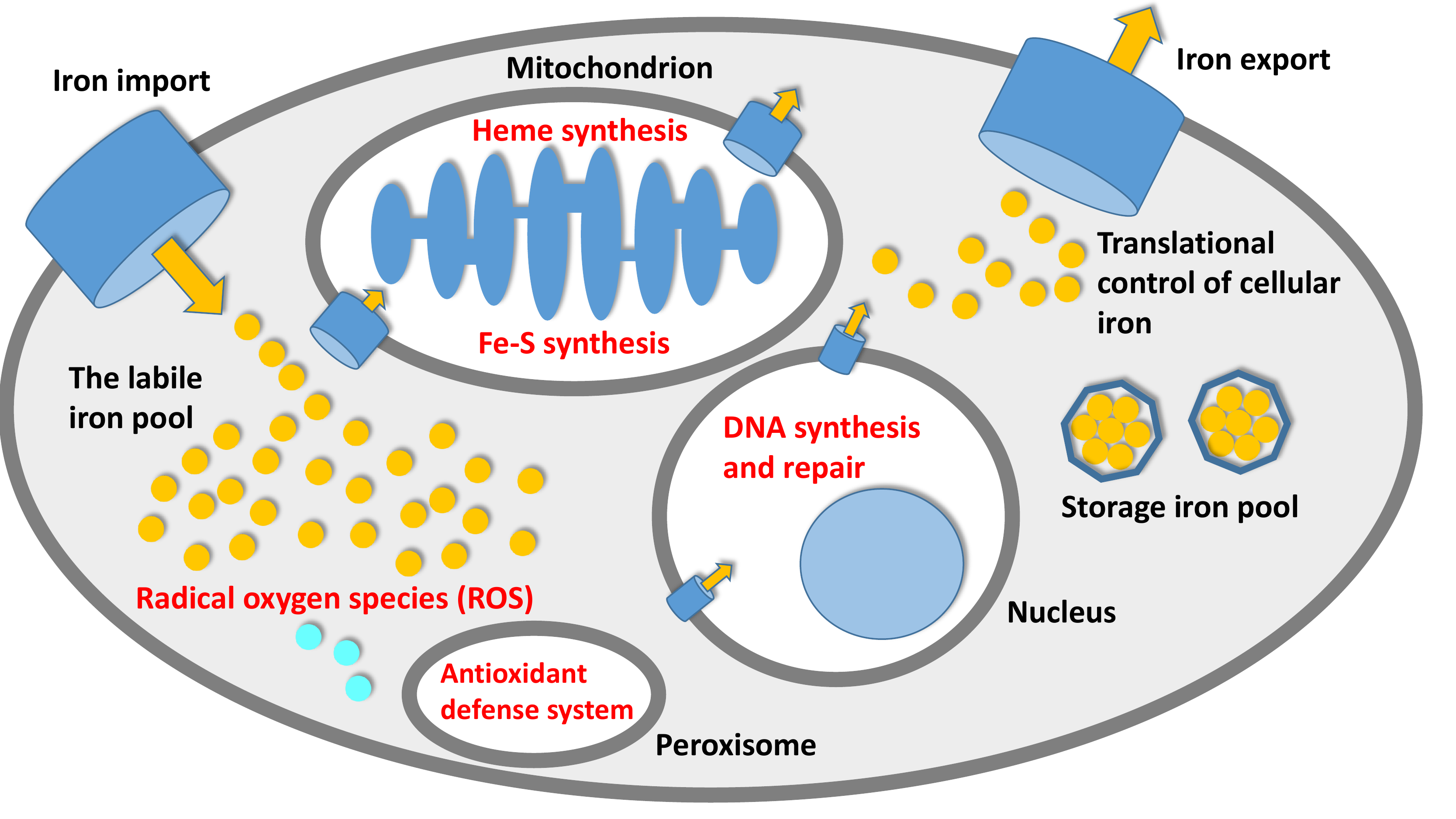}
\end{center}
\caption{
Iron is essential for cell survival, but toxic if not properly regulated. This diagram shows the complexity of human iron homeostasis already in its simplified form. Iron plays a decisive role in cell metabolism, cell death, and disease \cite{Lill2008,Bogdan2016}.  XFELO-M\"ossbauer Spectroscopy will access iron valence states and enable iron centered reactions inside cells to be monitored. This paves the way to the field of ironomics that tackles many aspects of human health and disease.}
\label{Fig9}
\end{figure}

The outstanding spectral brightness renders XFELO-M\"ossbauer spectroscopy applicable to iron containing proteins without requiring $^{57}$Fe enrichment. This will boost the interest of biological communities and generate a wealth of applications to reveal how important iron cofactors are formed inside cells. NRS experiments will be instrumental to follow iron dependent reactions inside cells, see Fig.\,\ref{Fig9}. In this way it will be possible, e.g., to highlight the iron sulfur assembly machinery not only in mammalian cells \cite{Jhurry2012}, but also in plant cells.

XFELO-M\"ossbauer spectroscopy with its outstanding sensitivity will also contribute to solving medical questions where Fe-containing proteins are present at only very low concentrations. For example, it could help to explore the general role of iron in neuro-degenerative diseases \cite{Zecca2001}, and it could be employed to investigate iron metabolism in cancer cells \cite{Torti2013}. Further, an XFELO could allow processes in iron-mediated biologial catalysis (e.g. nitrogen reduction, methane oxidation) to be studied at physiologically relevant conditions. 

In chemistry the identification of active iron-sites in heterogeneous catalysts is a demanding task. Supported catalysts often contain iron in many phases (e.g. iron oxide nanoparticles, as single ions, agglomerates of only a few iron centers) \cite{Velez2014,Varnell2016}. With XFELO M\"ossbauer spectroscopy it will not only be possible to acquire spectroscopic data with excellent statistics, but also to monitor the evolution of dynamical processes like catalyst activation and deactivation. 

\section{Non-linear X-ray Optics and Spectroscopy}
\label{sec:NLO}

The XFELO characteristics provide exciting opportunities in X-ray nonlinear optics and coherent spectroscopy. Its narrow bandwidth, high coherence, and widely tunable photon energy range are features of particular interest in nuclear nonlinear spectroscopy. The increased coherent bandwidth (i.e. shorter pulses) and peak power of an XFELO/MOPA would be of interest for coherent electronic and vibrational spectroscopies. These applications could exploit the possibility of tailoring the electric-field temporal profile spanning eV bandwidths.  We highlight some examples below.

\subsection{Probing low energy dynamics with atomic resolution through nonlinear X-ray mixing}

\begin{figure}
\begin{center}
   \includegraphics[width=0.7\columnwidth]{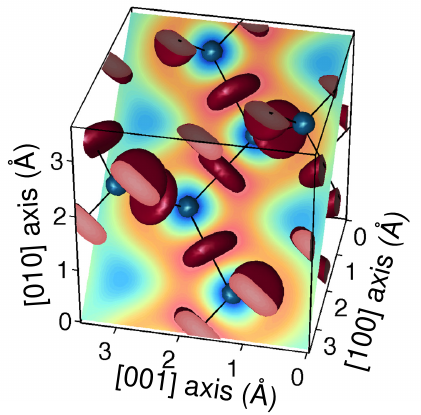}
\end{center}
\caption{Atomic-scale view of the XUV susceptibility of diamond at 60 eV from X-ray down conversion of $E_0 \sim\,$11 keV photons.  The image is obtained through the intensities of the signal photons at $E_0$ - 60 eV for different phase-matching conditions arising from various lattice planes.  The blue spheres respond in phase, and the red discs in the opposite phase, relative to the idler at 60 eV (which is absorbed by the crystal). Each carbon atom resides at the center of a blue sphere. The black lines indicate the bonding directions. From ref. \cite{Tamasaku2011}.}
\label{Fig10}
\end{figure}

Nonlinear mixing allows for the unique combination of high X-ray wave-vector for atomic spatial resolution with access to low-energy resonant excitations that can only be accessed at longer wavelength by single photon processes.  The XFELO characteristics are well suited for a variety of two-photon processes, including X-ray parametric down-conversion (PDC) \cite{Eisenberger1971} and various up-conversion processes \cite{Glover2012}. Both can involve lower - frequency components in addition to X-rays. This allows the electronic and/or lattice modulations induced through a lower-frequency resonance to be studied with atomic spatial resolution given by the X-ray wavelength because the frequency-shifted signal is diffracted from the crystal lattice. Tamasaku \textit{et al.} have exploited this using spontaneous parametric down-conversion to study the extreme ultraviolet (XUV) response of diamond with angstrom resolution without the need for irradiation or detection at XUV frequencies, see Fig.\,\ref{Fig10} \cite{Tamasaku2011}. These processes can also be stimulated in which case they are referred to as sum and difference frequency generation.  Glover et al. demonstrated sum-frequency generation in X-ray optical mixing experiments in diamond, using 8 keV X rays and 1.5 eV near-infrared light \cite{Glover2012}.  This provides access to the atomic-scale polarization arising from the induced local fields, and not just the long-wavelength dielectric response \cite{Freund1970,Eisenberger1971b,Glover2012}.  Another powerful probe of electronic excitations in solids will be X-ray PDC into the optical regime, which has been observed recently \cite{Schori2017}  

The high flux and coherence of the XFELO would be useful for probing low-energy coherent excitations in the time domain using pump-probe techniques, when the X-ray pulse duration is short compared to the period of oscillation, thus requiring a coherent bandwidth exceeding the frequency of the excitation.  This requires the MOPA configuration (see section \ref{sec:mopa}) for higher frequency phonons and electronic excitations. Momentum resolved X-ray diffuse scattering from correlated high-wavevector phonon pairs that were optically pumped was first measured in the time domain by Trigo et al. \cite{Trigo2013} and are a complementary approach to the frequency domain inelastic scattering measurements discussed above, applicable to both equilibrium and nonequlibrium dynamics.
 
The XFELO parameters will permit efficient focusing approaching the few nanometer scale that would facilitate a host of multiphoton X-ray interactions. Non-resonant X-ray nonlinearities may yield a complementary way to accesses low energy-scale electronic structure and dynamics.  Fuchs \textit{et al.} measured an anomalous red-shift in two-photon Compton scattering in which scattered photon energies are redshifted from twice the incident photon energy.  This was interpreted in terms of an enhanced nonlinear interaction with core-electrons \cite{Fuchs2015} that is not present in other X-ray nonlinearities such as second harmonic generation \cite{Shwartz2014}. 

The high brightness of the XFELO would also facilitate resonant X-ray nonlinearities. An example is X-ray two-photon absorption (TPA) \cite{Tamasaku2014,Ghimire2016}, which allows preferential access to dipole-forbidden transitions.  Stimulated resonant X-ray Raman scattering (SRXRS) using a narrow-band, tunable two-color X-ray source \cite{Weninger2013} could provide a wealth of information about core transitions of lighter atoms, valence electronic transitions and their coupling to atom-specific core transitions. With phase-related multiple X-ray pulses, 2D X-ray coherent Raman spectroscopy would be possible and would provide detailed information about correlations among multiple core transitions \cite{Mukamel2013}. 

\subsection{Two-photon correlation spectroscopy}

Unlike the coherent nonlinear mixing processes described above, "two photons in, two photons out" X-ray correlation spectroscopy would use two detectors for measurements of scattered photon pairs in order to determine 4-point space-time density correlation functions. Using a variant of the so-called impulse approximation, which allows standard Compton scattering data to be connected to the electron density in momentum space, two-photon correlation spectroscopy is expected to give access to 2-point momentum-space density correlations of electrons. Electron correlations in momentum space play a central role, for example, in superconductivity. Such spectroscopy can benefit from high spectral brightness associated with narrow bandwidth, and also from high repetition rate. 
 
A different type of two-photon correlation spectroscopy is ghost imaging which could yield atomic-scale resolution while avoiding radiation damage \cite{Pelliccia2016,Yu2016,Li2018}.  This and other spectroscopies involving entangled photon pairs generated through spontaneous parametric down conversion and higher-order optical coherence would benefit from the high repetition rate and spectral brightness of the XFELO.

\subsection{Quantum Optics and Nonlinear Spectroscopy with Nuclear Resonances}

The longitudinal coherence of optical fields is the core requisite to induce and control interference between different quantum pathways in atoms. In nuclei, similar developments so far were restricted to single photons interfering with themselves, due to the lack of sufficiently coherent photon sources. With an XFELO this situation will fundamentally change. Its full coherence and spectral brightness provides new avenues for studying the interaction between X-rays and nuclei under multiphoton excitation conditions, thus offering unique possibilities for nonlinear spectroscopy of the nucleus, as well as for novel approaches to nuclear state preparation and detection. 

For example, at low orders of nonlinearity quantum aspects involving X-ray photons could naturally be explored with two or more correlated photons. Potential approaches encompass both the generation of X-ray photon entanglement and its applications, and the exploration of quantum states in the nuclei by subsequent spectroscopic detection of scattered X-ray photons \cite{Heeg2016}. The availability of multiple coherent photons per pulse in turn enables detection of multiple correlated X-ray photons, providing access to higher-order correlation functions characterizing, e.g., density fluctuations, phonons or similar excitations. This will fuel the development of a broad class of new detection and analysis techniques.  With multiple potentially phase-locked driving fields, multi-dimensional spectroscopy techniques come within reach, providing further insight into the dynamics. 2D nuclear spectroscopy might reveal couplings among nuclear transitions that could provide fundamental insight into intra-nuclear interactions, analogous to what is revealed in two-dimensional spectroscopy throughout radio frequency to optical spectral ranges \cite{Bax1984,Cho2009}. This and the other 2D measurements mentioned above will require at least phase-related X-ray pulse pairs, which could be generated by splitting one XFELO output pulse by an X-ray split-and-delay line or by applying temporal control of resonantly scattered photons via ultrafast piezo modulation \cite{Heeg2017}.

Further progress is anticipated in the engineering of advanced nuclear level schemes. First steps have recently been demonstrated at 3rd generation light sources, by designing suitable target structures utilizing M\"ossbauer nuclei embedded in superlattices \cite{Haber2016} and planar X-ray cavities \cite{Roehlsberger2010,Heeg2013,Haber2017}. The XFELO will enhance these capabilities by its unique source properties, which, aside from the spectral brightness also includes coherent multi-pulse or multi-color excitation. The XFELO could also facilitate novel nuclear resonance excitation processes, such as non-linear two-photon excitation \cite{Doniach2000} or four-wave mixing. 

The pulse-to-pulse coherence of an energy stabilized XFELO enables one to realize a hard X-ray frequency comb (see section 6), facilitating ultrahigh-resolution X-ray spectroscopy of nuclear transitions. Examples include multi-level nuclear transition measurements, probing ultra-narrow X-ray M\"ossbauer resonances, dynamics of X-ray driven nuclear -- electronic transitions, and X-ray\,+\,laser double resonance experiments. Facilitated by X-ray comb spectroscopy, fascinating possibilities come into reach: X-ray frequency and wavelength metrology would be enabled by extending the optical frequency comb technologies and techniques to X-ray wavelengths. In addition to probing nuclear physics with unprecedented precision, linking nuclear transitions to the Cs standard can be used to search for the variation of fundamental constants with improved sensitivity \cite{Flambaum2006,Rellergert2010}. Nonlinear phase-coherent driving and probing at X-ray wavelengths will be possible over long times  $>$ 10 s. High-quality-factor nuclear transitions like the 12.4 keV level of $^{45}$Sc with a lifetime of $\sim$300 ms and $\Gamma_0/E_0 \sim 10^{-19}$ (see Fig.\,\ref{Fig8}) can be established as new and improved frequency standards. Importantly, the pulse-to-pulse coherence allows to excite these narrow resonances using a sequence of pulses, offering the possibility of orders of magnitude higher excitation fraction than expected from SASE \cite{Heeg2016}.

With sufficient temporal coherence and high intensity, coherent processes including nuclear coherent population transfer in the stimulated Raman adiabatic passage (STIRAP) technique \cite{Liao2011,Liao2013} or nuclear Rabi oscillations \cite{Buervenich2006} are rendered possible. Coherent population transfer would enable controlled pumping, storage and release of energy stored in long-lived nuclear excited states. In addition, also nuclear reactions  starting from excited nuclear states driven by the XFELO can be envisaged.

\section{X-ray Spectral Comb}
\label{sec:xraycomb}

Achieving high spectral resolution in the X-ray region will bring revolutionary impact to science, similar to what highly coherent lasers have done for atomic and molecular physics. Single-frequency narrow-linewidth X-ray lasers, which can be tuned to specific inner-shell quantum transition frequencies while providing a direct link from X-ray frequencies to radio frequency standards, would be the most desirable sources for spectroscopy. An X-ray frequency comb will have high spectral and temporal coherence. These properties will make it possible to test fundamental physics laws with unprecedented precision, obtain new information in order to finally resolve the structure and dynamics of the nucleus and create a clock with unrivaled precision together with a frequency standard in the X-ray frequency range. Furthermore, they will allow improvement of the sensitivity in the measurement of properties of solids and complex molecules.  

However, until now there is no known technique for obtaining such light sources. We consider the realization of an X-ray frequency comb to be feasible within the next decade, looking at possible routes to reach this goal \cite{Adams2015}.

\subsection{Applications of an X-ray Comb: Fundamental Physical Laws and Nuclear Structure}

Our best picture of the fundamental physical laws of the universe is the standard model, a paradigm of quantum field theory. 
Today, it is already known that the standard model is incomplete, and high-precision measurements are required to test candidate theories for its extention \cite{Safronova2018}. Such theories are also required to solve intriguing experimental disagreements such as the so-called proton radius puzzle \cite{Pohl2010, Antognini2013,Beyer2017}.

The most successful subset of the standard model, in terms of making accurate predictions that have been verified by experiment, is quantum electrodynamics (QED). Theoretical QED calculations can be performed most accurately for simple systems, e.g. few-electron atoms and ions. Since QED effects scale rapidly with nuclear charge, precision measurements in highly charged ions can provide the most stringent tests of QED  \cite{Vogel2013,Volotka2013,Crespo2016}. Such measurements require a stabilized laser source in the X-ray frequency regime. A compelling case for an XFELO would be high-resolution spectroscopy of highly charged, hydrogen-like heavy ions.

\begin{figure}
\begin{center}
    \includegraphics[width=\columnwidth]{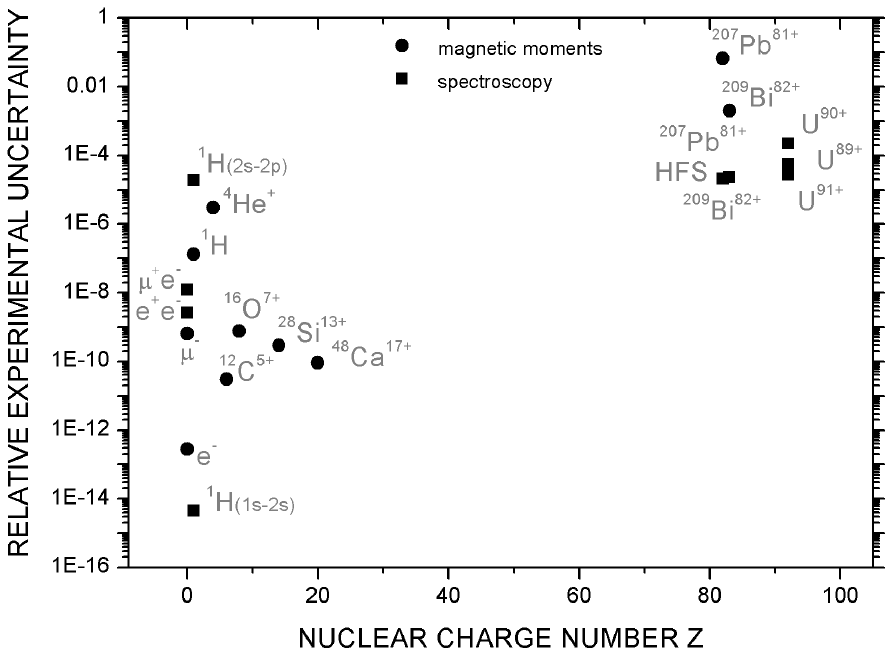}
\end{center}
\caption{Experimental precision in tests of QED in neutral atoms and highly charged ions. The  experimental uncertainties for the determination of the magnetic moments (circles) and other precision measurements (squares)  are displayed \cite{GSI2018}. Data were taken from \cite{Fee1993,Beier2005,Odom2006,Trassinelli2009,Matveev2013,Ullmann2017,Vogel2018}.}
\label{Fig11}
\end{figure}

As shown in Fig.\,\ref{Fig11}, for the low nuclear charge range the fractional frequency (energy) resolution is now better than 10$^{-14}$, providing some of the most precise measurements of the fundamental properties of matter. The precisely resolved transition energies in these low nuclear charge systems are at most a few eV, corresponding to photon energies available from optical lasers. Those achievements in precision laser spectroscopy have been enabled by frequency-stabilized lasers, optical frequency combs, and laser-cooled atoms and ions. For the shorter wavelengths and higher energies required for transitions in highly charged ions, the fractional spectral resolution currently achievable is only in the 10$^{-6}$ range \cite{Epp2011}. A highly coherent X-ray laser or X-ray comb would allow 10 orders of magnitude improvement on precision atomic spectroscopy in the X-ray region. Since QED effects in H-like ions scale with strong powers of the nuclear charge (e.g. the Lamb shift scales as $Z^4$), achieving precision on par with optical measurements provides much stronger tests of QED.

Probing high-energy transitions with high precision would also allow for a more precise measurement of fundamental constants, e.g., the fine structure constant. Moreover, it provides greater sensitivity in determining if the fundamental constants of nature are really constant, or vary with time \cite{Flambaum2006}. Precision measurements in the optical regime have placed severe constraints on this variation, but high-energy transitions, particularly nuclear transitions, give a much larger frequency shift for the same variation in the laws of nature due to the larger transition energy scale.

Higher precision spectroscopy of electronic transitions in the X-ray regime would also provide new information on the structure of the nucleus due to the impact of the latter on atomic structure. Notably, a complete understanding of the structure of the nucleus has not been achieved yet. Nuclear structure effects on the bound electron energy scale approximately as $Z^2/n^3$, where $n$ is the principle quantum number \cite{Noerterhaeuser2014,Campbell2016}. Therefore, low-lying electronic states in heavy atoms and ions, which require X-ray energies for excitation, are those most sensitive to nuclear structure. A stabilized X-ray laser can serve as a high resolution probe for nuclear charge distributions and perhaps nuclear magnetic distributions, bringing new data that could help resolve the long standing problem of nuclear structure effects in atomic transitions.

An exciting possibility that an X-ray frequency comb would enable is direct laser spectroscopy of nuclear transitions, a feat that has not been accomplished to date. This approach looks like a very promising route to explore coherent interactions between atomic nuclei and light. Particularly interesting would be to explore multi-photon nuclear excitation schemes to move into the field of nonlinear nuclear X-ray optics, as this would open new avenues for coherent control of x-ray optical properties. It should be noted that SASE XFELs are not coherent sources in the spirit of Glauber \cite{Glauber1963}, as they exhibit the Gaussian statistics of a chaotic source \cite{Singer2013}. Notwithstanding, present-day SASE XFELs exhibit a high photon degeneracy parameter and thus enable one to induce significant population changes of nuclear levels, as demonstrated in a recent experiment at SACLA \cite{Chumakov2018}. It remains to be seen whether \textit{seeded} SASE XFELs are fully coherent sources in all orders. At any rate, an XFELO-based frequency comb would constitute an outstanding source for laser spectroscopy of atomic and nuclear transitions in the regime of hard x-rays with ultimate energy resolution.

For instance, frequency comb spectroscopy of nuclear transitions could be used to look for changes in fundamental constants, with even higher precision than can be achieved with electronic transitions in highly charged ions, as these high energy transitions are extremely sensitive to the magnitude of fundamental constants \cite{Rellergert2010}. An interesting candidate in the energy range of the XFELO would be the 12.4 keV transition in $^{45}$Sc with a level width of 1 femto-eV ($\Delta E/E \approx 10^{-19}$), corresponding to a half-life of 280 ms. Moreover, within an extended photon energy range up to 60 - 80 keV that could be reached via HGHG (see section \ref{subsec:extended}) a large number of long-lived transitions in nuclear isomers \cite{Walker1999} can be reached, see Fig.\,\ref{Fig8}.
A far reaching application of direct frequency comb spectroscopy of nuclear resonances is the facilitation of a clock that is locked to a nuclear transition, similar to current state-of-the-art optical atomic clocks that are linked to radio frequency standards via optical frequency combs \cite{Ludlow2015}. A nuclear clock has the potential to provide a higher resonance line quality than an atomic clock \cite{Campbell2012}, since the nucleus is much less sensitive to perturbations from its macroscopic environment as compared to the electrons that are currently used in optical atomic clocks. 

\subsection{Technologies to enable phase-coherent X-ray combs}

Frequency combs are routinely available in the visible and near infrared spectral regions, providing a direct link between optical frequencies and radio frequency time standards. Even in the extreme ultraviolet (XUV) spectral region \cite{Benko2014} and at soft x-ray energies \cite{Cavaletto2014} frequency combs can be produced, and in the former case coherence times $> 1$ second have been demonstrated. The enabling technology is the high-order harmonic generation (HHG) process allowing to transfer infrared frequency combs via high-order wave mixing into the XUV. Extending this technology into the X-ray spectral region might be possible but the expected low average power levels make this approach very challenging. We instead propose two alternative directions that might enable X-ray frequency combs: (i) injection seeding of free electron laser (FELs) using an XUV comb seed and (ii) a frequency stabilized XFELO. Although promising, both approaches involve a number of technical challenges, which are briefly discussed below.

\subsubsection{Injection seeding of XFEL with XUV comb}

The technique of injection seeding of FELs with coherent femtosecond pulses is an established technique and great improvements of the longitudinal coherence of FEL pulses has been demonstrated \cite{Allaria2012}. Employing a frequency comb seed could increase the phase coherence of the generated X-ray radiation not only for individual pulses but over consecutive pulses, thus enabling the formation of an X-ray comb. Seeding near e.g. 100 eV could allow FEL emission at harmonics of the seed, possibly approaching 1 keV and maybe even beyond. A crucial parameter for frequency comb spectroscopy is the pulse repetition rate, defining the comb mode spacing, setting an upper limit for the broadest resolvable spectral feature (e.g., transition line width) without spectrally resolving the comb teeth. Most commonly, frequency combs operate with a repetition rate on the order of $f_{rep} \sim$100 MHz. Because of the limited laser pulse energy available at such a high repetition rate, frequency comb conversion via HHG into the XUV is today only possible via passive laser power enhancement inside a femtosecond enhancement cavity. Modern accelerator technology that provides the electron beams for FELs can operate at repetition rates on the order of 1\,MHz, a mode spacing which is low for frequency combs but still high enough to probe narrow X-ray transitions. While efficient HHG is easier at this repetition rate, passive power enhancement is challenging as the enhancement cavities must have correspondingly long lengths which makes them hard to stabilize. Instead, a single pass HHG scheme can be used requiring, however, a high average laser power. In order to allow the transfer of the high phase coherence from the seed to the emitted X-rays, the peak power of the seed $P_{seed}$ has to be much larger than the noise power $P_{noise}$. Taking into account $P_{seed}  > 100\, P_{noise}$ and considering typical FEL noise power levels at e.g. 100 eV (within a bandwidth of 1 nm), we obtain a required seed laser peak power on the order of 1 MW. For typical XUV comb parameters (laser pulse duration: 100 fs), and considering a conversion efficiency of 10$^{-6}$ from the infrared laser into the XUV at the desired photon energy and bandwidth, the required average power of the XUV comb driving laser would be on the order of 100 kW. This is already three orders of magnitude above today's high-power combs operating at the 100\,W level \cite{Ruehl2010,Emaury2015,Li2016,Luo2018} and would increase further if higher repetition rates are considered. While laser technology might advance to these levels, proof-of principle studies at lower photon energies where sufficient seed power is more easily available can help explore the general concept. We encourage establishing close collaboration between XFEL and XUV comb research groups to jointly explore this path and push the technology forward for a feasibility test in the near future. 

\subsubsection{Stabilizing an XFELO cavity with external stable references}

Another possible approach was proposed recently--phase-locking successive XFELO output pulses, using a narrow nuclear resonance such as $^{57}$Fe as a reference \cite{Adams2015}. The reference could also be based on existing optical stable laser technology. This concept might lead to higher repetition rates by operating the resonator in a "harmonic mode-locking" scheme, i.e. at multiples of the repetition frequency of the electron beam, provided the resonator loss is low enough. 

\begin{figure}
\begin{center}
    \includegraphics[width=\columnwidth]{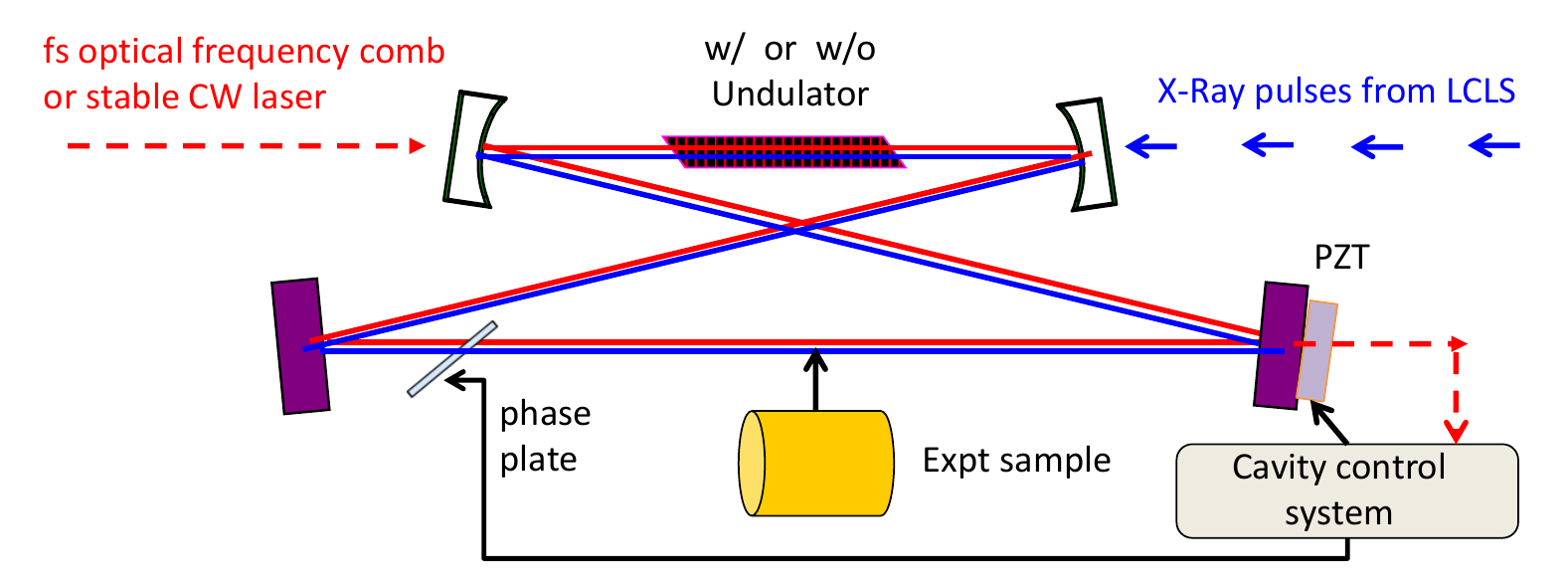}
\end{center}
\caption{Schematic of X-ray test resonator for studying undulator gain and dispersion, testing stabilization schemes and performing ring-down spectroscopy. The resonator will be pumped by an existing X-ray FEL and stabilized with an optical laser. The phase plate will be used to adjust the carrier envelope offset.}
\label{Fig12}
\end{figure}

In order to advance technology of the X-ray cavity stabilization, we propose setting up an X-ray test resonator sketched in Fig.\,\ref{Fig12}, which can be pumped, e.g., by an existing XFEL. This would allow testing of the basic XFELO concepts as, e.g., gain characteristics and dispersion characteristics of FEL undulators as well as stabilization schemes. It could also be used for X-ray intra-cavity spectroscopy (e.g. cavity ring-down) with scientifically interesting samples in the X-ray cavity.  An approach is illustrated below, showing the stabilization of an X-ray cavity to a stable optical frequency reference (either continuous wave or femtosecond comb). Stabilization could be achieved via a piezoelectric transducer (PZT) that controls the cavity length and using a dispersive element (phase plate) to adjust the carrier envelope offset. 

\section{Concluding Remarks}
\label{sec:conclusion}

With its unique radiation properties an XFELO reaches into uncharted regimes in space and time for high-resolution views on the relationship between structure and dynamics of matter. Recent advances in connection with various light source projects around the world indicate that the accelerator technology for an XFELO is essentially available. Moreover, encouraging progress has been made in the development of x-ray optics for a cavity that meets the XFELO specifications. Altogether, these technical developments build confidence in the feasibility and success of the XFELO project in general.

The output characteristics of an XFELO are complementary to those of high-gain type x-ray FELs, thus enriching the portfolio of hard x-ray FELs worldwide. X-ray methods that are currently limited by spectral flux, spatial and temporal coherence will benefit tremendously from an XFELO.
For inelastic x-ray scattering, an XFELO would allow for gains in resolution and flux that would turn this useful technique into one of the most powerful probes of condensed matter systems. With the picosecond-duration pulses and outstanding spectral flux of an XFELO it will be possible to extend scattering techniques to the fundamental timescales of correlated materials and to reveal the basic mechanisms of non-equilibrium processes and structural phase transitions. 

In addition to complete transverse coherence, extremely long longitudinal coherence and very short pulse duration, x-ray photon-correlation spectroscopy will take advantage of the three orders of magnitude increase in average coherent flux, thus closing the temporal gap between synchrotron radiation sources and x-ray lasers. Access to a wide range of new systems will be possible, e.g., to study biological dynamics in aqueous suspension or magnetic dynamics and spin transport in the picosecond regime.
Given a boost of several orders of magnitude in spectral flux at an XFELO compared to existing sources, it is anticipated that completely new research areas will open up for nuclear resonant scattering techniques. With several thousand nuclear resonant photons in one or a few modes, fundamentally new studies in the field of x-ray quantum optics will become possible, enabling to explore the foundations of collective and nonlinear light-matter interaction, the creation of nonclassical states of x-rays, etc.

With sufficient energy stabilization, e.g., via a nuclear resonance, an XFELO could even allow the realization of an x-ray comb with fundamental applications outside the realm of traditional x-ray physics, facilitating ultrahigh-resolution spectroscopy and extreme metrology that could outperform what comb technology has achieved at optical wavelengths.\\[4mm]

\begin{acknowledgments}
The authors of this paper are those who participated in the SLAC retreat on June 29 -- July 1, 2016 and those who contributed to the writing of this report. Jerry Hastings and Kwang-Je Kim edited the first draft from summaries of the working groups led by Alfred Baron (chapter \ref{sec:IXS}), Oleg Shpyrko and Mark Sutton (chapter \ref{sec:XPCS}), Ralf R\"ohlsberger (chapter \ref{sec:NRS}), David Reis and Keith Nelson (chapter \ref{sec:NLO}), and Leo Holberg and Jun Ye (chapter \ref{sec:xraycomb}).  Ralf R\"ohlsberger and Anders Madsen performed the final editing of the manuscript.
The retreat and writing of this report were supported by funding provided by U.S. Department of Energy, Office of Science, Office of Basic Energy Sciences, under Contract No. DE-AC02-06CH11357 (Advanced Photon Source, Argonne National Laboratory) and under Contract No. DE-AC02-76SF00515 (SLAC National Accelerator Laboratory). 
Work by Matthias Fuchs is supported by the U.S. Department of Energy, Office of Science, Basic Energy Sciences under Award DE-SC0016494
Christoph M. Heyl, Gil Porat, and Jun Ye acknowledge the support of the Air Force Office of Scientific Research grant
FA9550-15-1-0111, National Institute of Standards and Technology and the National Science Foundation Physics Frontier Center at JILA (PHY-1734006). Christoph M. Heyl was supported by the Swedish Research Council.
Zheng Li thanks the Volkswagen Foundation for a Peter Paul Ewald-fellowship.
Gerhard Gr\"ubel, Robin Santra and Ralf R\"ohlsberger thank the Hamburg Centre for Ultrafast Imaging (CUI) for financial support.
Gerhard Gr\"ubel and Robin Santra acknowledge financial support by the SFB 925 of the German Science Foundation.
WenTe Liao is supported by the Ministry of Science and Technology, Taiwan (Grant No. MOST 105-2112-M- 008-001-MY3 and 107-2112-M-008 -007 -MY3) and by the National Center for Theoretical Sciences, Taiwan.
Thomas Allison and Christopher Corder acknowledge support by the U.S. Air Force Office of Scientific Research under Award Nos. FA9550- 16-1-0164 and FA9550-13-1-0109.
Volker Sch\"unemann thanks the German Federal Ministery of Education and Research and the Spin+X SFB/TRR 173 of the German Science Foundation for financial support.
Stephen Cramer acknowledges funding from NIOH grant GM65440. We thank Thomas Pfeifer for helpful discussions.
\end{acknowledgments}
%
\input{xfelo_18Mar2019.bbl}
%
\end{document}

%% file: xfelo_18Mar2019.bbl
%

%% file: xfelo_25Mar2019.bbl
\begin{thebibliography}{160}%
\makeatletter
\providecommand \@ifxundefined [1]{%
 \@ifx{#1\undefined}
}%
\providecommand \@ifnum [1]{%
 \ifnum #1\expandafter \@firstoftwo
 \else \expandafter \@secondoftwo
 \fi
}%
\providecommand \@ifx [1]{%
 \ifx #1\expandafter \@firstoftwo
 \else \expandafter \@secondoftwo
 \fi
}%
\providecommand \natexlab [1]{#1}%
\providecommand \enquote  [1]{``#1''}%
\providecommand \bibnamefont  [1]{#1}%
\providecommand \bibfnamefont [1]{#1}%
\providecommand \citenamefont [1]{#1}%
\providecommand \href@noop [0]{\@secondoftwo}%
\providecommand \href [0]{\begingroup \@sanitize@url \@href}%
\providecommand \@href[1]{\@@startlink{#1}\@@href}%
\providecommand \@@href[1]{\endgroup#1\@@endlink}%
\providecommand \@sanitize@url [0]{\catcode `\\12\catcode `\$12\catcode
  `\&12\catcode `\#12\catcode `\^12\catcode `\_12\catcode `\%12\relax}%
\providecommand \@@startlink[1]{}%
\providecommand \@@endlink[0]{}%
\providecommand \url  [0]{\begingroup\@sanitize@url \@url }%
\providecommand \@url [1]{\endgroup\@href {#1}{\urlprefix }}%
\providecommand \urlprefix  [0]{URL }%
\providecommand \Eprint [0]{\href }%
\providecommand \doibase [0]{http://dx.doi.org/}%
\providecommand \selectlanguage [0]{\@gobble}%
\providecommand \bibinfo  [0]{\@secondoftwo}%
\providecommand \bibfield  [0]{\@secondoftwo}%
\providecommand \translation [1]{[#1]}%
\providecommand \BibitemOpen [0]{}%
\providecommand \bibitemStop [0]{}%
\providecommand \bibitemNoStop [0]{.\EOS\space}%
\providecommand \EOS [0]{\spacefactor3000\relax}%
\providecommand \BibitemShut  [1]{\csname bibitem#1\endcsname}%
\let\auto@bib@innerbib\@empty
\bibitem [{\citenamefont {Kolodziej}\ and\ \citenamefont
  {Maxwell}(2016)}]{Kolodziej2016}%
  \BibitemOpen
  \bibfield  {author} {\bibinfo {author} {\bibfnamefont {T.}~\bibnamefont
  {Kolodziej}}\ and\ \bibinfo {author} {\bibfnamefont {T.}~\bibnamefont
  {Maxwell}},\ }\href@noop {} {\bibfield  {journal} {\bibinfo  {journal}
  {Synchrotron Radiation News}\ }\textbf {\bibinfo {volume} {29}},\ \bibinfo
  {pages} {31} (\bibinfo {year} {2016})}\BibitemShut {NoStop}%
\bibitem [{\citenamefont {Kim}\ \emph {et~al.}(2008)\citenamefont {Kim},
  \citenamefont {Shvyd'ko},\ and\ \citenamefont {Reiche}}]{Kim2008}%
  \BibitemOpen
  \bibfield  {author} {\bibinfo {author} {\bibfnamefont {K.-J.}\ \bibnamefont
  {Kim}}, \bibinfo {author} {\bibfnamefont {Y.}~\bibnamefont {Shvyd'ko}}, \
  and\ \bibinfo {author} {\bibfnamefont {S.}~\bibnamefont {Reiche}},\ }\href
  {\doibase 10.1103/PhysRevLett.100.244802} {\bibfield  {journal} {\bibinfo
  {journal} {Phys. Rev. Lett.}\ }\textbf {\bibinfo {volume} {100}},\ \bibinfo
  {pages} {244802} (\bibinfo {year} {2008})}\BibitemShut {NoStop}%
\bibitem [{\citenamefont {Kim}\ and\ \citenamefont {Shvyd'ko}(2009)}]{Kim2009}%
  \BibitemOpen
  \bibfield  {author} {\bibinfo {author} {\bibfnamefont {K.-J.}\ \bibnamefont
  {Kim}}\ and\ \bibinfo {author} {\bibfnamefont {Y.~V.}\ \bibnamefont
  {Shvyd'ko}},\ }\href {\doibase 10.1103/PhysRevSTAB.12.030703} {\bibfield
  {journal} {\bibinfo  {journal} {Phys. Rev. ST Accel. Beams}\ }\textbf
  {\bibinfo {volume} {12}},\ \bibinfo {pages} {030703} (\bibinfo {year}
  {2009})}\BibitemShut {NoStop}%
\bibitem [{\citenamefont {Lindberg}\ \emph {et~al.}(2011)\citenamefont
  {Lindberg}, \citenamefont {Kim}, \citenamefont {Shvyd'ko},\ and\
  \citenamefont {Fawley}}]{Lindberg2011}%
  \BibitemOpen
  \bibfield  {author} {\bibinfo {author} {\bibfnamefont {R.~R.}\ \bibnamefont
  {Lindberg}}, \bibinfo {author} {\bibfnamefont {K.-J.}\ \bibnamefont {Kim}},
  \bibinfo {author} {\bibfnamefont {Y.}~\bibnamefont {Shvyd'ko}}, \ and\
  \bibinfo {author} {\bibfnamefont {W.~M.}\ \bibnamefont {Fawley}},\ }\href
  {\doibase 10.1103/PhysRevSTAB.14.010701} {\bibfield  {journal} {\bibinfo
  {journal} {Phys. Rev. ST Accel. Beams}\ }\textbf {\bibinfo {volume} {14}},\
  \bibinfo {pages} {010701} (\bibinfo {year} {2011})}\BibitemShut {NoStop}%
\bibitem [{\citenamefont {Shvyd'ko}\ \emph {et~al.}(2017)\citenamefont
  {Shvyd'ko}, \citenamefont {Blank},\ and\ \citenamefont
  {Terentyev}}]{Shvydko2017}%
  \BibitemOpen
  \bibfield  {author} {\bibinfo {author} {\bibfnamefont {Y.}~\bibnamefont
  {Shvyd'ko}}, \bibinfo {author} {\bibfnamefont {V.}~\bibnamefont {Blank}}, \
  and\ \bibinfo {author} {\bibfnamefont {S.}~\bibnamefont {Terentyev}},\ }\href
  {\doibase doi:10.1557/mrs.2017.119} {\bibfield  {journal} {\bibinfo
  {journal} {MRS Bulletin}\ }\textbf {\bibinfo {volume} {42}},\ \bibinfo
  {pages} {437} (\bibinfo {year} {2017})}\BibitemShut {NoStop}%
\bibitem [{\citenamefont {Kondratenko}\ and\ \citenamefont
  {Saldin}(1980)}]{Kondratenko1980}%
  \BibitemOpen
  \bibfield  {author} {\bibinfo {author} {\bibfnamefont {A.}~\bibnamefont
  {Kondratenko}}\ and\ \bibinfo {author} {\bibfnamefont {E.}~\bibnamefont
  {Saldin}},\ }in\ \href@noop {} {\emph {\bibinfo {booktitle} {Particle
  Accelerators}}},\ Vol.~\bibinfo {volume} {10}\ (\bibinfo  {publisher} {Gordon
  and Breach.},\ \bibinfo {year} {1980})\ p.\ \bibinfo {pages}
  {207}\BibitemShut {NoStop}%
\bibitem [{\citenamefont {Bonifacio}\ \emph {et~al.}(1984)\citenamefont
  {Bonifacio}, \citenamefont {Pellegrini},\ and\ \citenamefont
  {Narducci}}]{Bonifacio1984}%
  \BibitemOpen
  \bibfield  {author} {\bibinfo {author} {\bibfnamefont {R.}~\bibnamefont
  {Bonifacio}}, \bibinfo {author} {\bibfnamefont {C.}~\bibnamefont
  {Pellegrini}}, \ and\ \bibinfo {author} {\bibfnamefont {L.~M.}\ \bibnamefont
  {Narducci}},\ }\href {\doibase https://doi.org/10.1016/0030-4018(84)90105-6}
  {\bibfield  {journal} {\bibinfo  {journal} {Optics Communications}\ }\textbf
  {\bibinfo {volume} {50}},\ \bibinfo {pages} {373 } (\bibinfo {year}
  {1984})}\BibitemShut {NoStop}%
\bibitem [{\citenamefont {Huang}\ and\ \citenamefont {Kim}(2007)}]{Huang2007}%
  \BibitemOpen
  \bibfield  {author} {\bibinfo {author} {\bibfnamefont {Z.}~\bibnamefont
  {Huang}}\ and\ \bibinfo {author} {\bibfnamefont {K.-J.}\ \bibnamefont
  {Kim}},\ }\href {\doibase 10.1103/PhysRevSTAB.10.034801} {\bibfield
  {journal} {\bibinfo  {journal} {Phys. Rev. ST Accel. Beams}\ }\textbf
  {\bibinfo {volume} {10}},\ \bibinfo {pages} {034801} (\bibinfo {year}
  {2007})}\BibitemShut {NoStop}%
\bibitem [{\citenamefont {Kim}\ \emph {et~al.}(2012{\natexlab{a}})\citenamefont
  {Kim}, \citenamefont {Shvyd'ko},\ and\ \citenamefont {Lindberg}}]{Kim2012}%
  \BibitemOpen
  \bibfield  {author} {\bibinfo {author} {\bibfnamefont {K.-J.}\ \bibnamefont
  {Kim}}, \bibinfo {author} {\bibfnamefont {Y.~V.}\ \bibnamefont {Shvyd'ko}}, \
  and\ \bibinfo {author} {\bibfnamefont {R.~R.}\ \bibnamefont {Lindberg}},\
  }\href {\doibase 10.1080/08940886.2012.645421} {\bibfield  {journal}
  {\bibinfo  {journal} {Synchrotron Radiation News}\ }\textbf {\bibinfo
  {volume} {25}},\ \bibinfo {pages} {25} (\bibinfo {year}
  {2012}{\natexlab{a}})}\BibitemShut {NoStop}%
\bibitem [{\citenamefont {Kolodziej}\ \emph {et~al.}(2018)\citenamefont
  {Kolodziej}, \citenamefont {Shvyd'ko}, \citenamefont {Shu}, \citenamefont
  {Kearney}, \citenamefont {Stoupin}, \citenamefont {Liu}, \citenamefont {Gog},
  \citenamefont {Walko}, \citenamefont {Wang}, \citenamefont {Said},
  \citenamefont {Roberts}, \citenamefont {Goetze}, \citenamefont {Baldini},
  \citenamefont {Yang}, \citenamefont {Fister}, \citenamefont {Blank},
  \citenamefont {Terentyev},\ and\ \citenamefont {Kim}}]{Kolodziej2018}%
  \BibitemOpen
  \bibfield  {author} {\bibinfo {author} {\bibfnamefont {T.}~\bibnamefont
  {Kolodziej}}, \bibinfo {author} {\bibfnamefont {Y.}~\bibnamefont {Shvyd'ko}},
  \bibinfo {author} {\bibfnamefont {D.}~\bibnamefont {Shu}}, \bibinfo {author}
  {\bibfnamefont {S.}~\bibnamefont {Kearney}}, \bibinfo {author} {\bibfnamefont
  {S.}~\bibnamefont {Stoupin}}, \bibinfo {author} {\bibfnamefont
  {W.}~\bibnamefont {Liu}}, \bibinfo {author} {\bibfnamefont {T.}~\bibnamefont
  {Gog}}, \bibinfo {author} {\bibfnamefont {D.}~\bibnamefont {Walko}}, \bibinfo
  {author} {\bibfnamefont {J.}~\bibnamefont {Wang}}, \bibinfo {author}
  {\bibfnamefont {A.}~\bibnamefont {Said}}, \bibinfo {author} {\bibfnamefont
  {T.}~\bibnamefont {Roberts}}, \bibinfo {author} {\bibfnamefont
  {K.}~\bibnamefont {Goetze}}, \bibinfo {author} {\bibfnamefont
  {M.}~\bibnamefont {Baldini}}, \bibinfo {author} {\bibfnamefont
  {W.}~\bibnamefont {Yang}}, \bibinfo {author} {\bibfnamefont {T.}~\bibnamefont
  {Fister}}, \bibinfo {author} {\bibfnamefont {V.}~\bibnamefont {Blank}},
  \bibinfo {author} {\bibfnamefont {S.}~\bibnamefont {Terentyev}}, \ and\
  \bibinfo {author} {\bibfnamefont {K.-J.}\ \bibnamefont {Kim}},\ }\href@noop
  {} {\bibfield  {journal} {\bibinfo  {journal} {Journal of Synchrotron
  Radiation}\ }\textbf {\bibinfo {volume} {25}},\ \bibinfo {pages} {1022}
  (\bibinfo {year} {2018})}\BibitemShut {NoStop}%
\bibitem [{\citenamefont {Emma}\ \emph {et~al.}(2010)\citenamefont {Emma} \emph
  {et~al.}}]{Emma2010}%
  \BibitemOpen
  \bibfield  {author} {\bibinfo {author} {\bibfnamefont {P.}~\bibnamefont
  {Emma}} \emph {et~al.},\ }\href@noop {} {\bibfield  {journal} {\bibinfo
  {journal} {Nature Photonics}\ }\textbf {\bibinfo {volume} {4}},\ \bibinfo
  {pages} {641} (\bibinfo {year} {2010})}\BibitemShut {NoStop}%
\bibitem [{\citenamefont {Ishikawa}\ \emph {et~al.}(2012)\citenamefont
  {Ishikawa} \emph {et~al.}}]{Ishikawa2012}%
  \BibitemOpen
  \bibfield  {author} {\bibinfo {author} {\bibfnamefont {T.}~\bibnamefont
  {Ishikawa}} \emph {et~al.},\ }\href@noop {} {\bibfield  {journal} {\bibinfo
  {journal} {Nature Photonics}\ }\textbf {\bibinfo {volume} {6}},\ \bibinfo
  {pages} {540} (\bibinfo {year} {2012})}\BibitemShut {NoStop}%
\bibitem [{\citenamefont {Altarelli}\ \emph {et~al.}(2007)\citenamefont
  {Altarelli} \emph {et~al.}}]{Altarelli2007a}%
  \BibitemOpen
  \bibfield  {author} {\bibinfo {author} {\bibfnamefont {M.}~\bibnamefont
  {Altarelli}} \emph {et~al.},\ }\href@noop {} {\emph {\bibinfo {title} {The
  European X-Ray free-electron laser, Technical Design Report, DESY
  2006-097}}},\ \bibinfo {type} {Tech. Rep.}\ (\bibinfo  {institution}
  {Deutsches Elektronen-Synchrotron},\ \bibinfo {year} {2007})\BibitemShut
  {NoStop}%
\bibitem [{\citenamefont {Madsen}\ and\ \citenamefont
  {Sinn}(2017)}]{Madsen2017}%
  \BibitemOpen
  \bibfield  {author} {\bibinfo {author} {\bibfnamefont {A.}~\bibnamefont
  {Madsen}}\ and\ \bibinfo {author} {\bibfnamefont {H.}~\bibnamefont {Sinn}},\
  }\href@noop {} {\bibfield  {journal} {\bibinfo  {journal} {CERN Courier}\
  }\textbf {\bibinfo {volume} {57}},\ \bibinfo {pages} {19} (\bibinfo {year}
  {2017})}\BibitemShut {NoStop}%
\bibitem [{\citenamefont {Weise}\ and\ \citenamefont
  {Decking}(2017)}]{Weise2017}%
  \BibitemOpen
  \bibfield  {author} {\bibinfo {author} {\bibfnamefont {H.}~\bibnamefont
  {Weise}}\ and\ \bibinfo {author} {\bibfnamefont {W.}~\bibnamefont
  {Decking}},\ }\href@noop {} {\bibfield  {journal} {\bibinfo  {journal} {CERN
  Courier}\ }\textbf {\bibinfo {volume} {57}},\ \bibinfo {pages} {25} (\bibinfo
  {year} {2017})}\BibitemShut {NoStop}%
\bibitem [{\citenamefont {Tschentscher}\ \emph {et~al.}(2017)\citenamefont
  {Tschentscher}, \citenamefont {Bressler}, \citenamefont {Gr{\"u}nert},
  \citenamefont {Madsen}, \citenamefont {Mancuso}, \citenamefont {Meyer},
  \citenamefont {Scherz}, \citenamefont {Sinn},\ and\ \citenamefont
  {Zastrau}}]{Tschentscher2017}%
  \BibitemOpen
  \bibfield  {author} {\bibinfo {author} {\bibfnamefont {T.}~\bibnamefont
  {Tschentscher}}, \bibinfo {author} {\bibfnamefont {C.}~\bibnamefont
  {Bressler}}, \bibinfo {author} {\bibfnamefont {J.}~\bibnamefont
  {Gr{\"u}nert}}, \bibinfo {author} {\bibfnamefont {A.}~\bibnamefont {Madsen}},
  \bibinfo {author} {\bibfnamefont {A.~P.}\ \bibnamefont {Mancuso}}, \bibinfo
  {author} {\bibfnamefont {M.}~\bibnamefont {Meyer}}, \bibinfo {author}
  {\bibfnamefont {A.}~\bibnamefont {Scherz}}, \bibinfo {author} {\bibfnamefont
  {H.}~\bibnamefont {Sinn}}, \ and\ \bibinfo {author} {\bibfnamefont
  {U.}~\bibnamefont {Zastrau}},\ }\href {\doibase 10.3390/app7060592}
  {\bibfield  {journal} {\bibinfo  {journal} {Applied Sciences}\ }\textbf
  {\bibinfo {volume} {7}},\ \bibinfo {pages} {592} (\bibinfo {year}
  {2017})}\BibitemShut {NoStop}%
\bibitem [{\citenamefont {Schoenlein}\ \emph {et~al.}(2017)\citenamefont
  {Schoenlein}, \citenamefont {Boutet}, \citenamefont {Minitti},\ and\
  \citenamefont {Dunne}}]{Schoenlein2017}%
  \BibitemOpen
  \bibfield  {author} {\bibinfo {author} {\bibfnamefont {R.~W.}\ \bibnamefont
  {Schoenlein}}, \bibinfo {author} {\bibfnamefont {S.}~\bibnamefont {Boutet}},
  \bibinfo {author} {\bibfnamefont {M.~P.}\ \bibnamefont {Minitti}}, \ and\
  \bibinfo {author} {\bibfnamefont {A.}~\bibnamefont {Dunne}},\ }\href
  {\doibase 10.3390/app7080850} {\bibfield  {journal} {\bibinfo  {journal}
  {Applied Sciences}\ }\textbf {\bibinfo {volume} {7}},\ \bibinfo {pages} {850}
  (\bibinfo {year} {2017})}\BibitemShut {NoStop}%
\bibitem [{\citenamefont {Zhu}\ \emph {et~al.}(2017)\citenamefont {Zhu},
  \citenamefont {Zhao}, \citenamefont {Wang}, \citenamefont {Liu},
  \citenamefont {Li}, \citenamefont {Yin},\ and\ \citenamefont
  {Yang}}]{Zhu2017}%
  \BibitemOpen
  \bibfield  {author} {\bibinfo {author} {\bibfnamefont {Z.}~\bibnamefont
  {Zhu}}, \bibinfo {author} {\bibfnamefont {Z.}~\bibnamefont {Zhao}}, \bibinfo
  {author} {\bibfnamefont {D.}~\bibnamefont {Wang}}, \bibinfo {author}
  {\bibfnamefont {Z.}~\bibnamefont {Liu}}, \bibinfo {author} {\bibfnamefont
  {R.}~\bibnamefont {Li}}, \bibinfo {author} {\bibfnamefont {L.}~\bibnamefont
  {Yin}}, \ and\ \bibinfo {author} {\bibfnamefont {Z.}~\bibnamefont {Yang}},\
  }in\ \href@noop {} {\emph {\bibinfo {booktitle} {Proceedings of the 38th FEL
  Conference, Santa Fe, NM, USA}}}\ (\bibinfo {year} {2017})\BibitemShut
  {NoStop}%
\bibitem [{\citenamefont {Sekutowicz}\ \emph {et~al.}(2013)\citenamefont
  {Sekutowicz} \emph {et~al.}}]{Sekutowicz2013}%
  \BibitemOpen
  \bibfield  {author} {\bibinfo {author} {\bibfnamefont {J.}~\bibnamefont
  {Sekutowicz}} \emph {et~al.},\ }in\ \href@noop {} {\emph {\bibinfo
  {booktitle} {Proceedings of the 35th FEL Conference, New York, NY, USA}}}\
  (\bibinfo {year} {2013})\BibitemShut {NoStop}%
\bibitem [{\citenamefont {Qin}\ \emph {et~al.}(2017{\natexlab{a}})\citenamefont
  {Qin}, \citenamefont {Huang}, \citenamefont {Liu}, \citenamefont {Kim},
  \citenamefont {Lindberg}, \citenamefont {Ding}, \citenamefont {Huang},
  \citenamefont {Maxwell}, \citenamefont {Bane},\ and\ \citenamefont
  {Marcus}}]{Qin2017a}%
  \BibitemOpen
  \bibfield  {author} {\bibinfo {author} {\bibfnamefont {W.}~\bibnamefont
  {Qin}}, \bibinfo {author} {\bibfnamefont {S.}~\bibnamefont {Huang}}, \bibinfo
  {author} {\bibfnamefont {K.~X.}\ \bibnamefont {Liu}}, \bibinfo {author}
  {\bibfnamefont {K.-J.}\ \bibnamefont {Kim}}, \bibinfo {author} {\bibfnamefont
  {R.~R.}\ \bibnamefont {Lindberg}}, \bibinfo {author} {\bibfnamefont
  {Y.}~\bibnamefont {Ding}}, \bibinfo {author} {\bibfnamefont {Z.}~\bibnamefont
  {Huang}}, \bibinfo {author} {\bibfnamefont {T.}~\bibnamefont {Maxwell}},
  \bibinfo {author} {\bibfnamefont {K.}~\bibnamefont {Bane}}, \ and\ \bibinfo
  {author} {\bibfnamefont {G.}~\bibnamefont {Marcus}},\ }in\ \href@noop {}
  {\emph {\bibinfo {booktitle} {Proceedings of the 38th FEL Conference, Santa
  Fe, NM, USA}}}\ (\bibinfo {year} {2017})\BibitemShut {NoStop}%
\bibitem [{\citenamefont {Cai}\ \emph {et~al.}(2012)\citenamefont {Cai},
  \citenamefont {Bane}, \citenamefont {Hettel}, \citenamefont {Nosochkov},
  \citenamefont {Wang},\ and\ \citenamefont {Borland}}]{Cai2012}%
  \BibitemOpen
  \bibfield  {author} {\bibinfo {author} {\bibfnamefont {Y.}~\bibnamefont
  {Cai}}, \bibinfo {author} {\bibfnamefont {K.}~\bibnamefont {Bane}}, \bibinfo
  {author} {\bibfnamefont {R.}~\bibnamefont {Hettel}}, \bibinfo {author}
  {\bibfnamefont {Y.}~\bibnamefont {Nosochkov}}, \bibinfo {author}
  {\bibfnamefont {M.-H.}\ \bibnamefont {Wang}}, \ and\ \bibinfo {author}
  {\bibfnamefont {M.}~\bibnamefont {Borland}},\ }\href {\doibase
  10.1103/PhysRevSTAB.15.054002} {\bibfield  {journal} {\bibinfo  {journal}
  {Phys. Rev. ST Accel. Beams}\ }\textbf {\bibinfo {volume} {15}},\ \bibinfo
  {pages} {054002} (\bibinfo {year} {2012})}\BibitemShut {NoStop}%
\bibitem [{\citenamefont {Geloni}\ \emph {et~al.}(2011)\citenamefont {Geloni},
  \citenamefont {Kocharyan},\ and\ \citenamefont {Saldin}}]{Geloni2011}%
  \BibitemOpen
  \bibfield  {author} {\bibinfo {author} {\bibfnamefont {G.}~\bibnamefont
  {Geloni}}, \bibinfo {author} {\bibfnamefont {V.}~\bibnamefont {Kocharyan}}, \
  and\ \bibinfo {author} {\bibfnamefont {E.}~\bibnamefont {Saldin}},\ }\href
  {\doibase 10.1080/09500340.2011.586473} {\bibfield  {journal} {\bibinfo
  {journal} {Journal of Modern Optics}\ }\textbf {\bibinfo {volume} {58}},\
  \bibinfo {pages} {1391} (\bibinfo {year} {2011})}\BibitemShut {NoStop}%
\bibitem [{\citenamefont {Amann}\ \emph {et~al.}(2012)\citenamefont {Amann}
  \emph {et~al.}}]{Amann2012}%
  \BibitemOpen
  \bibfield  {author} {\bibinfo {author} {\bibfnamefont {J.}~\bibnamefont
  {Amann}} \emph {et~al.},\ }\href@noop {} {\bibfield  {journal} {\bibinfo
  {journal} {Nature Photonics}\ }\textbf {\bibinfo {volume} {6}},\ \bibinfo
  {pages} {693} (\bibinfo {year} {2012})}\BibitemShut {NoStop}%
\bibitem [{\citenamefont {Chubar}\ \emph {et~al.}(2016)\citenamefont {Chubar},
  \citenamefont {Geloni}, \citenamefont {Kocharyan}, \citenamefont {Madsen},
  \citenamefont {Saldin}, \citenamefont {Serkez}, \citenamefont {Shvyd'ko},\
  and\ \citenamefont {Sutter}}]{Chubar2016}%
  \BibitemOpen
  \bibfield  {author} {\bibinfo {author} {\bibfnamefont {O.}~\bibnamefont
  {Chubar}}, \bibinfo {author} {\bibfnamefont {G.}~\bibnamefont {Geloni}},
  \bibinfo {author} {\bibfnamefont {V.}~\bibnamefont {Kocharyan}}, \bibinfo
  {author} {\bibfnamefont {A.}~\bibnamefont {Madsen}}, \bibinfo {author}
  {\bibfnamefont {E.}~\bibnamefont {Saldin}}, \bibinfo {author} {\bibfnamefont
  {S.}~\bibnamefont {Serkez}}, \bibinfo {author} {\bibfnamefont
  {Y.}~\bibnamefont {Shvyd'ko}}, \ and\ \bibinfo {author} {\bibfnamefont
  {J.}~\bibnamefont {Sutter}},\ }\href {\doibase 10.1107/S1600577515024844}
  {\bibfield  {journal} {\bibinfo  {journal} {Journal of Synchrotron
  Radiation}\ }\textbf {\bibinfo {volume} {23}},\ \bibinfo {pages} {410}
  (\bibinfo {year} {2016})}\BibitemShut {NoStop}%
\bibitem [{\citenamefont {Lindberg}\ \emph {et~al.}(2013)\citenamefont
  {Lindberg}, \citenamefont {Kim}, \citenamefont {Cai}, \citenamefont {Ding},\
  and\ \citenamefont {Huang}}]{Lindberg2013}%
  \BibitemOpen
  \bibfield  {author} {\bibinfo {author} {\bibfnamefont {R.~R.}\ \bibnamefont
  {Lindberg}}, \bibinfo {author} {\bibfnamefont {K.~J.}\ \bibnamefont {Kim}},
  \bibinfo {author} {\bibfnamefont {Y.}~\bibnamefont {Cai}}, \bibinfo {author}
  {\bibfnamefont {Y.}~\bibnamefont {Ding}}, \ and\ \bibinfo {author}
  {\bibfnamefont {Z.}~\bibnamefont {Huang}},\ }in\ \href@noop {} {\emph
  {\bibinfo {booktitle} {Proceedings of the 35th FEL Conference, Manhattan, NY,
  USA}}}\ (\bibinfo {year} {2013})\BibitemShut {NoStop}%
\bibitem [{\citenamefont {Dai}\ \emph {et~al.}(2012)\citenamefont {Dai},
  \citenamefont {Deng},\ and\ \citenamefont {Dai}}]{Dai2012}%
  \BibitemOpen
  \bibfield  {author} {\bibinfo {author} {\bibfnamefont {J.}~\bibnamefont
  {Dai}}, \bibinfo {author} {\bibfnamefont {H.}~\bibnamefont {Deng}}, \ and\
  \bibinfo {author} {\bibfnamefont {Z.}~\bibnamefont {Dai}},\ }\href {\doibase
  10.1103/PhysRevLett.108.034802} {\bibfield  {journal} {\bibinfo  {journal}
  {Phys. Rev. Lett.}\ }\textbf {\bibinfo {volume} {108}},\ \bibinfo {pages}
  {034802} (\bibinfo {year} {2012})}\BibitemShut {NoStop}%
\bibitem [{\citenamefont {Yu}(1991)}]{Yu1991}%
  \BibitemOpen
  \bibfield  {author} {\bibinfo {author} {\bibfnamefont {L.~H.}\ \bibnamefont
  {Yu}},\ }\href {\doibase 10.1103/PhysRevA.44.5178} {\bibfield  {journal}
  {\bibinfo  {journal} {Phys. Rev. A}\ }\textbf {\bibinfo {volume} {44}},\
  \bibinfo {pages} {5178} (\bibinfo {year} {1991})}\BibitemShut {NoStop}%
\bibitem [{\citenamefont {Qin}\ \emph {et~al.}(2017{\natexlab{b}})\citenamefont
  {Qin}, \citenamefont {Kim}, \citenamefont {Lindberg},\ and\ \citenamefont
  {Wu}}]{Qin2017b}%
  \BibitemOpen
  \bibfield  {author} {\bibinfo {author} {\bibfnamefont {W.}~\bibnamefont
  {Qin}}, \bibinfo {author} {\bibfnamefont {K.-J.}\ \bibnamefont {Kim}},
  \bibinfo {author} {\bibfnamefont {R.~R.}\ \bibnamefont {Lindberg}}, \ and\
  \bibinfo {author} {\bibfnamefont {J.}~\bibnamefont {Wu}},\ }in\ \href@noop {}
  {\emph {\bibinfo {booktitle} {Proceedings of the 38th FEL Conference, Santa
  Fe, NM, USA}}}\ (\bibinfo {year} {2017})\BibitemShut {NoStop}%
\bibitem [{\citenamefont {Sheffield}\ \emph {et~al.}(2017)\citenamefont
  {Sheffield}, \citenamefont {Barnes},\ and\ \citenamefont
  {Tapia}}]{Sheffield2017}%
  \BibitemOpen
  \bibfield  {author} {\bibinfo {author} {\bibfnamefont {R.~L.}\ \bibnamefont
  {Sheffield}}, \bibinfo {author} {\bibfnamefont {C.~W.}\ \bibnamefont
  {Barnes}}, \ and\ \bibinfo {author} {\bibfnamefont {J.~P.}\ \bibnamefont
  {Tapia}},\ }in\ \href@noop {} {\emph {\bibinfo {booktitle} {Proceedings of
  the 38th FEL Conference, Santa Fe, NM, USA}}}\ (\bibinfo {year}
  {2017})\BibitemShut {NoStop}%
\bibitem [{\citenamefont {Adams}\ and\ \citenamefont {Kim}(2015)}]{Adams2015}%
  \BibitemOpen
  \bibfield  {author} {\bibinfo {author} {\bibfnamefont {B.~W.}\ \bibnamefont
  {Adams}}\ and\ \bibinfo {author} {\bibfnamefont {K.-J.}\ \bibnamefont
  {Kim}},\ }\href {\doibase 10.1103/PhysRevSTAB.18.030711} {\bibfield
  {journal} {\bibinfo  {journal} {Phys. Rev. ST Accel. Beams}\ }\textbf
  {\bibinfo {volume} {18}},\ \bibinfo {pages} {030711} (\bibinfo {year}
  {2015})}\BibitemShut {NoStop}%
\bibitem [{\citenamefont {Gerdau}\ and\ \citenamefont
  {de~Waard}(1999)}]{Gerdau1999}%
  \BibitemOpen
  \bibinfo {editor} {\bibfnamefont {E.}~\bibnamefont {Gerdau}}\ and\ \bibinfo
  {editor} {\bibfnamefont {H.}~\bibnamefont {de~Waard}},\ eds.,\ \href@noop {}
  {\emph {\bibinfo {title} {Nuclear Resonant Scattering of Synchrotron
  Radiation, Part A}}},\ Hyperfine Interactions, Vol. 123/124\ (\bibinfo
  {publisher} {Springer-Verlag},\ \bibinfo {year} {1999})\BibitemShut {NoStop}%
\bibitem [{\citenamefont {Gerdau}\ and\ \citenamefont
  {de~Waard}(2000)}]{Gerdau2000}%
  \BibitemOpen
  \bibinfo {editor} {\bibfnamefont {E.}~\bibnamefont {Gerdau}}\ and\ \bibinfo
  {editor} {\bibfnamefont {H.}~\bibnamefont {de~Waard}},\ eds.,\ \href@noop {}
  {\emph {\bibinfo {title} {Nuclear Resonant Scattering of Synchrotron
  Radiation, Part B}}},\ Hyperfine Interactions, Vol. 125\ (\bibinfo
  {publisher} {Springer-Verlag},\ \bibinfo {year} {2000})\BibitemShut {NoStop}%
\bibitem [{\citenamefont {R{\"o}hlsberger}(2004)}]{Roehlsberger2004}%
  \BibitemOpen
  \bibfield  {author} {\bibinfo {author} {\bibfnamefont {R.}~\bibnamefont
  {R{\"o}hlsberger}},\ }\href@noop {} {\emph {\bibinfo {title} {Nuclear
  Condensed Matter Physics with Synchrotron Radiation. Basic Principles,
  Methodology and Applications.}}},\ \bibinfo {series} {Springer Tracts in
  Modern Physics}\ No.\ \bibinfo {number} {208}\ (\bibinfo  {publisher}
  {Springer-Verlag Berlin Heidelberg},\ \bibinfo {year} {2004})\BibitemShut
  {NoStop}%
\bibitem [{\citenamefont {Madsen}\ \emph {et~al.}(2015)\citenamefont {Madsen},
  \citenamefont {Fluerasu},\ and\ \citenamefont {Ruta}}]{Madsen2015}%
  \BibitemOpen
  \bibfield  {author} {\bibinfo {author} {\bibfnamefont {A.}~\bibnamefont
  {Madsen}}, \bibinfo {author} {\bibfnamefont {A.}~\bibnamefont {Fluerasu}}, \
  and\ \bibinfo {author} {\bibfnamefont {B.}~\bibnamefont {Ruta}},\ }in\
  \href@noop {} {\emph {\bibinfo {booktitle} {Synchrotron Light Sources and
  Free-Electron Lasers. Accelerator Physics, Instrumentation and Science
  Applications}}},\ \bibinfo {editor} {edited by\ \bibinfo {editor}
  {\bibfnamefont {E.}~\bibnamefont {Jaeschke}}, \bibinfo {editor}
  {\bibfnamefont {S.}~\bibnamefont {Khan}}, \bibinfo {editor} {\bibfnamefont
  {J.~R.}\ \bibnamefont {Schneider}}, \ and\ \bibinfo {editor} {\bibfnamefont
  {J.~B.}\ \bibnamefont {Hastings}}}\ (\bibinfo  {publisher} {Springer
  Verlag},\ \bibinfo {year} {2015})\BibitemShut {NoStop}%
\bibitem [{\citenamefont {Sch{\"u}lke}(2007)}]{Schuelke2007}%
  \BibitemOpen
  \bibfield  {author} {\bibinfo {author} {\bibfnamefont {W.}~\bibnamefont
  {Sch{\"u}lke}},\ }\href@noop {} {\emph {\bibinfo {title} {Electron Dynamics
  by Inelastic X-ray Scattering}}},\ Oxford Series on Synchrotron Radiation\
  (\bibinfo  {publisher} {Oxford University Press},\ \bibinfo {year}
  {2007})\BibitemShut {NoStop}%
\bibitem [{\citenamefont {Baron}(2015)}]{Baron2015}%
  \BibitemOpen
  \bibfield  {author} {\bibinfo {author} {\bibfnamefont {A.~Q.~R.}\
  \bibnamefont {Baron}},\ }in\ \href@noop {} {\emph {\bibinfo {booktitle}
  {Synchrotron Light Sources and Free-Electron Lasers. Accelerator Physics,
  Instrumentation and Science Applications}}},\ \bibinfo {editor} {edited by\
  \bibinfo {editor} {\bibfnamefont {E.}~\bibnamefont {Jaeschke}}, \bibinfo
  {editor} {\bibfnamefont {S.}~\bibnamefont {Khan}}, \bibinfo {editor}
  {\bibfnamefont {J.~R.}\ \bibnamefont {Schneider}}, \ and\ \bibinfo {editor}
  {\bibfnamefont {J.~B.}\ \bibnamefont {Hastings}}}\ (\bibinfo  {publisher}
  {Springer Verlag},\ \bibinfo {year} {2015})\ \bibinfo {note} {high resolution
  IXS I {$\&$} II, see also arXiv:1504.01098}\BibitemShut {NoStop}%
\bibitem [{\citenamefont {Dziewonski}\ and\ \citenamefont
  {Anderson}(1981)}]{Dziewonski1981}%
  \BibitemOpen
  \bibfield  {author} {\bibinfo {author} {\bibfnamefont {A.~M.}\ \bibnamefont
  {Dziewonski}}\ and\ \bibinfo {author} {\bibfnamefont {D.~L.}\ \bibnamefont
  {Anderson}},\ }\href {\doibase https://doi.org/10.1016/0031-9201(81)90046-7}
  {\bibfield  {journal} {\bibinfo  {journal} {Physics of the Earth and
  Planetary Interiors}\ }\textbf {\bibinfo {volume} {25}},\ \bibinfo {pages}
  {297 } (\bibinfo {year} {1981})}\BibitemShut {NoStop}%
\bibitem [{\citenamefont {Nakajima}\ \emph {et~al.}(2015)\citenamefont
  {Nakajima}, \citenamefont {Imada}, \citenamefont {Hirose}, \citenamefont
  {Komabayashi}, \citenamefont {Ozawa}, \citenamefont {Tateno}, \citenamefont
  {Tsutsui}, \citenamefont {Kuwayama},\ and\ \citenamefont
  {Baron}}]{Nakajima2015}%
  \BibitemOpen
  \bibfield  {author} {\bibinfo {author} {\bibfnamefont {Y.}~\bibnamefont
  {Nakajima}}, \bibinfo {author} {\bibfnamefont {S.}~\bibnamefont {Imada}},
  \bibinfo {author} {\bibfnamefont {K.}~\bibnamefont {Hirose}}, \bibinfo
  {author} {\bibfnamefont {T.}~\bibnamefont {Komabayashi}}, \bibinfo {author}
  {\bibfnamefont {H.}~\bibnamefont {Ozawa}}, \bibinfo {author} {\bibfnamefont
  {S.}~\bibnamefont {Tateno}}, \bibinfo {author} {\bibfnamefont
  {S.}~\bibnamefont {Tsutsui}}, \bibinfo {author} {\bibfnamefont
  {Y.}~\bibnamefont {Kuwayama}}, \ and\ \bibinfo {author} {\bibfnamefont
  {A.~Q.~R.}\ \bibnamefont {Baron}},\ }\href@noop {} {\bibfield  {journal}
  {\bibinfo  {journal} {Nature Communications}\ }\textbf {\bibinfo {volume}
  {6}},\ \bibinfo {pages} {8942} (\bibinfo {year} {2015})}\BibitemShut
  {NoStop}%
\bibitem [{\citenamefont {Sakamaki}\ \emph {et~al.}(2016)\citenamefont
  {Sakamaki}, \citenamefont {Ohtani}, \citenamefont {Fukui}, \citenamefont
  {Kamada}, \citenamefont {Takahashi}, \citenamefont {Sakairi}, \citenamefont
  {Takahata}, \citenamefont {Sakai}, \citenamefont {Tsutsui}, \citenamefont
  {Ishikawa}, \citenamefont {Shiraishi}, \citenamefont {Seto}, \citenamefont
  {Tsuchiya},\ and\ \citenamefont {Baron}}]{Sakamaki2016}%
  \BibitemOpen
  \bibfield  {author} {\bibinfo {author} {\bibfnamefont {T.}~\bibnamefont
  {Sakamaki}}, \bibinfo {author} {\bibfnamefont {E.}~\bibnamefont {Ohtani}},
  \bibinfo {author} {\bibfnamefont {H.}~\bibnamefont {Fukui}}, \bibinfo
  {author} {\bibfnamefont {S.}~\bibnamefont {Kamada}}, \bibinfo {author}
  {\bibfnamefont {S.}~\bibnamefont {Takahashi}}, \bibinfo {author}
  {\bibfnamefont {T.}~\bibnamefont {Sakairi}}, \bibinfo {author} {\bibfnamefont
  {A.}~\bibnamefont {Takahata}}, \bibinfo {author} {\bibfnamefont
  {T.}~\bibnamefont {Sakai}}, \bibinfo {author} {\bibfnamefont
  {S.}~\bibnamefont {Tsutsui}}, \bibinfo {author} {\bibfnamefont
  {D.}~\bibnamefont {Ishikawa}}, \bibinfo {author} {\bibfnamefont
  {R.}~\bibnamefont {Shiraishi}}, \bibinfo {author} {\bibfnamefont
  {Y.}~\bibnamefont {Seto}}, \bibinfo {author} {\bibfnamefont {T.}~\bibnamefont
  {Tsuchiya}}, \ and\ \bibinfo {author} {\bibfnamefont {A.~Q.~R.}\ \bibnamefont
  {Baron}},\ }\href {\doibase 10.1126/sciadv.1500802} {\bibfield  {journal}
  {\bibinfo  {journal} {Science Advances}\ }\textbf {\bibinfo {volume} {2}},\
  \bibinfo {pages} {e1500802} (\bibinfo {year} {2016})}\BibitemShut {NoStop}%
\bibitem [{\citenamefont {Aynajian}\ \emph {et~al.}(2008)\citenamefont
  {Aynajian}, \citenamefont {Keller}, \citenamefont {Boeri}, \citenamefont
  {Shapiro}, \citenamefont {Habicht},\ and\ \citenamefont
  {Keimer}}]{Aynajian2008}%
  \BibitemOpen
  \bibfield  {author} {\bibinfo {author} {\bibfnamefont {P.}~\bibnamefont
  {Aynajian}}, \bibinfo {author} {\bibfnamefont {T.}~\bibnamefont {Keller}},
  \bibinfo {author} {\bibfnamefont {L.}~\bibnamefont {Boeri}}, \bibinfo
  {author} {\bibfnamefont {S.~M.}\ \bibnamefont {Shapiro}}, \bibinfo {author}
  {\bibfnamefont {K.}~\bibnamefont {Habicht}}, \ and\ \bibinfo {author}
  {\bibfnamefont {B.}~\bibnamefont {Keimer}},\ }\href {\doibase
  10.1126/science.1154115} {\bibfield  {journal} {\bibinfo  {journal}
  {Science}\ }\textbf {\bibinfo {volume} {319}},\ \bibinfo {pages} {1509}
  (\bibinfo {year} {2008})}\BibitemShut {NoStop}%
\bibitem [{\citenamefont {Ament}\ \emph {et~al.}(2011)\citenamefont {Ament},
  \citenamefont {van Veenendaal},\ and\ \citenamefont {van~den
  Brink}}]{Ament2011}%
  \BibitemOpen
  \bibfield  {author} {\bibinfo {author} {\bibfnamefont {L.~J.~P.}\
  \bibnamefont {Ament}}, \bibinfo {author} {\bibfnamefont {M.}~\bibnamefont
  {van Veenendaal}}, \ and\ \bibinfo {author} {\bibfnamefont {J.}~\bibnamefont
  {van~den Brink}},\ }\href@noop {} {\bibfield  {journal} {\bibinfo  {journal}
  {Europhysics Letters}\ }\textbf {\bibinfo {volume} {95}},\ \bibinfo {pages}
  {27008} (\bibinfo {year} {2011})}\BibitemShut {NoStop}%
\bibitem [{\citenamefont {Devereaux}\ \emph {et~al.}(2016)\citenamefont
  {Devereaux}, \citenamefont {Shvaika}, \citenamefont {Wu}, \citenamefont
  {Wohlfeld}, \citenamefont {Jia}, \citenamefont {Wang}, \citenamefont
  {Moritz}, \citenamefont {Chaix}, \citenamefont {Lee}, \citenamefont {Shen},
  \citenamefont {Ghiringhelli},\ and\ \citenamefont
  {Braicovich}}]{Deveraux2016}%
  \BibitemOpen
  \bibfield  {author} {\bibinfo {author} {\bibfnamefont {T.~P.}\ \bibnamefont
  {Devereaux}}, \bibinfo {author} {\bibfnamefont {A.~M.}\ \bibnamefont
  {Shvaika}}, \bibinfo {author} {\bibfnamefont {K.}~\bibnamefont {Wu}},
  \bibinfo {author} {\bibfnamefont {K.}~\bibnamefont {Wohlfeld}}, \bibinfo
  {author} {\bibfnamefont {C.~J.}\ \bibnamefont {Jia}}, \bibinfo {author}
  {\bibfnamefont {Y.}~\bibnamefont {Wang}}, \bibinfo {author} {\bibfnamefont
  {B.}~\bibnamefont {Moritz}}, \bibinfo {author} {\bibfnamefont
  {L.}~\bibnamefont {Chaix}}, \bibinfo {author} {\bibfnamefont {W.-S.}\
  \bibnamefont {Lee}}, \bibinfo {author} {\bibfnamefont {Z.-X.}\ \bibnamefont
  {Shen}}, \bibinfo {author} {\bibfnamefont {G.}~\bibnamefont {Ghiringhelli}},
  \ and\ \bibinfo {author} {\bibfnamefont {L.}~\bibnamefont {Braicovich}},\
  }\href {\doibase 10.1103/PhysRevX.6.041019} {\bibfield  {journal} {\bibinfo
  {journal} {Phys. Rev. X}\ }\textbf {\bibinfo {volume} {6}},\ \bibinfo {pages}
  {041019} (\bibinfo {year} {2016})}\BibitemShut {NoStop}%
\bibitem [{\citenamefont {Reyren}\ \emph {et~al.}(2007)\citenamefont {Reyren},
  \citenamefont {Thiel}, \citenamefont {Caviglia}, \citenamefont {Kourkoutis},
  \citenamefont {Hammerl}, \citenamefont {Richter}, \citenamefont {Schneider},
  \citenamefont {Kopp}, \citenamefont {R{\"u}etschi}, \citenamefont {Jaccard},
  \citenamefont {Gabay}, \citenamefont {Muller}, \citenamefont {Triscone},\
  and\ \citenamefont {Mannhart}}]{Reyren2007}%
  \BibitemOpen
  \bibfield  {author} {\bibinfo {author} {\bibfnamefont {N.}~\bibnamefont
  {Reyren}}, \bibinfo {author} {\bibfnamefont {S.}~\bibnamefont {Thiel}},
  \bibinfo {author} {\bibfnamefont {A.~D.}\ \bibnamefont {Caviglia}}, \bibinfo
  {author} {\bibfnamefont {L.~F.}\ \bibnamefont {Kourkoutis}}, \bibinfo
  {author} {\bibfnamefont {G.}~\bibnamefont {Hammerl}}, \bibinfo {author}
  {\bibfnamefont {C.}~\bibnamefont {Richter}}, \bibinfo {author} {\bibfnamefont
  {C.~W.}\ \bibnamefont {Schneider}}, \bibinfo {author} {\bibfnamefont
  {T.}~\bibnamefont {Kopp}}, \bibinfo {author} {\bibfnamefont {A.-S.}\
  \bibnamefont {R{\"u}etschi}}, \bibinfo {author} {\bibfnamefont
  {D.}~\bibnamefont {Jaccard}}, \bibinfo {author} {\bibfnamefont
  {M.}~\bibnamefont {Gabay}}, \bibinfo {author} {\bibfnamefont {D.~A.}\
  \bibnamefont {Muller}}, \bibinfo {author} {\bibfnamefont {J.-M.}\
  \bibnamefont {Triscone}}, \ and\ \bibinfo {author} {\bibfnamefont
  {J.}~\bibnamefont {Mannhart}},\ }\href {\doibase 10.1126/science.1146006}
  {\bibfield  {journal} {\bibinfo  {journal} {Science}\ }\textbf {\bibinfo
  {volume} {317}},\ \bibinfo {pages} {1196} (\bibinfo {year}
  {2007})}\BibitemShut {NoStop}%
\bibitem [{\citenamefont {Ge}\ \emph {et~al.}(2014)\citenamefont {Ge},
  \citenamefont {Liu}, \citenamefont {Liu}, \citenamefont {Gao}, \citenamefont
  {Qian}, \citenamefont {Xue}, \citenamefont {Liu},\ and\ \citenamefont
  {Jia}}]{Ge2014}%
  \BibitemOpen
  \bibfield  {author} {\bibinfo {author} {\bibfnamefont {J.-F.}\ \bibnamefont
  {Ge}}, \bibinfo {author} {\bibfnamefont {Z.-L.}\ \bibnamefont {Liu}},
  \bibinfo {author} {\bibfnamefont {C.}~\bibnamefont {Liu}}, \bibinfo {author}
  {\bibfnamefont {C.-L.}\ \bibnamefont {Gao}}, \bibinfo {author} {\bibfnamefont
  {D.}~\bibnamefont {Qian}}, \bibinfo {author} {\bibfnamefont {Q.-K.}\
  \bibnamefont {Xue}}, \bibinfo {author} {\bibfnamefont {Y.}~\bibnamefont
  {Liu}}, \ and\ \bibinfo {author} {\bibfnamefont {J.-F.}\ \bibnamefont
  {Jia}},\ }\href@noop {} {\bibfield  {journal} {\bibinfo  {journal} {Nature
  Materials}\ }\textbf {\bibinfo {volume} {14}},\ \bibinfo {pages} {285}
  (\bibinfo {year} {2014})}\BibitemShut {NoStop}%
\bibitem [{\citenamefont {Sette}\ \emph {et~al.}(1995)\citenamefont {Sette},
  \citenamefont {Ruocco}, \citenamefont {Krisch}, \citenamefont {Bergmann},
  \citenamefont {Masciovecchio}, \citenamefont {Mazzacurati}, \citenamefont
  {Signorelli},\ and\ \citenamefont {Verbeni}}]{Sette1995}%
  \BibitemOpen
  \bibfield  {author} {\bibinfo {author} {\bibfnamefont {F.}~\bibnamefont
  {Sette}}, \bibinfo {author} {\bibfnamefont {G.}~\bibnamefont {Ruocco}},
  \bibinfo {author} {\bibfnamefont {M.}~\bibnamefont {Krisch}}, \bibinfo
  {author} {\bibfnamefont {U.}~\bibnamefont {Bergmann}}, \bibinfo {author}
  {\bibfnamefont {C.}~\bibnamefont {Masciovecchio}}, \bibinfo {author}
  {\bibfnamefont {V.}~\bibnamefont {Mazzacurati}}, \bibinfo {author}
  {\bibfnamefont {G.}~\bibnamefont {Signorelli}}, \ and\ \bibinfo {author}
  {\bibfnamefont {R.}~\bibnamefont {Verbeni}},\ }\href {\doibase
  10.1103/PhysRevLett.75.850} {\bibfield  {journal} {\bibinfo  {journal} {Phys.
  Rev. Lett.}\ }\textbf {\bibinfo {volume} {75}},\ \bibinfo {pages} {850}
  (\bibinfo {year} {1995})}\BibitemShut {NoStop}%
\bibitem [{\citenamefont {Monaco}\ \emph {et~al.}(2001)\citenamefont {Monaco},
  \citenamefont {Nardone}, \citenamefont {Sette},\ and\ \citenamefont
  {Verbeni}}]{Monaco2001}%
  \BibitemOpen
  \bibfield  {author} {\bibinfo {author} {\bibfnamefont {G.}~\bibnamefont
  {Monaco}}, \bibinfo {author} {\bibfnamefont {M.}~\bibnamefont {Nardone}},
  \bibinfo {author} {\bibfnamefont {F.}~\bibnamefont {Sette}}, \ and\ \bibinfo
  {author} {\bibfnamefont {R.}~\bibnamefont {Verbeni}},\ }\href {\doibase
  10.1103/PhysRevB.64.212102} {\bibfield  {journal} {\bibinfo  {journal} {Phys.
  Rev. B}\ }\textbf {\bibinfo {volume} {64}},\ \bibinfo {pages} {212102}
  (\bibinfo {year} {2001})}\BibitemShut {NoStop}%
\bibitem [{\citenamefont {Scopigno}\ \emph {et~al.}(2003)\citenamefont
  {Scopigno}, \citenamefont {Pontecorvo}, \citenamefont {Di~Leonardo},
  \citenamefont {Krisch}, \citenamefont {Monaco}, \citenamefont {Ruocco},
  \citenamefont {Ruzicka},\ and\ \citenamefont {Sette}}]{Scopigno2003}%
  \BibitemOpen
  \bibfield  {author} {\bibinfo {author} {\bibfnamefont {T.}~\bibnamefont
  {Scopigno}}, \bibinfo {author} {\bibfnamefont {E.}~\bibnamefont
  {Pontecorvo}}, \bibinfo {author} {\bibfnamefont {R.}~\bibnamefont
  {Di~Leonardo}}, \bibinfo {author} {\bibfnamefont {M.}~\bibnamefont {Krisch}},
  \bibinfo {author} {\bibfnamefont {G.}~\bibnamefont {Monaco}}, \bibinfo
  {author} {\bibfnamefont {G.}~\bibnamefont {Ruocco}}, \bibinfo {author}
  {\bibfnamefont {B.}~\bibnamefont {Ruzicka}}, \ and\ \bibinfo {author}
  {\bibfnamefont {F.}~\bibnamefont {Sette}},\ }\href@noop {} {\bibfield
  {journal} {\bibinfo  {journal} {Journal of Physics: Condensed Matter}\
  }\textbf {\bibinfo {volume} {15}},\ \bibinfo {pages} {S1269} (\bibinfo {year}
  {2003})}\BibitemShut {NoStop}%
\bibitem [{\citenamefont {Drozdov}\ \emph {et~al.}(2015)\citenamefont
  {Drozdov}, \citenamefont {Eremets}, \citenamefont {Troyan}, \citenamefont
  {Ksenofontov},\ and\ \citenamefont {Shylin}}]{Drozdov2015}%
  \BibitemOpen
  \bibfield  {author} {\bibinfo {author} {\bibfnamefont {A.~P.}\ \bibnamefont
  {Drozdov}}, \bibinfo {author} {\bibfnamefont {M.~I.}\ \bibnamefont
  {Eremets}}, \bibinfo {author} {\bibfnamefont {I.~A.}\ \bibnamefont {Troyan}},
  \bibinfo {author} {\bibfnamefont {V.}~\bibnamefont {Ksenofontov}}, \ and\
  \bibinfo {author} {\bibfnamefont {S.~I.}\ \bibnamefont {Shylin}},\
  }\href@noop {} {\bibfield  {journal} {\bibinfo  {journal} {Nature}\ }\textbf
  {\bibinfo {volume} {525}},\ \bibinfo {pages} {73} (\bibinfo {year}
  {2015})}\BibitemShut {NoStop}%
\bibitem [{\citenamefont {Shvyd'ko}\ \emph {et~al.}(2013)\citenamefont
  {Shvyd'ko}, \citenamefont {Stoupin}, \citenamefont {Mundboth},\ and\
  \citenamefont {Kim}}]{Shvydko2013}%
  \BibitemOpen
  \bibfield  {author} {\bibinfo {author} {\bibfnamefont {Y.}~\bibnamefont
  {Shvyd'ko}}, \bibinfo {author} {\bibfnamefont {S.}~\bibnamefont {Stoupin}},
  \bibinfo {author} {\bibfnamefont {K.}~\bibnamefont {Mundboth}}, \ and\
  \bibinfo {author} {\bibfnamefont {J.}~\bibnamefont {Kim}},\ }\href@noop {}
  {\bibfield  {journal} {\bibinfo  {journal} {Phys. Rev. A}\ }\textbf {\bibinfo
  {volume} {87}},\ \bibinfo {pages} {043835} (\bibinfo {year}
  {2013})}\BibitemShut {NoStop}%
\bibitem [{\citenamefont {Shvyd'ko}(2015)}]{Shvydko2015}%
  \BibitemOpen
  \bibfield  {author} {\bibinfo {author} {\bibfnamefont {Y.}~\bibnamefont
  {Shvyd'ko}},\ }\href@noop {} {\bibfield  {journal} {\bibinfo  {journal}
  {Phys. Rev. A}\ }\textbf {\bibinfo {volume} {91}},\ \bibinfo {pages} {053817}
  (\bibinfo {year} {2015})}\BibitemShut {NoStop}%
\bibitem [{\citenamefont {Shvyd'ko}(2016)}]{Shvydko2016}%
  \BibitemOpen
  \bibfield  {author} {\bibinfo {author} {\bibfnamefont {Y.}~\bibnamefont
  {Shvyd'ko}},\ }\href@noop {} {\bibfield  {journal} {\bibinfo  {journal}
  {Phys. Rev. Lett.}\ }\textbf {\bibinfo {volume} {116}},\ \bibinfo {pages}
  {080801} (\bibinfo {year} {2016})}\BibitemShut {NoStop}%
\bibitem [{\citenamefont {Shvyd'ko}(2017)}]{Shvydko2017b}%
  \BibitemOpen
  \bibfield  {author} {\bibinfo {author} {\bibfnamefont {Y.}~\bibnamefont
  {Shvyd'ko}},\ }\href@noop {} {\bibfield  {journal} {\bibinfo  {journal}
  {Phys. Rev. A}\ }\textbf {\bibinfo {volume} {96}},\ \bibinfo {pages} {023804}
  (\bibinfo {year} {2017})}\BibitemShut {NoStop}%
\bibitem [{\citenamefont {Chumakov}\ \emph {et~al.}(1996)\citenamefont
  {Chumakov}, \citenamefont {Baron}, \citenamefont {R\"uffer}, \citenamefont
  {Gr\"unsteudel}, \citenamefont {Gr\"unsteudel},\ and\ \citenamefont
  {Meyer}}]{Chumakov1996}%
  \BibitemOpen
  \bibfield  {author} {\bibinfo {author} {\bibfnamefont {A.~I.}\ \bibnamefont
  {Chumakov}}, \bibinfo {author} {\bibfnamefont {A.~Q.~R.}\ \bibnamefont
  {Baron}}, \bibinfo {author} {\bibfnamefont {R.}~\bibnamefont {R\"uffer}},
  \bibinfo {author} {\bibfnamefont {H.}~\bibnamefont {Gr\"unsteudel}}, \bibinfo
  {author} {\bibfnamefont {H.~F.}\ \bibnamefont {Gr\"unsteudel}}, \ and\
  \bibinfo {author} {\bibfnamefont {A.}~\bibnamefont {Meyer}},\ }\href
  {\doibase 10.1103/PhysRevLett.76.4258} {\bibfield  {journal} {\bibinfo
  {journal} {Phys. Rev. Lett.}\ }\textbf {\bibinfo {volume} {76}},\ \bibinfo
  {pages} {4258} (\bibinfo {year} {1996})}\BibitemShut {NoStop}%
\bibitem [{\citenamefont {Baron}(2013)}]{Baron2013}%
  \BibitemOpen
  \bibfield  {author} {\bibinfo {author} {\bibfnamefont {A.~Q.~R.}\
  \bibnamefont {Baron}},\ }\href {\doibase 10.7566/JPSJS.82SA.SA029} {\bibfield
   {journal} {\bibinfo  {journal} {Journal of the Physical Society of Japan}\
  }\textbf {\bibinfo {volume} {82}},\ \bibinfo {pages} {SA029} (\bibinfo {year}
  {2013})}\BibitemShut {NoStop}%
\bibitem [{\citenamefont {Kim}\ \emph {et~al.}(2012{\natexlab{b}})\citenamefont
  {Kim}, \citenamefont {Casa}, \citenamefont {Upton}, \citenamefont {Gog},
  \citenamefont {Kim}, \citenamefont {Mitchell}, \citenamefont {van
  Veenendaal}, \citenamefont {Daghofer}, \citenamefont {van~den Brink},
  \citenamefont {Khaliullin},\ and\ \citenamefont {Kim}}]{Kim2012b}%
  \BibitemOpen
  \bibfield  {author} {\bibinfo {author} {\bibfnamefont {J.}~\bibnamefont
  {Kim}}, \bibinfo {author} {\bibfnamefont {D.}~\bibnamefont {Casa}}, \bibinfo
  {author} {\bibfnamefont {M.~H.}\ \bibnamefont {Upton}}, \bibinfo {author}
  {\bibfnamefont {T.}~\bibnamefont {Gog}}, \bibinfo {author} {\bibfnamefont
  {Y.-J.}\ \bibnamefont {Kim}}, \bibinfo {author} {\bibfnamefont {J.~F.}\
  \bibnamefont {Mitchell}}, \bibinfo {author} {\bibfnamefont {M.}~\bibnamefont
  {van Veenendaal}}, \bibinfo {author} {\bibfnamefont {M.}~\bibnamefont
  {Daghofer}}, \bibinfo {author} {\bibfnamefont {J.}~\bibnamefont {van~den
  Brink}}, \bibinfo {author} {\bibfnamefont {G.}~\bibnamefont {Khaliullin}}, \
  and\ \bibinfo {author} {\bibfnamefont {B.~J.}\ \bibnamefont {Kim}},\ }\href
  {\doibase 10.1103/PhysRevLett.108.177003} {\bibfield  {journal} {\bibinfo
  {journal} {Phys. Rev. Lett.}\ }\textbf {\bibinfo {volume} {108}},\ \bibinfo
  {pages} {177003} (\bibinfo {year} {2012}{\natexlab{b}})}\BibitemShut
  {NoStop}%
\bibitem [{\citenamefont {Yava\c{s}}\ \emph {et~al.}(2017)\citenamefont
  {Yava\c{s}}, \citenamefont {Sutter}, \citenamefont {Gog}, \citenamefont
  {Wille},\ and\ \citenamefont {Baron}}]{Yavas2017}%
  \BibitemOpen
  \bibfield  {author} {\bibinfo {author} {\bibfnamefont {H.}~\bibnamefont
  {Yava\c{s}}}, \bibinfo {author} {\bibfnamefont {J.~P.}\ \bibnamefont
  {Sutter}}, \bibinfo {author} {\bibfnamefont {T.}~\bibnamefont {Gog}},
  \bibinfo {author} {\bibfnamefont {H.-C.}\ \bibnamefont {Wille}}, \ and\
  \bibinfo {author} {\bibfnamefont {A.~Q.}\ \bibnamefont {Baron}},\ }\href
  {\doibase 10.1557/mrs.2017.94} {\bibfield  {journal} {\bibinfo  {journal}
  {MRS Bulletin}\ }\textbf {\bibinfo {volume} {42}},\ \bibinfo {pages} {424 }
  (\bibinfo {year} {2017})}\BibitemShut {NoStop}%
\bibitem [{\citenamefont {Larson}\ \emph {et~al.}(2007)\citenamefont {Larson},
  \citenamefont {Ku}, \citenamefont {Tischler}, \citenamefont {Lee},
  \citenamefont {Restrepo}, \citenamefont {Eguiluz}, \citenamefont {Zschack},\
  and\ \citenamefont {Finkelstein}}]{Larson2007}%
  \BibitemOpen
  \bibfield  {author} {\bibinfo {author} {\bibfnamefont {B.~C.}\ \bibnamefont
  {Larson}}, \bibinfo {author} {\bibfnamefont {W.}~\bibnamefont {Ku}}, \bibinfo
  {author} {\bibfnamefont {J.~Z.}\ \bibnamefont {Tischler}}, \bibinfo {author}
  {\bibfnamefont {C.-C.}\ \bibnamefont {Lee}}, \bibinfo {author} {\bibfnamefont
  {O.~D.}\ \bibnamefont {Restrepo}}, \bibinfo {author} {\bibfnamefont {A.~G.}\
  \bibnamefont {Eguiluz}}, \bibinfo {author} {\bibfnamefont {P.}~\bibnamefont
  {Zschack}}, \ and\ \bibinfo {author} {\bibfnamefont {K.~D.}\ \bibnamefont
  {Finkelstein}},\ }\href {\doibase 10.1103/PhysRevLett.99.026401} {\bibfield
  {journal} {\bibinfo  {journal} {Phys. Rev. Lett.}\ }\textbf {\bibinfo
  {volume} {99}},\ \bibinfo {pages} {026401} (\bibinfo {year}
  {2007})}\BibitemShut {NoStop}%
\bibitem [{\citenamefont {Sch{\"u}lke}(1982)}]{Schuelke1982}%
  \BibitemOpen
  \bibfield  {author} {\bibinfo {author} {\bibfnamefont {W.}~\bibnamefont
  {Sch{\"u}lke}},\ }\href {\doibase
  https://doi.org/10.1016/0038-1098(82)90856-0} {\bibfield  {journal} {\bibinfo
   {journal} {Solid State Communications}\ }\textbf {\bibinfo {volume} {43}},\
  \bibinfo {pages} {863 } (\bibinfo {year} {1982})}\BibitemShut {NoStop}%
\bibitem [{\citenamefont {Gan}\ \emph {et~al.}(2013)\citenamefont {Gan},
  \citenamefont {Kogar},\ and\ \citenamefont {Abbamonte}}]{Gan2013}%
  \BibitemOpen
  \bibfield  {author} {\bibinfo {author} {\bibfnamefont {Y.}~\bibnamefont
  {Gan}}, \bibinfo {author} {\bibfnamefont {A.}~\bibnamefont {Kogar}}, \ and\
  \bibinfo {author} {\bibfnamefont {P.}~\bibnamefont {Abbamonte}},\ }\href@noop
  {} {\bibfield  {journal} {\bibinfo  {journal} {Chemical Physics}\ }\textbf
  {\bibinfo {volume} {414}},\ \bibinfo {pages} {160} (\bibinfo {year}
  {2013})}\BibitemShut {NoStop}%
\bibitem [{\citenamefont {Sch\"ulke}\ and\ \citenamefont
  {Kaprolat}(1991)}]{Schuelke1991}%
  \BibitemOpen
  \bibfield  {author} {\bibinfo {author} {\bibfnamefont {W.}~\bibnamefont
  {Sch\"ulke}}\ and\ \bibinfo {author} {\bibfnamefont {A.}~\bibnamefont
  {Kaprolat}},\ }\href {\doibase 10.1103/PhysRevLett.67.879} {\bibfield
  {journal} {\bibinfo  {journal} {Phys. Rev. Lett.}\ }\textbf {\bibinfo
  {volume} {67}},\ \bibinfo {pages} {879} (\bibinfo {year} {1991})}\BibitemShut
  {NoStop}%
\bibitem [{\citenamefont {Kohl}(1985)}]{Kohl1985}%
  \BibitemOpen
  \bibfield  {author} {\bibinfo {author} {\bibfnamefont {H.}~\bibnamefont
  {Kohl}},\ }\href@noop {} {\bibfield  {journal} {\bibinfo  {journal} {Physica
  status solidi}\ }\textbf {\bibinfo {volume} {130}},\ \bibinfo {pages} {151}
  (\bibinfo {year} {1985})}\BibitemShut {NoStop}%
\bibitem [{\citenamefont {Van~Hove}(1954)}]{vanHove1954}%
  \BibitemOpen
  \bibfield  {author} {\bibinfo {author} {\bibfnamefont {L.}~\bibnamefont
  {Van~Hove}},\ }\href {\doibase 10.1103/PhysRev.95.249} {\bibfield  {journal}
  {\bibinfo  {journal} {Phys. Rev.}\ }\textbf {\bibinfo {volume} {95}},\
  \bibinfo {pages} {249} (\bibinfo {year} {1954})}\BibitemShut {NoStop}%
\bibitem [{\citenamefont {Iwashita}\ \emph {et~al.}(2017)\citenamefont
  {Iwashita}, \citenamefont {Wu}, \citenamefont {Chen}, \citenamefont
  {Tsutsui}, \citenamefont {Baron},\ and\ \citenamefont
  {Egami}}]{Iwashita2017}%
  \BibitemOpen
  \bibfield  {author} {\bibinfo {author} {\bibfnamefont {T.}~\bibnamefont
  {Iwashita}}, \bibinfo {author} {\bibfnamefont {B.}~\bibnamefont {Wu}},
  \bibinfo {author} {\bibfnamefont {W.-R.}\ \bibnamefont {Chen}}, \bibinfo
  {author} {\bibfnamefont {S.}~\bibnamefont {Tsutsui}}, \bibinfo {author}
  {\bibfnamefont {A.~Q.~R.}\ \bibnamefont {Baron}}, \ and\ \bibinfo {author}
  {\bibfnamefont {T.}~\bibnamefont {Egami}},\ }\href {\doibase
  10.1126/sciadv.1603079} {\bibfield  {journal} {\bibinfo  {journal} {Science
  Advances}\ }\textbf {\bibinfo {volume} {3}},\ \bibinfo {pages} {e1603079}
  (\bibinfo {year} {2017})}\BibitemShut {NoStop}%
\bibitem [{\citenamefont {Huotari}\ \emph {et~al.}(2000)\citenamefont
  {Huotari}, \citenamefont {H\"am\"al\"ainen}, \citenamefont {Manninen},
  \citenamefont {Kaprzyk}, \citenamefont {Bansil}, \citenamefont {Caliebe},
  \citenamefont {Buslaps}, \citenamefont {Honkim\"aki},\ and\ \citenamefont
  {Suortti}}]{Huotari2000}%
  \BibitemOpen
  \bibfield  {author} {\bibinfo {author} {\bibfnamefont {S.}~\bibnamefont
  {Huotari}}, \bibinfo {author} {\bibfnamefont {K.}~\bibnamefont
  {H\"am\"al\"ainen}}, \bibinfo {author} {\bibfnamefont {S.}~\bibnamefont
  {Manninen}}, \bibinfo {author} {\bibfnamefont {S.}~\bibnamefont {Kaprzyk}},
  \bibinfo {author} {\bibfnamefont {A.}~\bibnamefont {Bansil}}, \bibinfo
  {author} {\bibfnamefont {W.}~\bibnamefont {Caliebe}}, \bibinfo {author}
  {\bibfnamefont {T.}~\bibnamefont {Buslaps}}, \bibinfo {author} {\bibfnamefont
  {V.}~\bibnamefont {Honkim\"aki}}, \ and\ \bibinfo {author} {\bibfnamefont
  {P.}~\bibnamefont {Suortti}},\ }\href {\doibase 10.1103/PhysRevB.62.7956}
  {\bibfield  {journal} {\bibinfo  {journal} {Phys. Rev. B}\ }\textbf {\bibinfo
  {volume} {62}},\ \bibinfo {pages} {7956} (\bibinfo {year}
  {2000})}\BibitemShut {NoStop}%
\bibitem [{\citenamefont {Tanaka}\ \emph {et~al.}(2001)\citenamefont {Tanaka},
  \citenamefont {Sakurai}, \citenamefont {Stewart}, \citenamefont {Shiotani},
  \citenamefont {Mijnarends}, \citenamefont {Kaprzyk},\ and\ \citenamefont
  {Bansil}}]{Tanaka2001}%
  \BibitemOpen
  \bibfield  {author} {\bibinfo {author} {\bibfnamefont {Y.}~\bibnamefont
  {Tanaka}}, \bibinfo {author} {\bibfnamefont {Y.}~\bibnamefont {Sakurai}},
  \bibinfo {author} {\bibfnamefont {A.~T.}\ \bibnamefont {Stewart}}, \bibinfo
  {author} {\bibfnamefont {N.}~\bibnamefont {Shiotani}}, \bibinfo {author}
  {\bibfnamefont {P.~E.}\ \bibnamefont {Mijnarends}}, \bibinfo {author}
  {\bibfnamefont {S.}~\bibnamefont {Kaprzyk}}, \ and\ \bibinfo {author}
  {\bibfnamefont {A.}~\bibnamefont {Bansil}},\ }\href {\doibase
  10.1103/PhysRevB.63.045120} {\bibfield  {journal} {\bibinfo  {journal} {Phys.
  Rev. B}\ }\textbf {\bibinfo {volume} {63}},\ \bibinfo {pages} {045120}
  (\bibinfo {year} {2001})}\BibitemShut {NoStop}%
\bibitem [{\citenamefont {Falus}\ \emph {et~al.}(2006)\citenamefont {Falus},
  \citenamefont {Lurio},\ and\ \citenamefont {Mochrie}}]{Falus2006}%
  \BibitemOpen
  \bibfield  {author} {\bibinfo {author} {\bibfnamefont {P.}~\bibnamefont
  {Falus}}, \bibinfo {author} {\bibfnamefont {L.~B.}\ \bibnamefont {Lurio}}, \
  and\ \bibinfo {author} {\bibfnamefont {S.~G.~J.}\ \bibnamefont {Mochrie}},\
  }\href {\doibase 10.1107/S0909049506006789} {\bibfield  {journal} {\bibinfo
  {journal} {Journal of Synchrotron Radiation}\ }\textbf {\bibinfo {volume}
  {13}},\ \bibinfo {pages} {253} (\bibinfo {year} {2006})}\BibitemShut
  {NoStop}%
\bibitem [{\citenamefont {Shpyrko}(2014)}]{Shpyrko2014}%
  \BibitemOpen
  \bibfield  {author} {\bibinfo {author} {\bibfnamefont {O.~G.}\ \bibnamefont
  {Shpyrko}},\ }\href {\doibase 10.1107/S1600577514018232} {\bibfield
  {journal} {\bibinfo  {journal} {Journal of Synchrotron Radiation}\ }\textbf
  {\bibinfo {volume} {21}},\ \bibinfo {pages} {1057} (\bibinfo {year}
  {2014})}\BibitemShut {NoStop}%
\bibitem [{\citenamefont {Sinha}\ \emph {et~al.}(2014)\citenamefont {Sinha},
  \citenamefont {Jiang},\ and\ \citenamefont {Lurio}}]{Sinha2014}%
  \BibitemOpen
  \bibfield  {author} {\bibinfo {author} {\bibfnamefont {S.~K.}\ \bibnamefont
  {Sinha}}, \bibinfo {author} {\bibfnamefont {Z.}~\bibnamefont {Jiang}}, \ and\
  \bibinfo {author} {\bibfnamefont {L.~B.}\ \bibnamefont {Lurio}},\ }\href
  {\doibase 10.1002/adma.201401094} {\bibfield  {journal} {\bibinfo  {journal}
  {Advanced Materials}\ }\textbf {\bibinfo {volume} {26}},\ \bibinfo {pages}
  {7764} (\bibinfo {year} {2014})}\BibitemShut {NoStop}%
\bibitem [{\citenamefont {Sakurai}(2017)}]{Sakurai2017}%
  \BibitemOpen
  \bibfield  {author} {\bibinfo {author} {\bibfnamefont {S.}~\bibnamefont
  {Sakurai}},\ }\href {\doibase 10.1002/pi.5136} {\bibfield  {journal}
  {\bibinfo  {journal} {Polymer International}\ }\textbf {\bibinfo {volume}
  {66}},\ \bibinfo {pages} {237} (\bibinfo {year} {2017})}\BibitemShut
  {NoStop}%
\bibitem [{\citenamefont {Verwohlt}\ \emph {et~al.}(2018)\citenamefont
  {Verwohlt}, \citenamefont {Reiser}, \citenamefont {Randolph}, \citenamefont
  {Matic}, \citenamefont {Medina}, \citenamefont {Madsen}, \citenamefont
  {Sprung}, \citenamefont {Zozulya},\ and\ \citenamefont
  {Gutt}}]{Verwohlt2018}%
  \BibitemOpen
  \bibfield  {author} {\bibinfo {author} {\bibfnamefont {J.}~\bibnamefont
  {Verwohlt}}, \bibinfo {author} {\bibfnamefont {M.}~\bibnamefont {Reiser}},
  \bibinfo {author} {\bibfnamefont {L.}~\bibnamefont {Randolph}}, \bibinfo
  {author} {\bibfnamefont {A.}~\bibnamefont {Matic}}, \bibinfo {author}
  {\bibfnamefont {L.~A.}\ \bibnamefont {Medina}}, \bibinfo {author}
  {\bibfnamefont {A.}~\bibnamefont {Madsen}}, \bibinfo {author} {\bibfnamefont
  {M.}~\bibnamefont {Sprung}}, \bibinfo {author} {\bibfnamefont
  {A.}~\bibnamefont {Zozulya}}, \ and\ \bibinfo {author} {\bibfnamefont
  {C.}~\bibnamefont {Gutt}},\ }\href {\doibase 10.1103/PhysRevLett.120.168001}
  {\bibfield  {journal} {\bibinfo  {journal} {Phys. Rev. Lett.}\ }\textbf
  {\bibinfo {volume} {120}},\ \bibinfo {pages} {168001} (\bibinfo {year}
  {2018})}\BibitemShut {NoStop}%
\bibitem [{\citenamefont {Ruta}\ \emph {et~al.}(2012)\citenamefont {Ruta},
  \citenamefont {Chushkin}, \citenamefont {Monaco}, \citenamefont {Cipelletti},
  \citenamefont {Pineda}, \citenamefont {Bruna}, \citenamefont {Giordano},\
  and\ \citenamefont {Gonzalez-Silveira}}]{Ruta2012}%
  \BibitemOpen
  \bibfield  {author} {\bibinfo {author} {\bibfnamefont {B.}~\bibnamefont
  {Ruta}}, \bibinfo {author} {\bibfnamefont {Y.}~\bibnamefont {Chushkin}},
  \bibinfo {author} {\bibfnamefont {G.}~\bibnamefont {Monaco}}, \bibinfo
  {author} {\bibfnamefont {L.}~\bibnamefont {Cipelletti}}, \bibinfo {author}
  {\bibfnamefont {E.}~\bibnamefont {Pineda}}, \bibinfo {author} {\bibfnamefont
  {P.}~\bibnamefont {Bruna}}, \bibinfo {author} {\bibfnamefont {V.~M.}\
  \bibnamefont {Giordano}}, \ and\ \bibinfo {author} {\bibfnamefont
  {M.}~\bibnamefont {Gonzalez-Silveira}},\ }\href {\doibase
  10.1103/PhysRevLett.109.165701} {\bibfield  {journal} {\bibinfo  {journal}
  {Phys. Rev. Lett.}\ }\textbf {\bibinfo {volume} {109}},\ \bibinfo {pages}
  {165701} (\bibinfo {year} {2012})}\BibitemShut {NoStop}%
\bibitem [{\citenamefont {Hruszkewycz}\ \emph {et~al.}(2012)\citenamefont
  {Hruszkewycz}, \citenamefont {Sutton}, \citenamefont {Fuoss}, \citenamefont
  {Adams}, \citenamefont {Rosenkranz}, \citenamefont {Ludwig}, \citenamefont
  {Roseker}, \citenamefont {Fritz}, \citenamefont {Cammarata}, \citenamefont
  {Zhu}, \citenamefont {Lee}, \citenamefont {Lemke}, \citenamefont {Gutt},
  \citenamefont {Robert}, \citenamefont {Gr\"ubel},\ and\ \citenamefont
  {Stephenson}}]{Hruszkewycz2012}%
  \BibitemOpen
  \bibfield  {author} {\bibinfo {author} {\bibfnamefont {S.~O.}\ \bibnamefont
  {Hruszkewycz}}, \bibinfo {author} {\bibfnamefont {M.}~\bibnamefont {Sutton}},
  \bibinfo {author} {\bibfnamefont {P.~H.}\ \bibnamefont {Fuoss}}, \bibinfo
  {author} {\bibfnamefont {B.}~\bibnamefont {Adams}}, \bibinfo {author}
  {\bibfnamefont {S.}~\bibnamefont {Rosenkranz}}, \bibinfo {author}
  {\bibfnamefont {K.~F.}\ \bibnamefont {Ludwig}}, \bibinfo {author}
  {\bibfnamefont {W.}~\bibnamefont {Roseker}}, \bibinfo {author} {\bibfnamefont
  {D.}~\bibnamefont {Fritz}}, \bibinfo {author} {\bibfnamefont
  {M.}~\bibnamefont {Cammarata}}, \bibinfo {author} {\bibfnamefont
  {D.}~\bibnamefont {Zhu}}, \bibinfo {author} {\bibfnamefont {S.}~\bibnamefont
  {Lee}}, \bibinfo {author} {\bibfnamefont {H.}~\bibnamefont {Lemke}}, \bibinfo
  {author} {\bibfnamefont {C.}~\bibnamefont {Gutt}}, \bibinfo {author}
  {\bibfnamefont {A.}~\bibnamefont {Robert}}, \bibinfo {author} {\bibfnamefont
  {G.}~\bibnamefont {Gr\"ubel}}, \ and\ \bibinfo {author} {\bibfnamefont
  {G.~B.}\ \bibnamefont {Stephenson}},\ }\href {\doibase
  10.1103/PhysRevLett.109.185502} {\bibfield  {journal} {\bibinfo  {journal}
  {Phys. Rev. Lett.}\ }\textbf {\bibinfo {volume} {109}},\ \bibinfo {pages}
  {185502} (\bibinfo {year} {2012})}\BibitemShut {NoStop}%
\bibitem [{\citenamefont {Perakis}\ \emph {et~al.}(2018)\citenamefont {Perakis}
  \emph {et~al.}}]{Perakis2018}%
  \BibitemOpen
  \bibfield  {author} {\bibinfo {author} {\bibfnamefont {F.}~\bibnamefont
  {Perakis}} \emph {et~al.},\ }\href@noop {} {\bibfield  {journal} {\bibinfo
  {journal} {Nature Communications}\ }\textbf {\bibinfo {volume} {9}},\
  \bibinfo {pages} {1917} (\bibinfo {year} {2018})}\BibitemShut {NoStop}%
\bibitem [{\citenamefont {Gutt}\ \emph {et~al.}(2009)\citenamefont {Gutt},
  \citenamefont {Stadler}, \citenamefont {Duri}, \citenamefont {Autenrieth},
  \citenamefont {Leupold}, \citenamefont {Chushkin},\ and\ \citenamefont
  {Gr\"{u}bel}}]{Gutt2009}%
  \BibitemOpen
  \bibfield  {author} {\bibinfo {author} {\bibfnamefont {C.}~\bibnamefont
  {Gutt}}, \bibinfo {author} {\bibfnamefont {L.-M.}\ \bibnamefont {Stadler}},
  \bibinfo {author} {\bibfnamefont {A.}~\bibnamefont {Duri}}, \bibinfo {author}
  {\bibfnamefont {T.}~\bibnamefont {Autenrieth}}, \bibinfo {author}
  {\bibfnamefont {O.}~\bibnamefont {Leupold}}, \bibinfo {author} {\bibfnamefont
  {Y.}~\bibnamefont {Chushkin}}, \ and\ \bibinfo {author} {\bibfnamefont
  {G.}~\bibnamefont {Gr\"{u}bel}},\ }\href@noop {} {\bibfield  {journal}
  {\bibinfo  {journal} {Opt. Express}\ }\textbf {\bibinfo {volume} {17}},\
  \bibinfo {pages} {55} (\bibinfo {year} {2009})}\BibitemShut {NoStop}%
\bibitem [{\citenamefont {Osaka}\ \emph {et~al.}(2016)\citenamefont {Osaka},
  \citenamefont {Hirano}, \citenamefont {Sano}, \citenamefont {Inubushi},
  \citenamefont {Matsuyama}, \citenamefont {Tono}, \citenamefont {Ishikawa},
  \citenamefont {Yamauchi},\ and\ \citenamefont {Yabashi}}]{Osaka2016}%
  \BibitemOpen
  \bibfield  {author} {\bibinfo {author} {\bibfnamefont {T.}~\bibnamefont
  {Osaka}}, \bibinfo {author} {\bibfnamefont {T.}~\bibnamefont {Hirano}},
  \bibinfo {author} {\bibfnamefont {Y.}~\bibnamefont {Sano}}, \bibinfo {author}
  {\bibfnamefont {Y.}~\bibnamefont {Inubushi}}, \bibinfo {author}
  {\bibfnamefont {S.}~\bibnamefont {Matsuyama}}, \bibinfo {author}
  {\bibfnamefont {K.}~\bibnamefont {Tono}}, \bibinfo {author} {\bibfnamefont
  {T.}~\bibnamefont {Ishikawa}}, \bibinfo {author} {\bibfnamefont
  {K.}~\bibnamefont {Yamauchi}}, \ and\ \bibinfo {author} {\bibfnamefont
  {M.}~\bibnamefont {Yabashi}},\ }\href@noop {} {\bibfield  {journal} {\bibinfo
   {journal} {Opt. Express}\ }\textbf {\bibinfo {volume} {24}},\ \bibinfo
  {pages} {9187} (\bibinfo {year} {2016})}\BibitemShut {NoStop}%
\bibitem [{\citenamefont {Roseker}\ \emph {et~al.}(2018)\citenamefont
  {Roseker}, \citenamefont {Hruszkewycz}, \citenamefont {Lehmk\"uhler},
  \citenamefont {Walther}, \citenamefont {Schulte-Schrepping}, \citenamefont
  {Lee}, \citenamefont {Osaka}, \citenamefont {Str\"uder}, \citenamefont
  {Hartmann}, \citenamefont {Sikorski}, \citenamefont {Song}, \citenamefont
  {Robert}, \citenamefont {Fuoss}, \citenamefont {Sutton}, \citenamefont
  {Stephenson},\ and\ \citenamefont {Gr\"ubel}}]{Roseker2018}%
  \BibitemOpen
  \bibfield  {author} {\bibinfo {author} {\bibfnamefont {W.}~\bibnamefont
  {Roseker}}, \bibinfo {author} {\bibfnamefont {S.~O.}\ \bibnamefont
  {Hruszkewycz}}, \bibinfo {author} {\bibfnamefont {F.}~\bibnamefont
  {Lehmk\"uhler}}, \bibinfo {author} {\bibfnamefont {M.}~\bibnamefont
  {Walther}}, \bibinfo {author} {\bibfnamefont {H.}~\bibnamefont
  {Schulte-Schrepping}}, \bibinfo {author} {\bibfnamefont {S.}~\bibnamefont
  {Lee}}, \bibinfo {author} {\bibfnamefont {T.}~\bibnamefont {Osaka}}, \bibinfo
  {author} {\bibfnamefont {L.}~\bibnamefont {Str\"uder}}, \bibinfo {author}
  {\bibfnamefont {R.}~\bibnamefont {Hartmann}}, \bibinfo {author}
  {\bibfnamefont {M.}~\bibnamefont {Sikorski}}, \bibinfo {author}
  {\bibfnamefont {S.}~\bibnamefont {Song}}, \bibinfo {author} {\bibfnamefont
  {A.}~\bibnamefont {Robert}}, \bibinfo {author} {\bibfnamefont {P.~H.}\
  \bibnamefont {Fuoss}}, \bibinfo {author} {\bibfnamefont {M.}~\bibnamefont
  {Sutton}}, \bibinfo {author} {\bibfnamefont {G.~B.}\ \bibnamefont
  {Stephenson}}, \ and\ \bibinfo {author} {\bibfnamefont {G.}~\bibnamefont
  {Gr\"ubel}},\ }\href {\doibase 10.1038/s41467-018-04178-9} {\bibfield
  {journal} {\bibinfo  {journal} {Nature Communications}\ }\textbf {\bibinfo
  {volume} {9}},\ \bibinfo {pages} {1704} (\bibinfo {year} {2018})}\BibitemShut
  {NoStop}%
\bibitem [{\citenamefont {Allahgholi}\ \emph {et~al.}(2015)\citenamefont
  {Allahgholi} \emph {et~al.}}]{Allahgholi2015a}%
  \BibitemOpen
  \bibfield  {author} {\bibinfo {author} {\bibfnamefont {A.}~\bibnamefont
  {Allahgholi}} \emph {et~al.},\ }\href@noop {} {\bibfield  {journal} {\bibinfo
   {journal} {Journal of Instrumentation}\ }\textbf {\bibinfo {volume} {10}},\
  \bibinfo {pages} {C01023} (\bibinfo {year} {2015})}\BibitemShut {NoStop}%
\bibitem [{\citenamefont {Lehmk\"uhler}\ \emph {et~al.}(2015)\citenamefont
  {Lehmk\"uhler} \emph {et~al.}}]{Lehmkuehler2015}%
  \BibitemOpen
  \bibfield  {author} {\bibinfo {author} {\bibfnamefont {F.}~\bibnamefont
  {Lehmk\"uhler}} \emph {et~al.},\ }\href@noop {} {\bibfield  {journal}
  {\bibinfo  {journal} {Scientific Reports}\ }\textbf {\bibinfo {volume} {5}},\
  \bibinfo {pages} {17193} (\bibinfo {year} {2015})}\BibitemShut {NoStop}%
\bibitem [{\citenamefont {Lhermitte}\ \emph {et~al.}(2017)\citenamefont
  {Lhermitte}, \citenamefont {Rogers}, \citenamefont {Manet},\ and\
  \citenamefont {Sutton}}]{Lhermite2017}%
  \BibitemOpen
  \bibfield  {author} {\bibinfo {author} {\bibfnamefont {J.~R.~M.}\
  \bibnamefont {Lhermitte}}, \bibinfo {author} {\bibfnamefont {M.~C.}\
  \bibnamefont {Rogers}}, \bibinfo {author} {\bibfnamefont {S.}~\bibnamefont
  {Manet}}, \ and\ \bibinfo {author} {\bibfnamefont {M.}~\bibnamefont
  {Sutton}},\ }\href {\doibase 10.1063/1.4974099} {\bibfield  {journal}
  {\bibinfo  {journal} {Review of Scientific Instruments}\ }\textbf {\bibinfo
  {volume} {88}},\ \bibinfo {pages} {015112} (\bibinfo {year}
  {2017})}\BibitemShut {NoStop}%
\bibitem [{\citenamefont {Hastings}\ \emph {et~al.}(1991)\citenamefont
  {Hastings}, \citenamefont {Siddons}, \citenamefont {van B\"urck},
  \citenamefont {Hollatz},\ and\ \citenamefont {Bergmann}}]{Hastings1991}%
  \BibitemOpen
  \bibfield  {author} {\bibinfo {author} {\bibfnamefont {J.~B.}\ \bibnamefont
  {Hastings}}, \bibinfo {author} {\bibfnamefont {D.~P.}\ \bibnamefont
  {Siddons}}, \bibinfo {author} {\bibfnamefont {U.}~\bibnamefont {van
  B\"urck}}, \bibinfo {author} {\bibfnamefont {R.}~\bibnamefont {Hollatz}}, \
  and\ \bibinfo {author} {\bibfnamefont {U.}~\bibnamefont {Bergmann}},\ }\href
  {\doibase 10.1103/PhysRevLett.66.770} {\bibfield  {journal} {\bibinfo
  {journal} {Phys. Rev. Lett.}\ }\textbf {\bibinfo {volume} {66}},\ \bibinfo
  {pages} {770} (\bibinfo {year} {1991})}\BibitemShut {NoStop}%
\bibitem [{\citenamefont {Pineider}\ \emph {et~al.}(2013)\citenamefont
  {Pineider}, \citenamefont {de~Julian~Fernandez}, \citenamefont {Videtta},
  \citenamefont {Carlino}, \citenamefont {al~Hourani}, \citenamefont {Wilhelm},
  \citenamefont {Rogalev}, \citenamefont {Cozzoli}, \citenamefont {Ghigna},\
  and\ \citenamefont {Sangregorio}}]{Pineider2013}%
  \BibitemOpen
  \bibfield  {author} {\bibinfo {author} {\bibfnamefont {F.}~\bibnamefont
  {Pineider}}, \bibinfo {author} {\bibfnamefont {C.}~\bibnamefont
  {de~Julian~Fernandez}}, \bibinfo {author} {\bibfnamefont {V.}~\bibnamefont
  {Videtta}}, \bibinfo {author} {\bibfnamefont {E.}~\bibnamefont {Carlino}},
  \bibinfo {author} {\bibfnamefont {A.}~\bibnamefont {al~Hourani}}, \bibinfo
  {author} {\bibfnamefont {F.}~\bibnamefont {Wilhelm}}, \bibinfo {author}
  {\bibfnamefont {A.}~\bibnamefont {Rogalev}}, \bibinfo {author} {\bibfnamefont
  {P.~D.}\ \bibnamefont {Cozzoli}}, \bibinfo {author} {\bibfnamefont
  {P.}~\bibnamefont {Ghigna}}, \ and\ \bibinfo {author} {\bibfnamefont
  {C.}~\bibnamefont {Sangregorio}},\ }\href {\doibase 10.1021/nn305459m}
  {\bibfield  {journal} {\bibinfo  {journal} {ACS Nano}\ }\textbf {\bibinfo
  {volume} {7}},\ \bibinfo {pages} {857} (\bibinfo {year} {2013})}\BibitemShut
  {NoStop}%
\bibitem [{\citenamefont {Seto}\ \emph {et~al.}(1995)\citenamefont {Seto},
  \citenamefont {Yoda}, \citenamefont {Kikuta}, \citenamefont {Zhang},\ and\
  \citenamefont {Ando}}]{Seto1995}%
  \BibitemOpen
  \bibfield  {author} {\bibinfo {author} {\bibfnamefont {M.}~\bibnamefont
  {Seto}}, \bibinfo {author} {\bibfnamefont {Y.}~\bibnamefont {Yoda}}, \bibinfo
  {author} {\bibfnamefont {S.}~\bibnamefont {Kikuta}}, \bibinfo {author}
  {\bibfnamefont {X.~W.}\ \bibnamefont {Zhang}}, \ and\ \bibinfo {author}
  {\bibfnamefont {M.}~\bibnamefont {Ando}},\ }\href {\doibase
  10.1103/PhysRevLett.74.3828} {\bibfield  {journal} {\bibinfo  {journal}
  {Phys. Rev. Lett.}\ }\textbf {\bibinfo {volume} {74}},\ \bibinfo {pages}
  {3828} (\bibinfo {year} {1995})}\BibitemShut {NoStop}%
\bibitem [{\citenamefont {Sturhahn}\ \emph {et~al.}(1995)\citenamefont
  {Sturhahn}, \citenamefont {Toellner}, \citenamefont {Alp}, \citenamefont
  {Zhang}, \citenamefont {Ando}, \citenamefont {Yoda}, \citenamefont {Kikuta},
  \citenamefont {Seto}, \citenamefont {Kimball},\ and\ \citenamefont
  {Dabrowski}}]{Sturhahn1995}%
  \BibitemOpen
  \bibfield  {author} {\bibinfo {author} {\bibfnamefont {W.}~\bibnamefont
  {Sturhahn}}, \bibinfo {author} {\bibfnamefont {T.~S.}\ \bibnamefont
  {Toellner}}, \bibinfo {author} {\bibfnamefont {E.~E.}\ \bibnamefont {Alp}},
  \bibinfo {author} {\bibfnamefont {X.}~\bibnamefont {Zhang}}, \bibinfo
  {author} {\bibfnamefont {M.}~\bibnamefont {Ando}}, \bibinfo {author}
  {\bibfnamefont {Y.}~\bibnamefont {Yoda}}, \bibinfo {author} {\bibfnamefont
  {S.}~\bibnamefont {Kikuta}}, \bibinfo {author} {\bibfnamefont
  {M.}~\bibnamefont {Seto}}, \bibinfo {author} {\bibfnamefont {C.~W.}\
  \bibnamefont {Kimball}}, \ and\ \bibinfo {author} {\bibfnamefont
  {B.}~\bibnamefont {Dabrowski}},\ }\href {\doibase
  10.1103/PhysRevLett.74.3832} {\bibfield  {journal} {\bibinfo  {journal}
  {Phys. Rev. Lett.}\ }\textbf {\bibinfo {volume} {74}},\ \bibinfo {pages}
  {3832} (\bibinfo {year} {1995})}\BibitemShut {NoStop}%
\bibitem [{\citenamefont {Hoffman}\ \emph {et~al.}(2014)\citenamefont
  {Hoffman}, \citenamefont {Lukoyanov}, \citenamefont {Yang}, \citenamefont
  {Dean},\ and\ \citenamefont {Seefeldt}}]{Hoffman2014}%
  \BibitemOpen
  \bibfield  {author} {\bibinfo {author} {\bibfnamefont {B.~M.}\ \bibnamefont
  {Hoffman}}, \bibinfo {author} {\bibfnamefont {D.}~\bibnamefont {Lukoyanov}},
  \bibinfo {author} {\bibfnamefont {Z.-Y.}\ \bibnamefont {Yang}}, \bibinfo
  {author} {\bibfnamefont {D.~R.}\ \bibnamefont {Dean}}, \ and\ \bibinfo
  {author} {\bibfnamefont {L.~C.}\ \bibnamefont {Seefeldt}},\ }\href {\doibase
  10.1021/cr400641x} {\bibfield  {journal} {\bibinfo  {journal} {Chemical
  Reviews}\ }\textbf {\bibinfo {volume} {114}},\ \bibinfo {pages} {4041}
  (\bibinfo {year} {2014})}\BibitemShut {NoStop}%
\bibitem [{\citenamefont {von~der Wense}\ \emph {et~al.}(2016)\citenamefont
  {von~der Wense}, \citenamefont {Seiferle}, \citenamefont {Laatiaoui},
  \citenamefont {Neumayr}, \citenamefont {Maier}, \citenamefont {Wirth},
  \citenamefont {Mokry}, \citenamefont {Runke}, \citenamefont {Eberhardt},
  \citenamefont {D{\"u}llmann}, \citenamefont {Trautmann},\ and\ \citenamefont
  {Thirolf}}]{vonderWense2016}%
  \BibitemOpen
  \bibfield  {author} {\bibinfo {author} {\bibfnamefont {L.}~\bibnamefont
  {von~der Wense}}, \bibinfo {author} {\bibfnamefont {B.}~\bibnamefont
  {Seiferle}}, \bibinfo {author} {\bibfnamefont {M.}~\bibnamefont {Laatiaoui}},
  \bibinfo {author} {\bibfnamefont {J.~B.}\ \bibnamefont {Neumayr}}, \bibinfo
  {author} {\bibfnamefont {H.-J.}\ \bibnamefont {Maier}}, \bibinfo {author}
  {\bibfnamefont {H.-F.}\ \bibnamefont {Wirth}}, \bibinfo {author}
  {\bibfnamefont {C.}~\bibnamefont {Mokry}}, \bibinfo {author} {\bibfnamefont
  {J.}~\bibnamefont {Runke}}, \bibinfo {author} {\bibfnamefont
  {K.}~\bibnamefont {Eberhardt}}, \bibinfo {author} {\bibfnamefont {C.~E.}\
  \bibnamefont {D{\"u}llmann}}, \bibinfo {author} {\bibfnamefont {N.~G.}\
  \bibnamefont {Trautmann}}, \ and\ \bibinfo {author} {\bibfnamefont {P.~G.}\
  \bibnamefont {Thirolf}},\ }\href@noop {} {\bibfield  {journal} {\bibinfo
  {journal} {Nature}\ }\textbf {\bibinfo {volume} {533}},\ \bibinfo {pages}
  {47} (\bibinfo {year} {2016})}\BibitemShut {NoStop}%
\bibitem [{\citenamefont {Shenoy}\ and\ \citenamefont
  {R{\"o}hlsberger}(2008)}]{Shenoy2008}%
  \BibitemOpen
  \bibfield  {author} {\bibinfo {author} {\bibfnamefont {G.~K.}\ \bibnamefont
  {Shenoy}}\ and\ \bibinfo {author} {\bibfnamefont {R.}~\bibnamefont
  {R{\"o}hlsberger}},\ }\href {\doibase 10.1007/s10751-008-9720-y} {\bibfield
  {journal} {\bibinfo  {journal} {Hyperfine Interactions}\ }\textbf {\bibinfo
  {volume} {182}},\ \bibinfo {pages} {157} (\bibinfo {year}
  {2008})}\BibitemShut {NoStop}%
\bibitem [{\citenamefont {Siemens}\ \emph {et~al.}(2010)\citenamefont
  {Siemens}, \citenamefont {Li}, \citenamefont {Yang}, \citenamefont {Nelson},
  \citenamefont {Anderson}, \citenamefont {Murnane},\ and\ \citenamefont
  {Kapteyn}}]{Siemens2010}%
  \BibitemOpen
  \bibfield  {author} {\bibinfo {author} {\bibfnamefont {M.~E.}\ \bibnamefont
  {Siemens}}, \bibinfo {author} {\bibfnamefont {Q.}~\bibnamefont {Li}},
  \bibinfo {author} {\bibfnamefont {R.}~\bibnamefont {Yang}}, \bibinfo {author}
  {\bibfnamefont {K.~A.}\ \bibnamefont {Nelson}}, \bibinfo {author}
  {\bibfnamefont {E.~H.}\ \bibnamefont {Anderson}}, \bibinfo {author}
  {\bibfnamefont {M.~M.}\ \bibnamefont {Murnane}}, \ and\ \bibinfo {author}
  {\bibfnamefont {H.~C.}\ \bibnamefont {Kapteyn}},\ }\href@noop {} {\bibfield
  {journal} {\bibinfo  {journal} {Nature Materials}\ }\textbf {\bibinfo
  {volume} {9}},\ \bibinfo {pages} {26} (\bibinfo {year} {2010})}\BibitemShut
  {NoStop}%
\bibitem [{\citenamefont {Baron}\ \emph {et~al.}(1997)\citenamefont {Baron},
  \citenamefont {Franz}, \citenamefont {Meyer}, \citenamefont {R\"uffer},
  \citenamefont {Chumakov}, \citenamefont {Burkel},\ and\ \citenamefont
  {Petry}}]{Baron1997}%
  \BibitemOpen
  \bibfield  {author} {\bibinfo {author} {\bibfnamefont {A.~Q.~R.}\
  \bibnamefont {Baron}}, \bibinfo {author} {\bibfnamefont {H.}~\bibnamefont
  {Franz}}, \bibinfo {author} {\bibfnamefont {A.}~\bibnamefont {Meyer}},
  \bibinfo {author} {\bibfnamefont {R.}~\bibnamefont {R\"uffer}}, \bibinfo
  {author} {\bibfnamefont {A.~I.}\ \bibnamefont {Chumakov}}, \bibinfo {author}
  {\bibfnamefont {E.}~\bibnamefont {Burkel}}, \ and\ \bibinfo {author}
  {\bibfnamefont {W.}~\bibnamefont {Petry}},\ }\href {\doibase
  10.1103/PhysRevLett.79.2823} {\bibfield  {journal} {\bibinfo  {journal}
  {Phys. Rev. Lett.}\ }\textbf {\bibinfo {volume} {79}},\ \bibinfo {pages}
  {2823} (\bibinfo {year} {1997})}\BibitemShut {NoStop}%
\bibitem [{\citenamefont {R{\"o}hlsberger}\ \emph {et~al.}(1997)\citenamefont
  {R{\"o}hlsberger}, \citenamefont {Gerdau}, \citenamefont {R{\"u}ffer},
  \citenamefont {Sturhahn}, \citenamefont {Toellner}, \citenamefont
  {Chumakov},\ and\ \citenamefont {Alp}}]{Roehlsberger1997}%
  \BibitemOpen
  \bibfield  {author} {\bibinfo {author} {\bibfnamefont {R.}~\bibnamefont
  {R{\"o}hlsberger}}, \bibinfo {author} {\bibfnamefont {E.}~\bibnamefont
  {Gerdau}}, \bibinfo {author} {\bibfnamefont {R.}~\bibnamefont {R{\"u}ffer}},
  \bibinfo {author} {\bibfnamefont {W.}~\bibnamefont {Sturhahn}}, \bibinfo
  {author} {\bibfnamefont {T.}~\bibnamefont {Toellner}}, \bibinfo {author}
  {\bibfnamefont {A.}~\bibnamefont {Chumakov}}, \ and\ \bibinfo {author}
  {\bibfnamefont {E.}~\bibnamefont {Alp}},\ }\href {\doibase
  https://doi.org/10.1016/S0168-9002(97)00710-9} {\bibfield  {journal}
  {\bibinfo  {journal} {Nucl. Instrum. Meth. A:}\ }\textbf {\bibinfo {volume}
  {394}},\ \bibinfo {pages} {251 } (\bibinfo {year} {1997})}\BibitemShut
  {NoStop}%
\bibitem [{\citenamefont {R\"ohlsberger}\ \emph {et~al.}(2000)\citenamefont
  {R\"ohlsberger}, \citenamefont {Toellner}, \citenamefont {Sturhahn},
  \citenamefont {Quast}, \citenamefont {Alp}, \citenamefont {Bernhard},
  \citenamefont {Burkel}, \citenamefont {Leupold},\ and\ \citenamefont
  {Gerdau}}]{Roehlsberger2000}%
  \BibitemOpen
  \bibfield  {author} {\bibinfo {author} {\bibfnamefont {R.}~\bibnamefont
  {R\"ohlsberger}}, \bibinfo {author} {\bibfnamefont {T.~S.}\ \bibnamefont
  {Toellner}}, \bibinfo {author} {\bibfnamefont {W.}~\bibnamefont {Sturhahn}},
  \bibinfo {author} {\bibfnamefont {K.~W.}\ \bibnamefont {Quast}}, \bibinfo
  {author} {\bibfnamefont {E.~E.}\ \bibnamefont {Alp}}, \bibinfo {author}
  {\bibfnamefont {A.}~\bibnamefont {Bernhard}}, \bibinfo {author}
  {\bibfnamefont {E.}~\bibnamefont {Burkel}}, \bibinfo {author} {\bibfnamefont
  {O.}~\bibnamefont {Leupold}}, \ and\ \bibinfo {author} {\bibfnamefont
  {E.}~\bibnamefont {Gerdau}},\ }\href {\doibase 10.1103/PhysRevLett.84.1007}
  {\bibfield  {journal} {\bibinfo  {journal} {Phys. Rev. Lett.}\ }\textbf
  {\bibinfo {volume} {84}},\ \bibinfo {pages} {1007} (\bibinfo {year}
  {2000})}\BibitemShut {NoStop}%
\bibitem [{\citenamefont {Masuda}\ \emph {et~al.}(2009)\citenamefont {Masuda},
  \citenamefont {Mitsui}, \citenamefont {Kobayashi}, \citenamefont
  {Higashitaniguchi},\ and\ \citenamefont {Seto}}]{Masuda2009}%
  \BibitemOpen
  \bibfield  {author} {\bibinfo {author} {\bibfnamefont {R.}~\bibnamefont
  {Masuda}}, \bibinfo {author} {\bibfnamefont {T.}~\bibnamefont {Mitsui}},
  \bibinfo {author} {\bibfnamefont {Y.}~\bibnamefont {Kobayashi}}, \bibinfo
  {author} {\bibfnamefont {S.}~\bibnamefont {Higashitaniguchi}}, \ and\
  \bibinfo {author} {\bibfnamefont {M.}~\bibnamefont {Seto}},\ }\href@noop {}
  {\bibfield  {journal} {\bibinfo  {journal} {Japanese Journal of Applied
  Physics}\ }\textbf {\bibinfo {volume} {48}},\ \bibinfo {pages} {120221}
  (\bibinfo {year} {2009})}\BibitemShut {NoStop}%
\bibitem [{\citenamefont {Castrignano}\ and\ \citenamefont
  {Evers}(2018)}]{Castrignano2018}%
  \BibitemOpen
  \bibfield  {author} {\bibinfo {author} {\bibfnamefont {S.}~\bibnamefont
  {Castrignano}}\ and\ \bibinfo {author} {\bibfnamefont {J.}~\bibnamefont
  {Evers}},\ }\href@noop {} {\bibfield  {journal} {\bibinfo  {journal}
  {arXiv:1805.01672 [quant-ph]}\ } (\bibinfo {year} {2018})}\BibitemShut
  {NoStop}%
\bibitem [{\citenamefont {Jhurry}\ \emph {et~al.}(2012)\citenamefont {Jhurry},
  \citenamefont {Chakrabarti}, \citenamefont {McCormick}, \citenamefont
  {Holmes-Hampton},\ and\ \citenamefont {Lindahl}}]{Jhurry2012}%
  \BibitemOpen
  \bibfield  {author} {\bibinfo {author} {\bibfnamefont {N.~D.}\ \bibnamefont
  {Jhurry}}, \bibinfo {author} {\bibfnamefont {M.}~\bibnamefont {Chakrabarti}},
  \bibinfo {author} {\bibfnamefont {S.~P.}\ \bibnamefont {McCormick}}, \bibinfo
  {author} {\bibfnamefont {G.~P.}\ \bibnamefont {Holmes-Hampton}}, \ and\
  \bibinfo {author} {\bibfnamefont {P.~A.}\ \bibnamefont {Lindahl}},\ }\href
  {\doibase 10.1021/bi300382d} {\bibfield  {journal} {\bibinfo  {journal}
  {Biochemistry}\ }\textbf {\bibinfo {volume} {51}},\ \bibinfo {pages} {5276}
  (\bibinfo {year} {2012})}\BibitemShut {NoStop}%
\bibitem [{\citenamefont {Zecca}\ \emph {et~al.}(2001)\citenamefont {Zecca},
  \citenamefont {Gallorini}, \citenamefont {Sch\"unemann}, \citenamefont
  {Trautwein}, \citenamefont {Gerlach}, \citenamefont {Riederer}, \citenamefont
  {Vezzoni},\ and\ \citenamefont {Tampellini}}]{Zecca2001}%
  \BibitemOpen
  \bibfield  {author} {\bibinfo {author} {\bibfnamefont {L.}~\bibnamefont
  {Zecca}}, \bibinfo {author} {\bibfnamefont {M.}~\bibnamefont {Gallorini}},
  \bibinfo {author} {\bibfnamefont {V.}~\bibnamefont {Sch\"unemann}}, \bibinfo
  {author} {\bibfnamefont {A.~X.}\ \bibnamefont {Trautwein}}, \bibinfo {author}
  {\bibfnamefont {M.}~\bibnamefont {Gerlach}}, \bibinfo {author} {\bibfnamefont
  {P.}~\bibnamefont {Riederer}}, \bibinfo {author} {\bibfnamefont
  {P.}~\bibnamefont {Vezzoni}}, \ and\ \bibinfo {author} {\bibfnamefont
  {D.}~\bibnamefont {Tampellini}},\ }\href {\doibase
  10.1046/j.1471-4159.2001.00186.x} {\bibfield  {journal} {\bibinfo  {journal}
  {Journal of Neurochemistry}\ }\textbf {\bibinfo {volume} {76}},\ \bibinfo
  {pages} {1766} (\bibinfo {year} {2001})}\BibitemShut {NoStop}%
\bibitem [{\citenamefont {Torti}\ and\ \citenamefont
  {Torti}(2013)}]{Torti2013}%
  \BibitemOpen
  \bibfield  {author} {\bibinfo {author} {\bibfnamefont {S.~V.}\ \bibnamefont
  {Torti}}\ and\ \bibinfo {author} {\bibfnamefont {F.~M.}\ \bibnamefont
  {Torti}},\ }\href@noop {} {\bibfield  {journal} {\bibinfo  {journal} {Nature
  Reviews Cancer}\ }\textbf {\bibinfo {volume} {13}},\ \bibinfo {pages} {342}
  (\bibinfo {year} {2013})}\BibitemShut {NoStop}%
\bibitem [{\citenamefont {Perez~Velez}\ \emph {et~al.}(2014)\citenamefont
  {Perez~Velez}, \citenamefont {Ellmers}, \citenamefont {Huang}, \citenamefont
  {Bentrup}, \citenamefont {Sch{\"u}nemann}, \citenamefont {Gr{\"u}nert},\ and\
  \citenamefont {Br{\"u}ckner}}]{Velez2014}%
  \BibitemOpen
  \bibfield  {author} {\bibinfo {author} {\bibfnamefont {R.}~\bibnamefont
  {Perez~Velez}}, \bibinfo {author} {\bibfnamefont {I.}~\bibnamefont
  {Ellmers}}, \bibinfo {author} {\bibfnamefont {H.}~\bibnamefont {Huang}},
  \bibinfo {author} {\bibfnamefont {U.}~\bibnamefont {Bentrup}}, \bibinfo
  {author} {\bibfnamefont {V.}~\bibnamefont {Sch{\"u}nemann}}, \bibinfo
  {author} {\bibfnamefont {W.}~\bibnamefont {Gr{\"u}nert}}, \ and\ \bibinfo
  {author} {\bibfnamefont {A.}~\bibnamefont {Br{\"u}ckner}},\ }\href {\doibase
  https://doi.org/10.1016/j.jcat.2014.05.001} {\bibfield  {journal} {\bibinfo
  {journal} {Journal of Catalysis}\ }\textbf {\bibinfo {volume} {316}},\
  \bibinfo {pages} {103 } (\bibinfo {year} {2014})}\BibitemShut {NoStop}%
\bibitem [{\citenamefont {Varnell}\ \emph {et~al.}(2016)\citenamefont
  {Varnell}, \citenamefont {Tse}, \citenamefont {Schulz}, \citenamefont
  {Fister}, \citenamefont {Haasch}, \citenamefont {Timoshenko}, \citenamefont
  {Frenkel},\ and\ \citenamefont {Gewirth}}]{Varnell2016}%
  \BibitemOpen
  \bibfield  {author} {\bibinfo {author} {\bibfnamefont {J.~A.}\ \bibnamefont
  {Varnell}}, \bibinfo {author} {\bibfnamefont {E.~C.~M.}\ \bibnamefont {Tse}},
  \bibinfo {author} {\bibfnamefont {C.~E.}\ \bibnamefont {Schulz}}, \bibinfo
  {author} {\bibfnamefont {T.~T.}\ \bibnamefont {Fister}}, \bibinfo {author}
  {\bibfnamefont {R.~T.}\ \bibnamefont {Haasch}}, \bibinfo {author}
  {\bibfnamefont {J.}~\bibnamefont {Timoshenko}}, \bibinfo {author}
  {\bibfnamefont {A.~I.}\ \bibnamefont {Frenkel}}, \ and\ \bibinfo {author}
  {\bibfnamefont {A.~A.}\ \bibnamefont {Gewirth}},\ }\href@noop {} {\bibfield
  {journal} {\bibinfo  {journal} {Nature Communications}\ }\textbf {\bibinfo
  {volume} {7}},\ \bibinfo {pages} {12582} (\bibinfo {year}
  {2016})}\BibitemShut {NoStop}%
\bibitem [{\citenamefont {Lill}\ and\ \citenamefont
  {M{\"u}hlenhoff}(2008)}]{Lill2008}%
  \BibitemOpen
  \bibfield  {author} {\bibinfo {author} {\bibfnamefont {R.}~\bibnamefont
  {Lill}}\ and\ \bibinfo {author} {\bibfnamefont {U.}~\bibnamefont
  {M{\"u}hlenhoff}},\ }\href {\doibase
  10.1146/annurev.biochem.76.052705.162653} {\bibfield  {journal} {\bibinfo
  {journal} {Annual Review of Biochemistry}\ }\textbf {\bibinfo {volume}
  {77}},\ \bibinfo {pages} {669} (\bibinfo {year} {2008})}\BibitemShut
  {NoStop}%
\bibitem [{\citenamefont {Bogdan}\ \emph {et~al.}(2016)\citenamefont {Bogdan},
  \citenamefont {Miyazawa}, \citenamefont {Hashimoto},\ and\ \citenamefont
  {Tsuji}}]{Bogdan2016}%
  \BibitemOpen
  \bibfield  {author} {\bibinfo {author} {\bibfnamefont {A.~R.}\ \bibnamefont
  {Bogdan}}, \bibinfo {author} {\bibfnamefont {M.}~\bibnamefont {Miyazawa}},
  \bibinfo {author} {\bibfnamefont {K.}~\bibnamefont {Hashimoto}}, \ and\
  \bibinfo {author} {\bibfnamefont {Y.}~\bibnamefont {Tsuji}},\ }\href
  {\doibase https://doi.org/10.1016/j.tibs.2015.11.012} {\bibfield  {journal}
  {\bibinfo  {journal} {Trends in Biochemical Sciences}\ }\textbf {\bibinfo
  {volume} {41}},\ \bibinfo {pages} {274 } (\bibinfo {year}
  {2016})}\BibitemShut {NoStop}%
\bibitem [{\citenamefont {Eisenberger}\ and\ \citenamefont
  {McCall}(1971{\natexlab{a}})}]{Eisenberger1971}%
  \BibitemOpen
  \bibfield  {author} {\bibinfo {author} {\bibfnamefont {P.}~\bibnamefont
  {Eisenberger}}\ and\ \bibinfo {author} {\bibfnamefont {S.~L.}\ \bibnamefont
  {McCall}},\ }\href {\doibase 10.1103/PhysRevLett.26.684} {\bibfield
  {journal} {\bibinfo  {journal} {Phys. Rev. Lett.}\ }\textbf {\bibinfo
  {volume} {26}},\ \bibinfo {pages} {684} (\bibinfo {year}
  {1971}{\natexlab{a}})}\BibitemShut {NoStop}%
\bibitem [{\citenamefont {Glover}\ \emph {et~al.}(2012)\citenamefont {Glover},
  \citenamefont {Fritz}, \citenamefont {Cammarata}, \citenamefont {Allison},
  \citenamefont {Coh}, \citenamefont {Feldkamp}, \citenamefont {Lemke},
  \citenamefont {Zhu}, \citenamefont {Feng}, \citenamefont {Coffee},
  \citenamefont {Fuchs}, \citenamefont {Ghimire}, \citenamefont {Chen},
  \citenamefont {Shwartz}, \citenamefont {Reis}, \citenamefont {Harris},\ and\
  \citenamefont {Hastings}}]{Glover2012}%
  \BibitemOpen
  \bibfield  {author} {\bibinfo {author} {\bibfnamefont {T.~E.}\ \bibnamefont
  {Glover}}, \bibinfo {author} {\bibfnamefont {D.~M.}\ \bibnamefont {Fritz}},
  \bibinfo {author} {\bibfnamefont {M.}~\bibnamefont {Cammarata}}, \bibinfo
  {author} {\bibfnamefont {T.~K.}\ \bibnamefont {Allison}}, \bibinfo {author}
  {\bibfnamefont {S.}~\bibnamefont {Coh}}, \bibinfo {author} {\bibfnamefont
  {J.~M.}\ \bibnamefont {Feldkamp}}, \bibinfo {author} {\bibfnamefont
  {H.}~\bibnamefont {Lemke}}, \bibinfo {author} {\bibfnamefont
  {D.}~\bibnamefont {Zhu}}, \bibinfo {author} {\bibfnamefont {Y.}~\bibnamefont
  {Feng}}, \bibinfo {author} {\bibfnamefont {R.~N.}\ \bibnamefont {Coffee}},
  \bibinfo {author} {\bibfnamefont {M.}~\bibnamefont {Fuchs}}, \bibinfo
  {author} {\bibfnamefont {S.}~\bibnamefont {Ghimire}}, \bibinfo {author}
  {\bibfnamefont {J.}~\bibnamefont {Chen}}, \bibinfo {author} {\bibfnamefont
  {S.}~\bibnamefont {Shwartz}}, \bibinfo {author} {\bibfnamefont {D.~A.}\
  \bibnamefont {Reis}}, \bibinfo {author} {\bibfnamefont {S.~E.}\ \bibnamefont
  {Harris}}, \ and\ \bibinfo {author} {\bibfnamefont {J.~B.}\ \bibnamefont
  {Hastings}},\ }\href@noop {} {\bibfield  {journal} {\bibinfo  {journal}
  {Nature}\ }\textbf {\bibinfo {volume} {488}},\ \bibinfo {pages} {603}
  (\bibinfo {year} {2012})}\BibitemShut {NoStop}%
\bibitem [{\citenamefont {Tamasaku}\ \emph {et~al.}(2011)\citenamefont
  {Tamasaku}, \citenamefont {Sawada}, \citenamefont {Nishibori},\ and\
  \citenamefont {Ishikawa}}]{Tamasaku2011}%
  \BibitemOpen
  \bibfield  {author} {\bibinfo {author} {\bibfnamefont {K.}~\bibnamefont
  {Tamasaku}}, \bibinfo {author} {\bibfnamefont {K.}~\bibnamefont {Sawada}},
  \bibinfo {author} {\bibfnamefont {E.}~\bibnamefont {Nishibori}}, \ and\
  \bibinfo {author} {\bibfnamefont {T.}~\bibnamefont {Ishikawa}},\ }\href@noop
  {} {\bibfield  {journal} {\bibinfo  {journal} {Nature Physics}\ }\textbf
  {\bibinfo {volume} {7}},\ \bibinfo {pages} {705} (\bibinfo {year}
  {2011})}\BibitemShut {NoStop}%
\bibitem [{\citenamefont {Freund}\ and\ \citenamefont
  {Levine}(1970)}]{Freund1970}%
  \BibitemOpen
  \bibfield  {author} {\bibinfo {author} {\bibfnamefont {I.}~\bibnamefont
  {Freund}}\ and\ \bibinfo {author} {\bibfnamefont {B.~F.}\ \bibnamefont
  {Levine}},\ }\href@noop {} {\bibfield  {journal} {\bibinfo  {journal} {Phys.
  Rev. Lett.}\ }\textbf {\bibinfo {volume} {25}},\ \bibinfo {pages} {1241}
  (\bibinfo {year} {1970})}\BibitemShut {NoStop}%
\bibitem [{\citenamefont {Eisenberger}\ and\ \citenamefont
  {McCall}(1971{\natexlab{b}})}]{Eisenberger1971b}%
  \BibitemOpen
  \bibfield  {author} {\bibinfo {author} {\bibfnamefont {P.~M.}\ \bibnamefont
  {Eisenberger}}\ and\ \bibinfo {author} {\bibfnamefont {S.~L.}\ \bibnamefont
  {McCall}},\ }\href {\doibase 10.1103/PhysRevA.3.1145} {\bibfield  {journal}
  {\bibinfo  {journal} {Phys. Rev. A}\ }\textbf {\bibinfo {volume} {3}},\
  \bibinfo {pages} {1145} (\bibinfo {year} {1971}{\natexlab{b}})}\BibitemShut
  {NoStop}%
\bibitem [{\citenamefont {Schori}\ \emph {et~al.}(2017)\citenamefont {Schori},
  \citenamefont {B\"omer}, \citenamefont {Borodin}, \citenamefont {Collins},
  \citenamefont {Detlefs}, \citenamefont {Moretti~Sala}, \citenamefont
  {Yudovich},\ and\ \citenamefont {Shwartz}}]{Schori2017}%
  \BibitemOpen
  \bibfield  {author} {\bibinfo {author} {\bibfnamefont {A.}~\bibnamefont
  {Schori}}, \bibinfo {author} {\bibfnamefont {C.}~\bibnamefont {B\"omer}},
  \bibinfo {author} {\bibfnamefont {D.}~\bibnamefont {Borodin}}, \bibinfo
  {author} {\bibfnamefont {S.~P.}\ \bibnamefont {Collins}}, \bibinfo {author}
  {\bibfnamefont {B.}~\bibnamefont {Detlefs}}, \bibinfo {author} {\bibfnamefont
  {M.}~\bibnamefont {Moretti~Sala}}, \bibinfo {author} {\bibfnamefont
  {S.}~\bibnamefont {Yudovich}}, \ and\ \bibinfo {author} {\bibfnamefont
  {S.}~\bibnamefont {Shwartz}},\ }\href@noop {} {\bibfield  {journal} {\bibinfo
   {journal} {Phys. Rev. Lett.}\ }\textbf {\bibinfo {volume} {119}},\ \bibinfo
  {pages} {253902} (\bibinfo {year} {2017})}\BibitemShut {NoStop}%
\bibitem [{\citenamefont {Trigo}\ \emph {et~al.}(2013)\citenamefont {Trigo}
  \emph {et~al.}}]{Trigo2013}%
  \BibitemOpen
  \bibfield  {author} {\bibinfo {author} {\bibfnamefont {M.}~\bibnamefont
  {Trigo}} \emph {et~al.},\ }\href@noop {} {\bibfield  {journal} {\bibinfo
  {journal} {Nature Physics}\ }\textbf {\bibinfo {volume} {9}},\ \bibinfo
  {pages} {790} (\bibinfo {year} {2013})}\BibitemShut {NoStop}%
\bibitem [{\citenamefont {Fuchs}\ \emph {et~al.}(2015)\citenamefont {Fuchs}
  \emph {et~al.}}]{Fuchs2015}%
  \BibitemOpen
  \bibfield  {author} {\bibinfo {author} {\bibfnamefont {M.}~\bibnamefont
  {Fuchs}} \emph {et~al.},\ }\href@noop {} {\bibfield  {journal} {\bibinfo
  {journal} {Nature Physics}\ }\textbf {\bibinfo {volume} {11}},\ \bibinfo
  {pages} {964} (\bibinfo {year} {2015})}\BibitemShut {NoStop}%
\bibitem [{\citenamefont {Shwartz}\ \emph {et~al.}(2014)\citenamefont
  {Shwartz}, \citenamefont {Fuchs}, \citenamefont {Hastings}, \citenamefont
  {Inubushi}, \citenamefont {Ishikawa}, \citenamefont {Katayama}, \citenamefont
  {Reis}, \citenamefont {Sato}, \citenamefont {Tono}, \citenamefont {Yabashi},
  \citenamefont {Yudovich},\ and\ \citenamefont {Harris}}]{Shwartz2014}%
  \BibitemOpen
  \bibfield  {author} {\bibinfo {author} {\bibfnamefont {S.}~\bibnamefont
  {Shwartz}}, \bibinfo {author} {\bibfnamefont {M.}~\bibnamefont {Fuchs}},
  \bibinfo {author} {\bibfnamefont {J.~B.}\ \bibnamefont {Hastings}}, \bibinfo
  {author} {\bibfnamefont {Y.}~\bibnamefont {Inubushi}}, \bibinfo {author}
  {\bibfnamefont {T.}~\bibnamefont {Ishikawa}}, \bibinfo {author}
  {\bibfnamefont {T.}~\bibnamefont {Katayama}}, \bibinfo {author}
  {\bibfnamefont {D.~A.}\ \bibnamefont {Reis}}, \bibinfo {author}
  {\bibfnamefont {T.}~\bibnamefont {Sato}}, \bibinfo {author} {\bibfnamefont
  {K.}~\bibnamefont {Tono}}, \bibinfo {author} {\bibfnamefont {M.}~\bibnamefont
  {Yabashi}}, \bibinfo {author} {\bibfnamefont {S.}~\bibnamefont {Yudovich}}, \
  and\ \bibinfo {author} {\bibfnamefont {S.~E.}\ \bibnamefont {Harris}},\
  }\href {\doibase 10.1103/PhysRevLett.112.163901} {\bibfield  {journal}
  {\bibinfo  {journal} {Phys. Rev. Lett.}\ }\textbf {\bibinfo {volume} {112}},\
  \bibinfo {pages} {163901} (\bibinfo {year} {2014})}\BibitemShut {NoStop}%
\bibitem [{\citenamefont {Tamasaku}\ \emph {et~al.}(2014)\citenamefont
  {Tamasaku}, \citenamefont {Shigemasa}, \citenamefont {Inubushi},
  \citenamefont {Katayama}, \citenamefont {Sawada}, \citenamefont {Yumoto},
  \citenamefont {Ohashi}, \citenamefont {Mimura}, \citenamefont {Yabashi},
  \citenamefont {Yamauchi},\ and\ \citenamefont {Ishikawa}}]{Tamasaku2014}%
  \BibitemOpen
  \bibfield  {author} {\bibinfo {author} {\bibfnamefont {K.}~\bibnamefont
  {Tamasaku}}, \bibinfo {author} {\bibfnamefont {E.}~\bibnamefont {Shigemasa}},
  \bibinfo {author} {\bibfnamefont {Y.}~\bibnamefont {Inubushi}}, \bibinfo
  {author} {\bibfnamefont {T.}~\bibnamefont {Katayama}}, \bibinfo {author}
  {\bibfnamefont {K.}~\bibnamefont {Sawada}}, \bibinfo {author} {\bibfnamefont
  {H.}~\bibnamefont {Yumoto}}, \bibinfo {author} {\bibfnamefont
  {H.}~\bibnamefont {Ohashi}}, \bibinfo {author} {\bibfnamefont
  {H.}~\bibnamefont {Mimura}}, \bibinfo {author} {\bibfnamefont
  {M.}~\bibnamefont {Yabashi}}, \bibinfo {author} {\bibfnamefont
  {K.}~\bibnamefont {Yamauchi}}, \ and\ \bibinfo {author} {\bibfnamefont
  {T.}~\bibnamefont {Ishikawa}},\ }\href@noop {} {\bibfield  {journal}
  {\bibinfo  {journal} {Nature Photonics}\ }\textbf {\bibinfo {volume} {8}},\
  \bibinfo {pages} {313} (\bibinfo {year} {2014})}\BibitemShut {NoStop}%
\bibitem [{\citenamefont {Ghimire}\ \emph {et~al.}(2016)\citenamefont
  {Ghimire}, \citenamefont {Fuchs}, \citenamefont {Hastings}, \citenamefont
  {Herrmann}, \citenamefont {Inubushi}, \citenamefont {Pines}, \citenamefont
  {Shwartz}, \citenamefont {Yabashi},\ and\ \citenamefont
  {Reis}}]{Ghimire2016}%
  \BibitemOpen
  \bibfield  {author} {\bibinfo {author} {\bibfnamefont {S.}~\bibnamefont
  {Ghimire}}, \bibinfo {author} {\bibfnamefont {M.}~\bibnamefont {Fuchs}},
  \bibinfo {author} {\bibfnamefont {J.}~\bibnamefont {Hastings}}, \bibinfo
  {author} {\bibfnamefont {S.~C.}\ \bibnamefont {Herrmann}}, \bibinfo {author}
  {\bibfnamefont {Y.}~\bibnamefont {Inubushi}}, \bibinfo {author}
  {\bibfnamefont {J.}~\bibnamefont {Pines}}, \bibinfo {author} {\bibfnamefont
  {S.}~\bibnamefont {Shwartz}}, \bibinfo {author} {\bibfnamefont
  {M.}~\bibnamefont {Yabashi}}, \ and\ \bibinfo {author} {\bibfnamefont
  {D.~A.}\ \bibnamefont {Reis}},\ }\href {\doibase 10.1103/PhysRevA.94.043418}
  {\bibfield  {journal} {\bibinfo  {journal} {Phys. Rev. A}\ }\textbf {\bibinfo
  {volume} {94}},\ \bibinfo {pages} {043418} (\bibinfo {year}
  {2016})}\BibitemShut {NoStop}%
\bibitem [{\citenamefont {Weninger}\ \emph {et~al.}(2013)\citenamefont
  {Weninger}, \citenamefont {Purvis}, \citenamefont {Ryan}, \citenamefont
  {London}, \citenamefont {Bozek}, \citenamefont {Bostedt}, \citenamefont
  {Graf}, \citenamefont {Brown}, \citenamefont {Rocca},\ and\ \citenamefont
  {Rohringer}}]{Weninger2013}%
  \BibitemOpen
  \bibfield  {author} {\bibinfo {author} {\bibfnamefont {C.}~\bibnamefont
  {Weninger}}, \bibinfo {author} {\bibfnamefont {M.}~\bibnamefont {Purvis}},
  \bibinfo {author} {\bibfnamefont {D.}~\bibnamefont {Ryan}}, \bibinfo {author}
  {\bibfnamefont {R.~A.}\ \bibnamefont {London}}, \bibinfo {author}
  {\bibfnamefont {J.~D.}\ \bibnamefont {Bozek}}, \bibinfo {author}
  {\bibfnamefont {C.}~\bibnamefont {Bostedt}}, \bibinfo {author} {\bibfnamefont
  {A.}~\bibnamefont {Graf}}, \bibinfo {author} {\bibfnamefont {G.}~\bibnamefont
  {Brown}}, \bibinfo {author} {\bibfnamefont {J.~J.}\ \bibnamefont {Rocca}}, \
  and\ \bibinfo {author} {\bibfnamefont {N.}~\bibnamefont {Rohringer}},\ }\href
  {\doibase 10.1103/PhysRevLett.111.233902} {\bibfield  {journal} {\bibinfo
  {journal} {Phys. Rev. Lett.}\ }\textbf {\bibinfo {volume} {111}},\ \bibinfo
  {pages} {233902} (\bibinfo {year} {2013})}\BibitemShut {NoStop}%
\bibitem [{\citenamefont {Mukamel}\ \emph {et~al.}(2013)\citenamefont
  {Mukamel}, \citenamefont {Healion}, \citenamefont {Zhang},\ and\
  \citenamefont {Biggs}}]{Mukamel2013}%
  \BibitemOpen
  \bibfield  {author} {\bibinfo {author} {\bibfnamefont {S.}~\bibnamefont
  {Mukamel}}, \bibinfo {author} {\bibfnamefont {D.}~\bibnamefont {Healion}},
  \bibinfo {author} {\bibfnamefont {Y.}~\bibnamefont {Zhang}}, \ and\ \bibinfo
  {author} {\bibfnamefont {J.~D.}\ \bibnamefont {Biggs}},\ }\href {\doibase
  10.1146/annurev-physchem-040412-110021} {\bibfield  {journal} {\bibinfo
  {journal} {Annual Review of Physical Chemistry}\ }\textbf {\bibinfo {volume}
  {64}},\ \bibinfo {pages} {101} (\bibinfo {year} {2013})}\BibitemShut
  {NoStop}%
\bibitem [{\citenamefont {Pelliccia}\ \emph {et~al.}(2016)\citenamefont
  {Pelliccia}, \citenamefont {Rack}, \citenamefont {Scheel}, \citenamefont
  {Cantelli},\ and\ \citenamefont {Paganin}}]{Pelliccia2016}%
  \BibitemOpen
  \bibfield  {author} {\bibinfo {author} {\bibfnamefont {D.}~\bibnamefont
  {Pelliccia}}, \bibinfo {author} {\bibfnamefont {A.}~\bibnamefont {Rack}},
  \bibinfo {author} {\bibfnamefont {M.}~\bibnamefont {Scheel}}, \bibinfo
  {author} {\bibfnamefont {V.}~\bibnamefont {Cantelli}}, \ and\ \bibinfo
  {author} {\bibfnamefont {D.~M.}\ \bibnamefont {Paganin}},\ }\href {\doibase
  10.1103/PhysRevLett.117.113902} {\bibfield  {journal} {\bibinfo  {journal}
  {Phys. Rev. Lett.}\ }\textbf {\bibinfo {volume} {117}},\ \bibinfo {pages}
  {113902} (\bibinfo {year} {2016})}\BibitemShut {NoStop}%
\bibitem [{\citenamefont {Yu}\ \emph {et~al.}(2016)\citenamefont {Yu},
  \citenamefont {Lu}, \citenamefont {Han}, \citenamefont {Xie}, \citenamefont
  {Du}, \citenamefont {Xiao},\ and\ \citenamefont {Zhu}}]{Yu2016}%
  \BibitemOpen
  \bibfield  {author} {\bibinfo {author} {\bibfnamefont {H.}~\bibnamefont
  {Yu}}, \bibinfo {author} {\bibfnamefont {R.}~\bibnamefont {Lu}}, \bibinfo
  {author} {\bibfnamefont {S.}~\bibnamefont {Han}}, \bibinfo {author}
  {\bibfnamefont {H.}~\bibnamefont {Xie}}, \bibinfo {author} {\bibfnamefont
  {G.}~\bibnamefont {Du}}, \bibinfo {author} {\bibfnamefont {T.}~\bibnamefont
  {Xiao}}, \ and\ \bibinfo {author} {\bibfnamefont {D.}~\bibnamefont {Zhu}},\
  }\href {\doibase 10.1103/PhysRevLett.117.113901} {\bibfield  {journal}
  {\bibinfo  {journal} {Phys. Rev. Lett.}\ }\textbf {\bibinfo {volume} {117}},\
  \bibinfo {pages} {113901} (\bibinfo {year} {2016})}\BibitemShut {NoStop}%
\bibitem [{\citenamefont {Li}\ \emph {et~al.}(2018)\citenamefont {Li},
  \citenamefont {Medvedev}, \citenamefont {Chapman},\ and\ \citenamefont
  {Shih}}]{Li2018}%
  \BibitemOpen
  \bibfield  {author} {\bibinfo {author} {\bibfnamefont {Z.}~\bibnamefont
  {Li}}, \bibinfo {author} {\bibfnamefont {N.}~\bibnamefont {Medvedev}},
  \bibinfo {author} {\bibfnamefont {H.~N.}\ \bibnamefont {Chapman}}, \ and\
  \bibinfo {author} {\bibfnamefont {Y.}~\bibnamefont {Shih}},\ }\href@noop {}
  {\bibfield  {journal} {\bibinfo  {journal} {Journal of Physics B: Atomic,
  Molecular and Optical Physics}\ }\textbf {\bibinfo {volume} {51}},\ \bibinfo
  {pages} {025503} (\bibinfo {year} {2018})}\BibitemShut {NoStop}%
\bibitem [{\citenamefont {Heeg}\ \emph {et~al.}(2016)\citenamefont {Heeg},
  \citenamefont {Keitel},\ and\ \citenamefont {Evers}}]{Heeg2016}%
  \BibitemOpen
  \bibfield  {author} {\bibinfo {author} {\bibfnamefont {K.~P.}\ \bibnamefont
  {Heeg}}, \bibinfo {author} {\bibfnamefont {C.~H.}\ \bibnamefont {Keitel}}, \
  and\ \bibinfo {author} {\bibfnamefont {J.}~\bibnamefont {Evers}},\
  }\href@noop {} {\bibfield  {journal} {\bibinfo  {journal} {arXiv:1607.04116
  [quant-ph]}\ } (\bibinfo {year} {2016})}\BibitemShut {NoStop}%
\bibitem [{\citenamefont {Bax}\ \emph {et~al.}(1984)\citenamefont {Bax},
  \citenamefont {Hawkins},\ and\ \citenamefont {Szeverenyi}}]{Bax1984}%
  \BibitemOpen
  \bibfield  {author} {\bibinfo {author} {\bibfnamefont {B.~L.}\ \bibnamefont
  {Bax}}, \bibinfo {author} {\bibfnamefont {G.~E.~M.}\ \bibnamefont {Hawkins}},
  \ and\ \bibinfo {author} {\bibfnamefont {N.~M.}\ \bibnamefont {Szeverenyi}},\
  }in\ \href@noop {} {\emph {\bibinfo {booktitle} {Magnetic Resonance}}},\
  \bibinfo {series} {NATO ASI series}, Vol.\ \bibinfo {volume} {124},\ \bibinfo
  {editor} {edited by\ \bibinfo {editor} {\bibfnamefont {L.}~\bibnamefont
  {Petrakis}}\ and\ \bibinfo {editor} {\bibfnamefont {J.~P.}\ \bibnamefont
  {Fraissard}}}\ (\bibinfo  {publisher} {Springer-Verlag},\ \bibinfo {year}
  {1984})\BibitemShut {NoStop}%
\bibitem [{\citenamefont {Cho}(2009)}]{Cho2009}%
  \BibitemOpen
  \bibfield  {author} {\bibinfo {author} {\bibfnamefont {M.}~\bibnamefont
  {Cho}},\ }\href@noop {} {\emph {\bibinfo {title} {Two-Dimensional Optical
  Spectroscopy}}}\ (\bibinfo  {publisher} {CRC Press},\ \bibinfo {year}
  {2009})\BibitemShut {NoStop}%
\bibitem [{\citenamefont {Heeg}\ \emph {et~al.}(2017)\citenamefont {Heeg},
  \citenamefont {Kaldun}, \citenamefont {Strohm}, \citenamefont {Reiser},
  \citenamefont {Ott}, \citenamefont {Subramanian}, \citenamefont {Lentrodt},
  \citenamefont {Haber}, \citenamefont {Wille}, \citenamefont {Goerttler},
  \citenamefont {R{\"u}ffer}, \citenamefont {Keitel}, \citenamefont
  {R{\"o}hlsberger}, \citenamefont {Pfeifer},\ and\ \citenamefont
  {Evers}}]{Heeg2017}%
  \BibitemOpen
  \bibfield  {author} {\bibinfo {author} {\bibfnamefont {K.~P.}\ \bibnamefont
  {Heeg}}, \bibinfo {author} {\bibfnamefont {A.}~\bibnamefont {Kaldun}},
  \bibinfo {author} {\bibfnamefont {C.}~\bibnamefont {Strohm}}, \bibinfo
  {author} {\bibfnamefont {P.}~\bibnamefont {Reiser}}, \bibinfo {author}
  {\bibfnamefont {C.}~\bibnamefont {Ott}}, \bibinfo {author} {\bibfnamefont
  {R.}~\bibnamefont {Subramanian}}, \bibinfo {author} {\bibfnamefont
  {D.}~\bibnamefont {Lentrodt}}, \bibinfo {author} {\bibfnamefont
  {J.}~\bibnamefont {Haber}}, \bibinfo {author} {\bibfnamefont {H.-C.}\
  \bibnamefont {Wille}}, \bibinfo {author} {\bibfnamefont {S.}~\bibnamefont
  {Goerttler}}, \bibinfo {author} {\bibfnamefont {R.}~\bibnamefont
  {R{\"u}ffer}}, \bibinfo {author} {\bibfnamefont {C.~H.}\ \bibnamefont
  {Keitel}}, \bibinfo {author} {\bibfnamefont {R.}~\bibnamefont
  {R{\"o}hlsberger}}, \bibinfo {author} {\bibfnamefont {T.}~\bibnamefont
  {Pfeifer}}, \ and\ \bibinfo {author} {\bibfnamefont {J.}~\bibnamefont
  {Evers}},\ }\href {\doibase 10.1126/science.aan3512} {\bibfield  {journal}
  {\bibinfo  {journal} {Science}\ }\textbf {\bibinfo {volume} {357}},\ \bibinfo
  {pages} {375} (\bibinfo {year} {2017})}\BibitemShut {NoStop}%
\bibitem [{\citenamefont {Haber}\ \emph {et~al.}(2016)\citenamefont {Haber},
  \citenamefont {Schulze}, \citenamefont {Schlage}, \citenamefont {Loetzsch},
  \citenamefont {Bocklage}, \citenamefont {Gurieva}, \citenamefont {Bernhardt},
  \citenamefont {Wille}, \citenamefont {R{\"u}ffer}, \citenamefont {Uschmann},
  \citenamefont {Paulus},\ and\ \citenamefont {R{\"o}hlsberger}}]{Haber2016}%
  \BibitemOpen
  \bibfield  {author} {\bibinfo {author} {\bibfnamefont {J.}~\bibnamefont
  {Haber}}, \bibinfo {author} {\bibfnamefont {K.~S.}\ \bibnamefont {Schulze}},
  \bibinfo {author} {\bibfnamefont {K.}~\bibnamefont {Schlage}}, \bibinfo
  {author} {\bibfnamefont {R.}~\bibnamefont {Loetzsch}}, \bibinfo {author}
  {\bibfnamefont {L.}~\bibnamefont {Bocklage}}, \bibinfo {author}
  {\bibfnamefont {T.}~\bibnamefont {Gurieva}}, \bibinfo {author} {\bibfnamefont
  {H.}~\bibnamefont {Bernhardt}}, \bibinfo {author} {\bibfnamefont {H.-C.}\
  \bibnamefont {Wille}}, \bibinfo {author} {\bibfnamefont {R.}~\bibnamefont
  {R{\"u}ffer}}, \bibinfo {author} {\bibfnamefont {I.}~\bibnamefont
  {Uschmann}}, \bibinfo {author} {\bibfnamefont {G.~G.}\ \bibnamefont
  {Paulus}}, \ and\ \bibinfo {author} {\bibfnamefont {R.}~\bibnamefont
  {R{\"o}hlsberger}},\ }\href@noop {} {\bibfield  {journal} {\bibinfo
  {journal} {Nature Photonics}\ }\textbf {\bibinfo {volume} {10}},\ \bibinfo
  {pages} {445} (\bibinfo {year} {2016})}\BibitemShut {NoStop}%
\bibitem [{\citenamefont {R\"ohlsberger}\ \emph {et~al.}(2012)\citenamefont
  {R\"ohlsberger}, \citenamefont {Wille}, \citenamefont {Schlage},\ and\
  \citenamefont {Sahoo}}]{Roehlsberger2010}%
  \BibitemOpen
  \bibfield  {author} {\bibinfo {author} {\bibfnamefont {R.}~\bibnamefont
  {R\"ohlsberger}}, \bibinfo {author} {\bibfnamefont {H.-C.}\ \bibnamefont
  {Wille}}, \bibinfo {author} {\bibfnamefont {K.}~\bibnamefont {Schlage}}, \
  and\ \bibinfo {author} {\bibfnamefont {B.}~\bibnamefont {Sahoo}},\
  }\href@noop {} {\bibfield  {journal} {\bibinfo  {journal} {Nature}\ }\textbf
  {\bibinfo {volume} {482}},\ \bibinfo {pages} {199} (\bibinfo {year}
  {2012})}\BibitemShut {NoStop}%
\bibitem [{\citenamefont {Heeg}\ \emph {et~al.}(2013)\citenamefont {Heeg},
  \citenamefont {Wille}, \citenamefont {Schlage}, \citenamefont {Guryeva},
  \citenamefont {Schumacher}, \citenamefont {Uschmann}, \citenamefont
  {Schulze}, \citenamefont {Marx}, \citenamefont {K\"ampfer}, \citenamefont
  {Paulus}, \citenamefont {R\"ohlsberger},\ and\ \citenamefont
  {Evers}}]{Heeg2013}%
  \BibitemOpen
  \bibfield  {author} {\bibinfo {author} {\bibfnamefont {K.~P.}\ \bibnamefont
  {Heeg}}, \bibinfo {author} {\bibfnamefont {H.-C.}\ \bibnamefont {Wille}},
  \bibinfo {author} {\bibfnamefont {K.}~\bibnamefont {Schlage}}, \bibinfo
  {author} {\bibfnamefont {T.}~\bibnamefont {Guryeva}}, \bibinfo {author}
  {\bibfnamefont {D.}~\bibnamefont {Schumacher}}, \bibinfo {author}
  {\bibfnamefont {I.}~\bibnamefont {Uschmann}}, \bibinfo {author}
  {\bibfnamefont {K.~S.}\ \bibnamefont {Schulze}}, \bibinfo {author}
  {\bibfnamefont {B.}~\bibnamefont {Marx}}, \bibinfo {author} {\bibfnamefont
  {T.}~\bibnamefont {K\"ampfer}}, \bibinfo {author} {\bibfnamefont {G.~G.}\
  \bibnamefont {Paulus}}, \bibinfo {author} {\bibfnamefont {R.}~\bibnamefont
  {R\"ohlsberger}}, \ and\ \bibinfo {author} {\bibfnamefont {J.}~\bibnamefont
  {Evers}},\ }\href {\doibase 10.1103/PhysRevLett.111.073601} {\bibfield
  {journal} {\bibinfo  {journal} {Phys. Rev. Lett.}\ }\textbf {\bibinfo
  {volume} {111}},\ \bibinfo {pages} {073601} (\bibinfo {year}
  {2013})}\BibitemShut {NoStop}%
\bibitem [{\citenamefont {Haber}\ \emph {et~al.}(2017)\citenamefont {Haber},
  \citenamefont {Kong}, \citenamefont {Strohm}, \citenamefont {Willing},
  \citenamefont {Gollwitzer}, \citenamefont {Bocklage}, \citenamefont
  {R{\"u}ffer}, \citenamefont {P{\'a}lffy},\ and\ \citenamefont
  {R{\"o}hlsberger}}]{Haber2017}%
  \BibitemOpen
  \bibfield  {author} {\bibinfo {author} {\bibfnamefont {J.}~\bibnamefont
  {Haber}}, \bibinfo {author} {\bibfnamefont {X.}~\bibnamefont {Kong}},
  \bibinfo {author} {\bibfnamefont {C.}~\bibnamefont {Strohm}}, \bibinfo
  {author} {\bibfnamefont {S.}~\bibnamefont {Willing}}, \bibinfo {author}
  {\bibfnamefont {J.}~\bibnamefont {Gollwitzer}}, \bibinfo {author}
  {\bibfnamefont {L.}~\bibnamefont {Bocklage}}, \bibinfo {author}
  {\bibfnamefont {R.}~\bibnamefont {R{\"u}ffer}}, \bibinfo {author}
  {\bibfnamefont {A.}~\bibnamefont {P{\'a}lffy}}, \ and\ \bibinfo {author}
  {\bibfnamefont {R.}~\bibnamefont {R{\"o}hlsberger}},\ }\href {\doibase
  10.1038/s41566-017-0013-3} {\bibfield  {journal} {\bibinfo  {journal} {Nature
  Photonics}\ }\textbf {\bibinfo {volume} {11}},\ \bibinfo {pages} {720}
  (\bibinfo {year} {2017})}\BibitemShut {NoStop}%
\bibitem [{\citenamefont {Doniach}(2000)}]{Doniach2000}%
  \BibitemOpen
  \bibfield  {author} {\bibinfo {author} {\bibfnamefont {S.}~\bibnamefont
  {Doniach}},\ }\href {\doibase 10.1107/S0909049500004143} {\bibfield
  {journal} {\bibinfo  {journal} {Journal of Synchrotron Radiation}\ }\textbf
  {\bibinfo {volume} {7}},\ \bibinfo {pages} {116} (\bibinfo {year}
  {2000})}\BibitemShut {NoStop}%
\bibitem [{\citenamefont {Flambaum}(2006)}]{Flambaum2006}%
  \BibitemOpen
  \bibfield  {author} {\bibinfo {author} {\bibfnamefont {V.~V.}\ \bibnamefont
  {Flambaum}},\ }\href@noop {} {\bibfield  {journal} {\bibinfo  {journal} {AIP
  Conference Proceedings}\ }\textbf {\bibinfo {volume} {869}},\ \bibinfo
  {pages} {29} (\bibinfo {year} {2006})}\BibitemShut {NoStop}%
\bibitem [{\citenamefont {Rellergert}\ \emph {et~al.}(2010)\citenamefont
  {Rellergert}, \citenamefont {DeMille}, \citenamefont {Greco}, \citenamefont
  {Hehlen}, \citenamefont {Torgerson},\ and\ \citenamefont
  {Hudson}}]{Rellergert2010}%
  \BibitemOpen
  \bibfield  {author} {\bibinfo {author} {\bibfnamefont {W.~G.}\ \bibnamefont
  {Rellergert}}, \bibinfo {author} {\bibfnamefont {D.}~\bibnamefont {DeMille}},
  \bibinfo {author} {\bibfnamefont {R.~R.}\ \bibnamefont {Greco}}, \bibinfo
  {author} {\bibfnamefont {M.~P.}\ \bibnamefont {Hehlen}}, \bibinfo {author}
  {\bibfnamefont {J.~R.}\ \bibnamefont {Torgerson}}, \ and\ \bibinfo {author}
  {\bibfnamefont {E.~R.}\ \bibnamefont {Hudson}},\ }\href@noop {} {\bibfield
  {journal} {\bibinfo  {journal} {Phys. Rev. Lett.}\ }\textbf {\bibinfo
  {volume} {104}},\ \bibinfo {pages} {200802} (\bibinfo {year}
  {2010})}\BibitemShut {NoStop}%
\bibitem [{\citenamefont {Liao}\ \emph {et~al.}(2011)\citenamefont {Liao},
  \citenamefont {P{\'a}lffy},\ and\ \citenamefont {Keitel}}]{Liao2011}%
  \BibitemOpen
  \bibfield  {author} {\bibinfo {author} {\bibfnamefont {W.-T.}\ \bibnamefont
  {Liao}}, \bibinfo {author} {\bibfnamefont {A.}~\bibnamefont {P{\'a}lffy}}, \
  and\ \bibinfo {author} {\bibfnamefont {C.~H.}\ \bibnamefont {Keitel}},\
  }\href {\doibase https://doi.org/10.1016/j.physletb.2011.09.107} {\bibfield
  {journal} {\bibinfo  {journal} {Physics Letters B}\ }\textbf {\bibinfo
  {volume} {705}},\ \bibinfo {pages} {134 } (\bibinfo {year}
  {2011})}\BibitemShut {NoStop}%
\bibitem [{\citenamefont {Liao}\ \emph {et~al.}(2013)\citenamefont {Liao},
  \citenamefont {P{\'a}lffy},\ and\ \citenamefont {Keitel}}]{Liao2013}%
  \BibitemOpen
  \bibfield  {author} {\bibinfo {author} {\bibfnamefont {W.-T.}\ \bibnamefont
  {Liao}}, \bibinfo {author} {\bibfnamefont {A.}~\bibnamefont {P{\'a}lffy}}, \
  and\ \bibinfo {author} {\bibfnamefont {C.~H.}\ \bibnamefont {Keitel}},\
  }\href@noop {} {\bibfield  {journal} {\bibinfo  {journal} {Phys. Rev. C}\
  }\textbf {\bibinfo {volume} {87}},\ \bibinfo {pages} {054609} (\bibinfo
  {year} {2013})}\BibitemShut {NoStop}%
\bibitem [{\citenamefont {B\"urvenich}\ \emph {et~al.}(2006)\citenamefont
  {B\"urvenich}, \citenamefont {Evers},\ and\ \citenamefont
  {Keitel}}]{Buervenich2006}%
  \BibitemOpen
  \bibfield  {author} {\bibinfo {author} {\bibfnamefont {T.~J.}\ \bibnamefont
  {B\"urvenich}}, \bibinfo {author} {\bibfnamefont {J.}~\bibnamefont {Evers}},
  \ and\ \bibinfo {author} {\bibfnamefont {C.~H.}\ \bibnamefont {Keitel}},\
  }\href {\doibase 10.1103/PhysRevLett.96.142501} {\bibfield  {journal}
  {\bibinfo  {journal} {Phys. Rev. Lett.}\ }\textbf {\bibinfo {volume} {96}},\
  \bibinfo {pages} {142501} (\bibinfo {year} {2006})}\BibitemShut {NoStop}%
\bibitem [{\citenamefont {Safronova}\ \emph {et~al.}(2018)\citenamefont
  {Safronova}, \citenamefont {Budker}, \citenamefont {DeMille}, \citenamefont
  {Kimball}, \citenamefont {Derevianko},\ and\ \citenamefont
  {Clark}}]{Safronova2018}%
  \BibitemOpen
  \bibfield  {author} {\bibinfo {author} {\bibfnamefont {M.~S.}\ \bibnamefont
  {Safronova}}, \bibinfo {author} {\bibfnamefont {D.}~\bibnamefont {Budker}},
  \bibinfo {author} {\bibfnamefont {D.}~\bibnamefont {DeMille}}, \bibinfo
  {author} {\bibfnamefont {D.~F.~J.}\ \bibnamefont {Kimball}}, \bibinfo
  {author} {\bibfnamefont {A.}~\bibnamefont {Derevianko}}, \ and\ \bibinfo
  {author} {\bibfnamefont {C.~W.}\ \bibnamefont {Clark}},\ }\href {\doibase
  10.1103/RevModPhys.90.025008} {\bibfield  {journal} {\bibinfo  {journal}
  {Rev. Mod. Phys.}\ }\textbf {\bibinfo {volume} {90}},\ \bibinfo {pages}
  {025008} (\bibinfo {year} {2018})}\BibitemShut {NoStop}%
\bibitem [{\citenamefont {Pohl}\ \emph {et~al.}(2010)\citenamefont {Pohl} \emph
  {et~al.}}]{Pohl2010}%
  \BibitemOpen
  \bibfield  {author} {\bibinfo {author} {\bibfnamefont {R.}~\bibnamefont
  {Pohl}} \emph {et~al.},\ }\href@noop {} {\bibfield  {journal} {\bibinfo
  {journal} {Nature}\ }\textbf {\bibinfo {volume} {466}},\ \bibinfo {pages}
  {213} (\bibinfo {year} {2010})}\BibitemShut {NoStop}%
\bibitem [{\citenamefont {Antognini}\ \emph {et~al.}(2013)\citenamefont
  {Antognini} \emph {et~al.}}]{Antognini2013}%
  \BibitemOpen
  \bibfield  {author} {\bibinfo {author} {\bibfnamefont {A.}~\bibnamefont
  {Antognini}} \emph {et~al.},\ }\href {\doibase 10.1126/science.1230016}
  {\bibfield  {journal} {\bibinfo  {journal} {Science}\ }\textbf {\bibinfo
  {volume} {339}},\ \bibinfo {pages} {417} (\bibinfo {year}
  {2013})}\BibitemShut {NoStop}%
\bibitem [{\citenamefont {Beyer}\ \emph {et~al.}(2017)\citenamefont {Beyer},
  \citenamefont {Maisenbacher}, \citenamefont {Matveev}, \citenamefont {Pohl},
  \citenamefont {Khabarova}, \citenamefont {Grinin}, \citenamefont {Lamour},
  \citenamefont {Yost}, \citenamefont {H{\"a}nsch}, \citenamefont
  {Kolachevsky},\ and\ \citenamefont {Udem}}]{Beyer2017}%
  \BibitemOpen
  \bibfield  {author} {\bibinfo {author} {\bibfnamefont {A.}~\bibnamefont
  {Beyer}}, \bibinfo {author} {\bibfnamefont {L.}~\bibnamefont {Maisenbacher}},
  \bibinfo {author} {\bibfnamefont {A.}~\bibnamefont {Matveev}}, \bibinfo
  {author} {\bibfnamefont {R.}~\bibnamefont {Pohl}}, \bibinfo {author}
  {\bibfnamefont {K.}~\bibnamefont {Khabarova}}, \bibinfo {author}
  {\bibfnamefont {A.}~\bibnamefont {Grinin}}, \bibinfo {author} {\bibfnamefont
  {T.}~\bibnamefont {Lamour}}, \bibinfo {author} {\bibfnamefont {D.~C.}\
  \bibnamefont {Yost}}, \bibinfo {author} {\bibfnamefont {T.~W.}\ \bibnamefont
  {H{\"a}nsch}}, \bibinfo {author} {\bibfnamefont {N.}~\bibnamefont
  {Kolachevsky}}, \ and\ \bibinfo {author} {\bibfnamefont {T.}~\bibnamefont
  {Udem}},\ }\href {\doibase 10.1126/science.aah6677} {\bibfield  {journal}
  {\bibinfo  {journal} {Science}\ }\textbf {\bibinfo {volume} {358}},\ \bibinfo
  {pages} {79} (\bibinfo {year} {2017})}\BibitemShut {NoStop}%
\bibitem [{\citenamefont {Vogel}\ and\ \citenamefont
  {Quint}(2013)}]{Vogel2013}%
  \BibitemOpen
  \bibfield  {author} {\bibinfo {author} {\bibfnamefont {M.}~\bibnamefont
  {Vogel}}\ and\ \bibinfo {author} {\bibfnamefont {W.}~\bibnamefont {Quint}},\
  }\href {\doibase 10.1002/andp.201300032} {\bibfield  {journal} {\bibinfo
  {journal} {Annalen der Physik}\ }\textbf {\bibinfo {volume} {525}},\ \bibinfo
  {pages} {505} (\bibinfo {year} {2013})}\BibitemShut {NoStop}%
\bibitem [{\citenamefont {Volotka}\ \emph {et~al.}(2013)\citenamefont
  {Volotka}, \citenamefont {Glazov}, \citenamefont {Plunien},\ and\
  \citenamefont {Shabaev}}]{Volotka2013}%
  \BibitemOpen
  \bibfield  {author} {\bibinfo {author} {\bibfnamefont {A.~V.}\ \bibnamefont
  {Volotka}}, \bibinfo {author} {\bibfnamefont {D.~A.}\ \bibnamefont {Glazov}},
  \bibinfo {author} {\bibfnamefont {G.}~\bibnamefont {Plunien}}, \ and\
  \bibinfo {author} {\bibfnamefont {V.~M.}\ \bibnamefont {Shabaev}},\ }\href
  {\doibase 10.1002/andp.201300079} {\bibfield  {journal} {\bibinfo  {journal}
  {Annalen der Physik}\ }\textbf {\bibinfo {volume} {525}},\ \bibinfo {pages}
  {636} (\bibinfo {year} {2013})}\BibitemShut {NoStop}%
\bibitem [{\citenamefont {Crespo L\'opez-Urrutia}(2016)}]{Crespo2016}%
  \BibitemOpen
  \bibfield  {author} {\bibinfo {author} {\bibfnamefont {J.~R.}\ \bibnamefont
  {Crespo L\'opez-Urrutia}},\ }\href@noop {} {\bibfield  {journal} {\bibinfo
  {journal} {Journal of Physics: Conference Series}\ }\textbf {\bibinfo
  {volume} {723}},\ \bibinfo {pages} {012052} (\bibinfo {year}
  {2016})}\BibitemShut {NoStop}%
\bibitem [{\citenamefont {Epp}\ \emph {et~al.}(2010)\citenamefont {Epp},
  \citenamefont {L\'opez-Urrutia}, \citenamefont {Simon}, \citenamefont
  {Baumann}, \citenamefont {Brenner}, \citenamefont {Ginzel}, \citenamefont
  {Guerassimova}, \citenamefont {M\"ackel}, \citenamefont {Mokler},
  \citenamefont {Schmitt}, \citenamefont {Tawara},\ and\ \citenamefont
  {Ullrich}}]{Epp2011}%
  \BibitemOpen
  \bibfield  {author} {\bibinfo {author} {\bibfnamefont {S.~W.}\ \bibnamefont
  {Epp}}, \bibinfo {author} {\bibfnamefont {J.~R.~C.}\ \bibnamefont
  {L\'opez-Urrutia}}, \bibinfo {author} {\bibfnamefont {M.~C.}\ \bibnamefont
  {Simon}}, \bibinfo {author} {\bibfnamefont {T.}~\bibnamefont {Baumann}},
  \bibinfo {author} {\bibfnamefont {G.}~\bibnamefont {Brenner}}, \bibinfo
  {author} {\bibfnamefont {R.}~\bibnamefont {Ginzel}}, \bibinfo {author}
  {\bibfnamefont {N.}~\bibnamefont {Guerassimova}}, \bibinfo {author}
  {\bibfnamefont {V.}~\bibnamefont {M\"ackel}}, \bibinfo {author}
  {\bibfnamefont {P.~H.}\ \bibnamefont {Mokler}}, \bibinfo {author}
  {\bibfnamefont {B.~L.}\ \bibnamefont {Schmitt}}, \bibinfo {author}
  {\bibfnamefont {H.}~\bibnamefont {Tawara}}, \ and\ \bibinfo {author}
  {\bibfnamefont {J.}~\bibnamefont {Ullrich}},\ }\href@noop {} {\bibfield
  {journal} {\bibinfo  {journal} {Journal of Physics B: Atomic, Molecular and
  Optical Physics}\ }\textbf {\bibinfo {volume} {43}},\ \bibinfo {pages}
  {194008} (\bibinfo {year} {2010})}\BibitemShut {NoStop}%
\bibitem [{\citenamefont {N\"ortersh\"auser}\ and\ \citenamefont
  {Geppert}(2014)}]{Noerterhaeuser2014}%
  \BibitemOpen
  \bibfield  {author} {\bibinfo {author} {\bibfnamefont {W.}~\bibnamefont
  {N\"ortersh\"auser}}\ and\ \bibinfo {author} {\bibfnamefont {C.}~\bibnamefont
  {Geppert}},\ }in\ \href@noop {} {\emph {\bibinfo {booktitle} {The Euroschool
  on Exotic Beams, Vol. IV}}},\ \bibinfo {series} {Lecture Notes in Physics},
  Vol.\ \bibinfo {volume} {879}\ (\bibinfo  {publisher} {Springer Verlag
  Berlin-Heidelberg},\ \bibinfo {year} {2014})\BibitemShut {NoStop}%
\bibitem [{\citenamefont {Campbell}\ \emph {et~al.}(2016)\citenamefont
  {Campbell}, \citenamefont {Moore},\ and\ \citenamefont
  {Pearson}}]{Campbell2016}%
  \BibitemOpen
  \bibfield  {author} {\bibinfo {author} {\bibfnamefont {P.}~\bibnamefont
  {Campbell}}, \bibinfo {author} {\bibfnamefont {I.}~\bibnamefont {Moore}}, \
  and\ \bibinfo {author} {\bibfnamefont {M.}~\bibnamefont {Pearson}},\ }\href
  {\doibase https://doi.org/10.1016/j.ppnp.2015.09.003} {\bibfield  {journal}
  {\bibinfo  {journal} {Progress in Particle and Nuclear Physics}\ }\textbf
  {\bibinfo {volume} {86}},\ \bibinfo {pages} {127 } (\bibinfo {year}
  {2016})}\BibitemShut {NoStop}%
\bibitem [{\citenamefont {Vogel}\ and\ \citenamefont
  {St{\"o}hlker}(2018)}]{GSI2018}%
  \BibitemOpen
  \bibfield  {author} {\bibinfo {author} {\bibfnamefont {M.}~\bibnamefont
  {Vogel}}\ and\ \bibinfo {author} {\bibfnamefont {T.}~\bibnamefont
  {St{\"o}hlker}},\ }\href@noop {} {} (\bibinfo {year} {2018}),\ \bibinfo
  {note} {this figure was provided by Manuel Vogel and Thomas St\"ohlker,
  Gesellschaft f\"ur Schwerionenforschung GSI, Planckstra\ss e 1, 64291
  Darmstadt, Germany}\BibitemShut {NoStop}%
\bibitem [{\citenamefont {Fee}\ \emph {et~al.}(1993)\citenamefont {Fee},
  \citenamefont {Mills}, \citenamefont {Chu}, \citenamefont {Shaw},
  \citenamefont {Danzmann}, \citenamefont {Chichester},\ and\ \citenamefont
  {Zuckerman}}]{Fee1993}%
  \BibitemOpen
  \bibfield  {author} {\bibinfo {author} {\bibfnamefont {M.~S.}\ \bibnamefont
  {Fee}}, \bibinfo {author} {\bibfnamefont {A.~P.}\ \bibnamefont {Mills}},
  \bibinfo {author} {\bibfnamefont {S.}~\bibnamefont {Chu}}, \bibinfo {author}
  {\bibfnamefont {E.~D.}\ \bibnamefont {Shaw}}, \bibinfo {author}
  {\bibfnamefont {K.}~\bibnamefont {Danzmann}}, \bibinfo {author}
  {\bibfnamefont {R.~J.}\ \bibnamefont {Chichester}}, \ and\ \bibinfo {author}
  {\bibfnamefont {D.~M.}\ \bibnamefont {Zuckerman}},\ }\href {\doibase
  10.1103/PhysRevLett.70.1397} {\bibfield  {journal} {\bibinfo  {journal}
  {Phys. Rev. Lett.}\ }\textbf {\bibinfo {volume} {70}},\ \bibinfo {pages}
  {1397} (\bibinfo {year} {1993})}\BibitemShut {NoStop}%
\bibitem [{\citenamefont {Beier}\ \emph {et~al.}(2005)\citenamefont {Beier},
  \citenamefont {Dahl}, \citenamefont {Kluge}, \citenamefont {Kozhuhaorv},\
  and\ \citenamefont {Quint}}]{Beier2005}%
  \BibitemOpen
  \bibfield  {author} {\bibinfo {author} {\bibfnamefont {T.}~\bibnamefont
  {Beier}}, \bibinfo {author} {\bibfnamefont {L.}~\bibnamefont {Dahl}},
  \bibinfo {author} {\bibfnamefont {H.-J.}\ \bibnamefont {Kluge}}, \bibinfo
  {author} {\bibfnamefont {C.}~\bibnamefont {Kozhuhaorv}}, \ and\ \bibinfo
  {author} {\bibfnamefont {W.}~\bibnamefont {Quint}},\ }\href {\doibase
  https://doi.org/10.1016/j.nimb.2005.03.193} {\bibfield  {journal} {\bibinfo
  {journal} {Nuclear Instruments and Methods in Physics Research Section B}\
  }\textbf {\bibinfo {volume} {235}},\ \bibinfo {pages} {473 } (\bibinfo {year}
  {2005})}\BibitemShut {NoStop}%
\bibitem [{\citenamefont {Odom}\ \emph {et~al.}(2006)\citenamefont {Odom},
  \citenamefont {Hanneke}, \citenamefont {D'Urso},\ and\ \citenamefont
  {Gabrielse}}]{Odom2006}%
  \BibitemOpen
  \bibfield  {author} {\bibinfo {author} {\bibfnamefont {B.}~\bibnamefont
  {Odom}}, \bibinfo {author} {\bibfnamefont {D.}~\bibnamefont {Hanneke}},
  \bibinfo {author} {\bibfnamefont {B.}~\bibnamefont {D'Urso}}, \ and\ \bibinfo
  {author} {\bibfnamefont {G.}~\bibnamefont {Gabrielse}},\ }\href {\doibase
  10.1103/PhysRevLett.97.030801} {\bibfield  {journal} {\bibinfo  {journal}
  {Phys. Rev. Lett.}\ }\textbf {\bibinfo {volume} {97}},\ \bibinfo {pages}
  {030801} (\bibinfo {year} {2006})}\BibitemShut {NoStop}%
\bibitem [{\citenamefont {Trassinelli}\ \emph {et~al.}(2009)\citenamefont
  {Trassinelli} \emph {et~al.}}]{Trassinelli2009}%
  \BibitemOpen
  \bibfield  {author} {\bibinfo {author} {\bibfnamefont {M.}~\bibnamefont
  {Trassinelli}} \emph {et~al.},\ }\href@noop {} {\bibfield  {journal}
  {\bibinfo  {journal} {Europhysics Letters}\ }\textbf {\bibinfo {volume}
  {87}},\ \bibinfo {pages} {63001} (\bibinfo {year} {2009})}\BibitemShut
  {NoStop}%
\bibitem [{\citenamefont {Matveev}\ \emph {et~al.}(2013)\citenamefont {Matveev}
  \emph {et~al.}}]{Matveev2013}%
  \BibitemOpen
  \bibfield  {author} {\bibinfo {author} {\bibfnamefont {A.}~\bibnamefont
  {Matveev}} \emph {et~al.},\ }\href {\doibase 10.1103/PhysRevLett.110.230801}
  {\bibfield  {journal} {\bibinfo  {journal} {Phys. Rev. Lett.}\ }\textbf
  {\bibinfo {volume} {110}},\ \bibinfo {pages} {230801} (\bibinfo {year}
  {2013})}\BibitemShut {NoStop}%
\bibitem [{\citenamefont {Ullmann}\ \emph {et~al.}(2017)\citenamefont {Ullmann}
  \emph {et~al.}}]{Ullmann2017}%
  \BibitemOpen
  \bibfield  {author} {\bibinfo {author} {\bibfnamefont {J.}~\bibnamefont
  {Ullmann}} \emph {et~al.},\ }\href@noop {} {\bibfield  {journal} {\bibinfo
  {journal} {Nature Communications}\ }\textbf {\bibinfo {volume} {8}},\
  \bibinfo {pages} {15484} (\bibinfo {year} {2017})}\BibitemShut {NoStop}%
\bibitem [{\citenamefont {Vogel}(2018)}]{Vogel2018}%
  \BibitemOpen
  \bibfield  {author} {\bibinfo {author} {\bibfnamefont {M.}~\bibnamefont
  {Vogel}},\ }in\ \href {\doibase 10.1007/978-3-319-76264-7_22} {\emph
  {\bibinfo {booktitle} {Particle Confinement in Penning Traps: An
  Introduction}}}\ (\bibinfo  {publisher} {Springer International Publishing},\
  \bibinfo {year} {2018})\ pp.\ \bibinfo {pages} {335--345}\BibitemShut
  {NoStop}%
\bibitem [{\citenamefont {Glauber}(1963)}]{Glauber1963}%
  \BibitemOpen
  \bibfield  {author} {\bibinfo {author} {\bibfnamefont {R.~J.}\ \bibnamefont
  {Glauber}},\ }\href {\doibase 10.1103/PhysRev.130.2529} {\bibfield  {journal}
  {\bibinfo  {journal} {Phys. Rev.}\ }\textbf {\bibinfo {volume} {130}},\
  \bibinfo {pages} {2529} (\bibinfo {year} {1963})}\BibitemShut {NoStop}%
\bibitem [{\citenamefont {Singer}\ \emph {et~al.}(2013)\citenamefont {Singer},
  \citenamefont {Lorenz}, \citenamefont {Sorgenfrei}, \citenamefont
  {Gerasimova}, \citenamefont {Gulden}, \citenamefont {Yefanov}, \citenamefont
  {Kurta}, \citenamefont {Shabalin}, \citenamefont {Dronyak}, \citenamefont
  {Treusch}, \citenamefont {Kocharyan}, \citenamefont {Weckert}, \citenamefont
  {Wurth},\ and\ \citenamefont {Vartanyants}}]{Singer2013}%
  \BibitemOpen
  \bibfield  {author} {\bibinfo {author} {\bibfnamefont {A.}~\bibnamefont
  {Singer}}, \bibinfo {author} {\bibfnamefont {U.}~\bibnamefont {Lorenz}},
  \bibinfo {author} {\bibfnamefont {F.}~\bibnamefont {Sorgenfrei}}, \bibinfo
  {author} {\bibfnamefont {N.}~\bibnamefont {Gerasimova}}, \bibinfo {author}
  {\bibfnamefont {J.}~\bibnamefont {Gulden}}, \bibinfo {author} {\bibfnamefont
  {O.~M.}\ \bibnamefont {Yefanov}}, \bibinfo {author} {\bibfnamefont {R.~P.}\
  \bibnamefont {Kurta}}, \bibinfo {author} {\bibfnamefont {A.}~\bibnamefont
  {Shabalin}}, \bibinfo {author} {\bibfnamefont {R.}~\bibnamefont {Dronyak}},
  \bibinfo {author} {\bibfnamefont {R.}~\bibnamefont {Treusch}}, \bibinfo
  {author} {\bibfnamefont {V.}~\bibnamefont {Kocharyan}}, \bibinfo {author}
  {\bibfnamefont {E.}~\bibnamefont {Weckert}}, \bibinfo {author} {\bibfnamefont
  {W.}~\bibnamefont {Wurth}}, \ and\ \bibinfo {author} {\bibfnamefont {I.~A.}\
  \bibnamefont {Vartanyants}},\ }\href {\doibase
  10.1103/PhysRevLett.111.034802} {\bibfield  {journal} {\bibinfo  {journal}
  {Phys. Rev. Lett.}\ }\textbf {\bibinfo {volume} {111}},\ \bibinfo {pages}
  {034802} (\bibinfo {year} {2013})}\BibitemShut {NoStop}%
\bibitem [{\citenamefont {Chumakov}\ \emph {et~al.}(2018)\citenamefont
  {Chumakov}, \citenamefont {Baron}, \citenamefont {Sergueev}, \citenamefont
  {Strohm}, \citenamefont {Leupold}, \citenamefont {Shvyd'ko}, \citenamefont
  {Smirnov}, \citenamefont {R{\"u}ffer}, \citenamefont {Inubushi},
  \citenamefont {Yabashi}, \citenamefont {Tono}, \citenamefont {Kudo},\ and\
  \citenamefont {Ishikawa}}]{Chumakov2018}%
  \BibitemOpen
  \bibfield  {author} {\bibinfo {author} {\bibfnamefont {A.~I.}\ \bibnamefont
  {Chumakov}}, \bibinfo {author} {\bibfnamefont {A.~Q.~R.}\ \bibnamefont
  {Baron}}, \bibinfo {author} {\bibfnamefont {I.}~\bibnamefont {Sergueev}},
  \bibinfo {author} {\bibfnamefont {C.}~\bibnamefont {Strohm}}, \bibinfo
  {author} {\bibfnamefont {O.}~\bibnamefont {Leupold}}, \bibinfo {author}
  {\bibfnamefont {Y.}~\bibnamefont {Shvyd'ko}}, \bibinfo {author}
  {\bibfnamefont {G.~V.}\ \bibnamefont {Smirnov}}, \bibinfo {author}
  {\bibfnamefont {R.}~\bibnamefont {R{\"u}ffer}}, \bibinfo {author}
  {\bibfnamefont {Y.}~\bibnamefont {Inubushi}}, \bibinfo {author}
  {\bibfnamefont {M.}~\bibnamefont {Yabashi}}, \bibinfo {author} {\bibfnamefont
  {K.}~\bibnamefont {Tono}}, \bibinfo {author} {\bibfnamefont {T.}~\bibnamefont
  {Kudo}}, \ and\ \bibinfo {author} {\bibfnamefont {T.}~\bibnamefont
  {Ishikawa}},\ }\href {\doibase 10.1038/s41567-017-0001-z} {\bibfield
  {journal} {\bibinfo  {journal} {Nature Physics}\ }\textbf {\bibinfo {volume}
  {14}},\ \bibinfo {pages} {261} (\bibinfo {year} {2018})}\BibitemShut
  {NoStop}%
\bibitem [{\citenamefont {Walker}\ and\ \citenamefont
  {Dracoulis}(1999)}]{Walker1999}%
  \BibitemOpen
  \bibfield  {author} {\bibinfo {author} {\bibfnamefont {P.}~\bibnamefont
  {Walker}}\ and\ \bibinfo {author} {\bibfnamefont {G.}~\bibnamefont
  {Dracoulis}},\ }\href@noop {} {\bibfield  {journal} {\bibinfo  {journal}
  {Nature}\ }\textbf {\bibinfo {volume} {399}},\ \bibinfo {pages} {35}
  (\bibinfo {year} {1999})}\BibitemShut {NoStop}%
\bibitem [{\citenamefont {Ludlow}\ \emph {et~al.}(2015)\citenamefont {Ludlow},
  \citenamefont {Boyd}, \citenamefont {Ye}, \citenamefont {Peik},\ and\
  \citenamefont {Schmidt}}]{Ludlow2015}%
  \BibitemOpen
  \bibfield  {author} {\bibinfo {author} {\bibfnamefont {A.~D.}\ \bibnamefont
  {Ludlow}}, \bibinfo {author} {\bibfnamefont {M.~M.}\ \bibnamefont {Boyd}},
  \bibinfo {author} {\bibfnamefont {J.}~\bibnamefont {Ye}}, \bibinfo {author}
  {\bibfnamefont {E.}~\bibnamefont {Peik}}, \ and\ \bibinfo {author}
  {\bibfnamefont {P.~O.}\ \bibnamefont {Schmidt}},\ }\href {\doibase
  10.1103/RevModPhys.87.637} {\bibfield  {journal} {\bibinfo  {journal} {Rev.
  Mod. Phys.}\ }\textbf {\bibinfo {volume} {87}},\ \bibinfo {pages} {637}
  (\bibinfo {year} {2015})}\BibitemShut {NoStop}%
\bibitem [{\citenamefont {Campbell}\ \emph {et~al.}(2012)\citenamefont
  {Campbell}, \citenamefont {Radnaev}, \citenamefont {Kuzmich}, \citenamefont
  {Dzuba}, \citenamefont {Flambaum},\ and\ \citenamefont
  {Derevianko}}]{Campbell2012}%
  \BibitemOpen
  \bibfield  {author} {\bibinfo {author} {\bibfnamefont {C.~J.}\ \bibnamefont
  {Campbell}}, \bibinfo {author} {\bibfnamefont {A.~G.}\ \bibnamefont
  {Radnaev}}, \bibinfo {author} {\bibfnamefont {A.}~\bibnamefont {Kuzmich}},
  \bibinfo {author} {\bibfnamefont {V.~A.}\ \bibnamefont {Dzuba}}, \bibinfo
  {author} {\bibfnamefont {V.~V.}\ \bibnamefont {Flambaum}}, \ and\ \bibinfo
  {author} {\bibfnamefont {A.}~\bibnamefont {Derevianko}},\ }\href {\doibase
  10.1103/PhysRevLett.108.120802} {\bibfield  {journal} {\bibinfo  {journal}
  {Phys. Rev. Lett.}\ }\textbf {\bibinfo {volume} {108}},\ \bibinfo {pages}
  {120802} (\bibinfo {year} {2012})}\BibitemShut {NoStop}%
\bibitem [{\citenamefont {Benko}\ \emph {et~al.}(2014)\citenamefont {Benko},
  \citenamefont {Allison}, \citenamefont {Cing{\"o}z}, \citenamefont {Hua},
  \citenamefont {Labaye}, \citenamefont {Yost},\ and\ \citenamefont
  {Ye}}]{Benko2014}%
  \BibitemOpen
  \bibfield  {author} {\bibinfo {author} {\bibfnamefont {C.}~\bibnamefont
  {Benko}}, \bibinfo {author} {\bibfnamefont {T.~K.}\ \bibnamefont {Allison}},
  \bibinfo {author} {\bibfnamefont {A.}~\bibnamefont {Cing{\"o}z}}, \bibinfo
  {author} {\bibfnamefont {L.}~\bibnamefont {Hua}}, \bibinfo {author}
  {\bibfnamefont {F.}~\bibnamefont {Labaye}}, \bibinfo {author} {\bibfnamefont
  {D.~C.}\ \bibnamefont {Yost}}, \ and\ \bibinfo {author} {\bibfnamefont
  {J.}~\bibnamefont {Ye}},\ }\href@noop {} {\bibfield  {journal} {\bibinfo
  {journal} {Nature Photonics}\ }\textbf {\bibinfo {volume} {8}},\ \bibinfo
  {pages} {530} (\bibinfo {year} {2014})}\BibitemShut {NoStop}%
\bibitem [{\citenamefont {Cavaletto}\ \emph {et~al.}(2014)\citenamefont
  {Cavaletto}, \citenamefont {Harman}, \citenamefont {Ott}, \citenamefont
  {Buth}, \citenamefont {Pfeifer},\ and\ \citenamefont
  {Keitel}}]{Cavaletto2014}%
  \BibitemOpen
  \bibfield  {author} {\bibinfo {author} {\bibfnamefont {S.~M.}\ \bibnamefont
  {Cavaletto}}, \bibinfo {author} {\bibfnamefont {Z.}~\bibnamefont {Harman}},
  \bibinfo {author} {\bibfnamefont {C.}~\bibnamefont {Ott}}, \bibinfo {author}
  {\bibfnamefont {C.}~\bibnamefont {Buth}}, \bibinfo {author} {\bibfnamefont
  {T.}~\bibnamefont {Pfeifer}}, \ and\ \bibinfo {author} {\bibfnamefont
  {C.~H.}\ \bibnamefont {Keitel}},\ }\href@noop {} {\bibfield  {journal}
  {\bibinfo  {journal} {Nature Photonics}\ }\textbf {\bibinfo {volume} {8}},\
  \bibinfo {pages} {520} (\bibinfo {year} {2014})}\BibitemShut {NoStop}%
\bibitem [{\citenamefont {Allaria}\ \emph {et~al.}(2012)\citenamefont {Allaria}
  \emph {et~al.}}]{Allaria2012}%
  \BibitemOpen
  \bibfield  {author} {\bibinfo {author} {\bibfnamefont {E.}~\bibnamefont
  {Allaria}} \emph {et~al.},\ }\href@noop {} {\bibfield  {journal} {\bibinfo
  {journal} {Nature Photonics}\ }\textbf {\bibinfo {volume} {6}},\ \bibinfo
  {pages} {699} (\bibinfo {year} {2012})}\BibitemShut {NoStop}%
\bibitem [{\citenamefont {Ruehl}\ \emph {et~al.}(2010)\citenamefont {Ruehl},
  \citenamefont {Marcinkevicius}, \citenamefont {Fermann},\ and\ \citenamefont
  {Hartl}}]{Ruehl2010}%
  \BibitemOpen
  \bibfield  {author} {\bibinfo {author} {\bibfnamefont {A.}~\bibnamefont
  {Ruehl}}, \bibinfo {author} {\bibfnamefont {A.}~\bibnamefont
  {Marcinkevicius}}, \bibinfo {author} {\bibfnamefont {M.~E.}\ \bibnamefont
  {Fermann}}, \ and\ \bibinfo {author} {\bibfnamefont {I.}~\bibnamefont
  {Hartl}},\ }\href {\doibase 10.1364/OL.35.003015} {\bibfield  {journal}
  {\bibinfo  {journal} {Opt. Lett.}\ }\textbf {\bibinfo {volume} {35}},\
  \bibinfo {pages} {3015} (\bibinfo {year} {2010})}\BibitemShut {NoStop}%
\bibitem [{\citenamefont {Emaury}\ \emph {et~al.}(2015)\citenamefont {Emaury},
  \citenamefont {Diebold}, \citenamefont {Klenner}, \citenamefont {Saraceno},
  \citenamefont {Schilt}, \citenamefont {S\"{u}dmeyer},\ and\ \citenamefont
  {Keller}}]{Emaury2015}%
  \BibitemOpen
  \bibfield  {author} {\bibinfo {author} {\bibfnamefont {F.}~\bibnamefont
  {Emaury}}, \bibinfo {author} {\bibfnamefont {A.}~\bibnamefont {Diebold}},
  \bibinfo {author} {\bibfnamefont {A.}~\bibnamefont {Klenner}}, \bibinfo
  {author} {\bibfnamefont {C.}~\bibnamefont {Saraceno}}, \bibinfo {author}
  {\bibfnamefont {S.}~\bibnamefont {Schilt}}, \bibinfo {author} {\bibfnamefont
  {T.}~\bibnamefont {S\"{u}dmeyer}}, \ and\ \bibinfo {author} {\bibfnamefont
  {U.}~\bibnamefont {Keller}},\ }\href {\doibase 10.1364/OE.23.021836}
  {\bibfield  {journal} {\bibinfo  {journal} {Opt. Express}\ }\textbf {\bibinfo
  {volume} {23}},\ \bibinfo {pages} {21836} (\bibinfo {year}
  {2015})}\BibitemShut {NoStop}%
\bibitem [{\citenamefont {Li}\ \emph {et~al.}(2016)\citenamefont {Li},
  \citenamefont {Reber}, \citenamefont {Corder}, \citenamefont {Chen},
  \citenamefont {Zhao},\ and\ \citenamefont {Allison}}]{Li2016}%
  \BibitemOpen
  \bibfield  {author} {\bibinfo {author} {\bibfnamefont {X.}~\bibnamefont
  {Li}}, \bibinfo {author} {\bibfnamefont {M.~A.~R.}\ \bibnamefont {Reber}},
  \bibinfo {author} {\bibfnamefont {C.}~\bibnamefont {Corder}}, \bibinfo
  {author} {\bibfnamefont {Y.}~\bibnamefont {Chen}}, \bibinfo {author}
  {\bibfnamefont {P.}~\bibnamefont {Zhao}}, \ and\ \bibinfo {author}
  {\bibfnamefont {T.~K.}\ \bibnamefont {Allison}},\ }\href {\doibase
  10.1063/1.4962867} {\bibfield  {journal} {\bibinfo  {journal} {Review of
  Scientific Instruments}\ }\textbf {\bibinfo {volume} {87}},\ \bibinfo {pages}
  {093114} (\bibinfo {year} {2016})}\BibitemShut {NoStop}%
\bibitem [{\citenamefont {Luo}\ \emph {et~al.}(2018)\citenamefont {Luo},
  \citenamefont {Liu}, \citenamefont {Gu}, \citenamefont {Wang}, \citenamefont
  {Zhu}, \citenamefont {Zhang}, \citenamefont {Deng}, \citenamefont {Zhou},
  \citenamefont {Li},\ and\ \citenamefont {Zeng}}]{Luo2018}%
  \BibitemOpen
  \bibfield  {author} {\bibinfo {author} {\bibfnamefont {D.}~\bibnamefont
  {Luo}}, \bibinfo {author} {\bibfnamefont {Y.}~\bibnamefont {Liu}}, \bibinfo
  {author} {\bibfnamefont {C.}~\bibnamefont {Gu}}, \bibinfo {author}
  {\bibfnamefont {C.}~\bibnamefont {Wang}}, \bibinfo {author} {\bibfnamefont
  {Z.}~\bibnamefont {Zhu}}, \bibinfo {author} {\bibfnamefont {W.}~\bibnamefont
  {Zhang}}, \bibinfo {author} {\bibfnamefont {Z.}~\bibnamefont {Deng}},
  \bibinfo {author} {\bibfnamefont {L.}~\bibnamefont {Zhou}}, \bibinfo {author}
  {\bibfnamefont {W.}~\bibnamefont {Li}}, \ and\ \bibinfo {author}
  {\bibfnamefont {H.}~\bibnamefont {Zeng}},\ }\href {\doibase
  10.1063/1.5012100} {\bibfield  {journal} {\bibinfo  {journal} {Applied
  Physics Letters}\ }\textbf {\bibinfo {volume} {112}},\ \bibinfo {pages}
  {061106} (\bibinfo {year} {2018})}\BibitemShut {NoStop}%
\end{thebibliography}
